\documentclass[12pt]{article}
\usepackage[latin9]{inputenc}
\usepackage{geometry}
\usepackage{float}
\usepackage{mathtools}
\usepackage{amsmath}
\usepackage{amsthm}
\usepackage{amssymb}
\usepackage{graphicx}
\usepackage{setspace}

\usepackage{psfrag}
\usepackage[authoryear]{natbib}
\usepackage{rotating}
\usepackage{rotfloat}
\usepackage{bbold}
\makeatletter

\providecommand{\tabularnewline}{\\}
\floatstyle{ruled}
\newfloat{algorithm}{tbp}{loa}
\providecommand{\algorithmname}{Algorithm}
\floatname{algorithm}{\protect\algorithmname}

  \theoremstyle{plain}
  \newtheorem{lem}{\protect\lemmaname}
  \theoremstyle{plain}
  \newtheorem{assumption}{\protect\assumptionname}

\usepackage{latexsym}
\usepackage{amsthm}\usepackage{amsfonts}\usepackage{graphicx}\usepackage{epsfig}
\usepackage{bm}
\newcommand*{\patchAmsMathEnvironmentForLineno}[1]{%
      \expandafter\let\csname old#1\expandafter\endcsname\csname #1\endcsname
      \expandafter\let\csname oldend#1\expandafter\endcsname\csname end#1\endcsname
      \renewenvironment{#1}%
         {\linenomath\csname old#1\endcsname}%
         {\csname oldend#1\endcsname\endlinenomath}}%
    \newcommand*{\patchBothAmsMathEnvironmentsForLineno}[1]{%
      \patchAmsMathEnvironmentForLineno{#1}%
      \patchAmsMathEnvironmentForLineno{#1*}}%
    \AtBeginDocument{%
    \patchBothAmsMathEnvironmentsForLineno{equation}%
    \patchBothAmsMathEnvironmentsForLineno{align}%
    \patchBothAmsMathEnvironmentsForLineno{flalign}%
    \patchBothAmsMathEnvironmentsForLineno{alignat}%
    \patchBothAmsMathEnvironmentsForLineno{gather}%
    \patchBothAmsMathEnvironmentsForLineno{multline}%
    }
\usepackage[mathlines,displaymath]{lineno}

\include{def}
\def\dispmuskip{\thinmuskip= 3mu plus 0mu minus 2mu \medmuskip=  4mu plus 2mu minus 2mu \thickmuskip=5mu plus 5mu minus 2mu}
\def\textmuskip{\thinmuskip= 0mu                    \medmuskip=  1mu plus 1mu minus 1mu \thickmuskip=2mu plus 3mu minus 1mu}
\def\beq{\dispmuskip\begin{equation}}    \def\eeq{\end{equation}\textmuskip}
\def\beqn{\dispmuskip\begin{displaymath}}\def\eeqn{\end{displaymath}\textmuskip}
\def\bea{\dispmuskip\begin{eqnarray}}    \def\eea{\end{eqnarray}\textmuskip}
\def\bean{\dispmuskip\begin{eqnarray*}}  \def\eean{\end{eqnarray*}\textmuskip}

\newcommand{\wt}{\widetilde}
\newcommand{\blind}{0}

\def\E{{\mathbb E}}                         

\def\M{{\cal M}}
\def\N{{\cal N}}

\usepackage[mathlines,displaymath]{lineno}
\usepackage{authblk}
\makeatother

\providecommand{\assumptionname}{Assumption}
\providecommand{\lemmaname}{Lemma}

\renewcommand{\eqref}[1]{Eq.~(\ref{#1})}
\newcommand{\secref}[1]{Section~\ref{#1}}

\newcommand{\algref}[1]{Alg.~\ref{#1}}

\newcommand{\tabref}[1]{Table~\ref{#1}}
\newcommand{\figref}[1]{Fig.~\ref{#1}}

\newcommand{\comment}[1]{}

\usepackage{xr}
\makeatletter
\newcommand*{\addFileDependency}[1]{
  \typeout{(#1)}
  \@addtofilelist{#1}
  \IfFileExists{#1}{}{\typeout{No file #1.}}
}
\makeatother

\newcommand*{\myexternaldocument}[1]{
    \externaldocument{#1}
    \addFileDependency{#1.tex}
    \addFileDependency{#1.aux}
}

\myexternaldocument{reply2refereesEcoSta}
\myexternaldocument{Supplementary}

\def\E{\mathrm{E}}

\def\({\Big ( }
\def\){\Big )}

\def\bs{\boldsymbol}

\begin{document}
\if0\blind
{
\title{\bf Flexible and Robust Particle Tempering for State Space Models}	
\author[2,4]{David Gunawan}
\author[1,4]{Robert Kohn\footnote{Corresponding Author: Robert Kohn, School of Economics, UNSW Business School, University of New South Wales, Email: {r.kohn@unsw.edu.au}}}
\author[3,4]{Minh Ngoc Tran}
\affil[1]{School of Economics, UNSW Business School, University of New South Wales}
\affil[2] {School of Mathematics and Applied Statistics, University of Wollongong}
\affil[3]{Discipline of Business Analytics, University of Sydney}
\affil[4]{Australian Center of Excellence for Mathematical and Statistical Frontiers}

\renewcommand\Authands{ and }
  \maketitle
} \fi

\if1\blind
{
     \title{\bf Flexible and Robust Particle Density Tempering for State Space Models}		
\author{}
\maketitle
} \fi
\bigskip

\maketitle






\begin{abstract}
Density tempering (also called density annealing) 
is a sequential Monte Carlo  approach to Bayesian inference for general state models; it is an alternative to Markov chain Monte Carlo. When applied to state space models, it moves a collection of parameters and latent states (which are called particles) through a number of stages, with each stage having its own target distribution. The particles are initially generated from a distribution that
is easy to sample from, e.g. the prior;
the target at the final stage is the posterior distribution. 
Tempering is usually  carried out either in batch mode,
involving all the data at
each stage, or sequentially  with
observations added at each stage, which is called data tempering. 
Our paper proposes efficient Markov moves for generating
 the parameters and states for each stage of  particle based density tempering. This allows the proposed SMC methods to increase (scale up) the number of parameters and states that can be handled. Most of the current literature uses a
 pseudo-marginal Markov move step with the states \lq\lq integrated out,\rq\rq 
  and the parameters generated by a random walk proposal;
although this strategy is general, it is very 
inefficient when the states or parameters are high dimensional.
We also build on the work of \citet{dufays2016evolutionary} and make  data tempering more robust
to outliers and structural changes for models with intractable likelihoods by adding
 batch tempering at each stage.
The performance of the proposed methods is evaluated 
 using univariate stochastic volatility models with outliers and structural breaks and
high dimensional factor stochastic volatility
models having both many parameters and many latent states.
\end{abstract}

Keywords: Factor stochastic volatility model; Hamiltonian Monte Carlo; Outliers; Structural change; Particle Markov chain Monte Carlo; Diffusion process


\section{Introduction \label{sec:Introduction}}

%
%

Joint inference over both the parameters and latent states in nonlinear and non-Gaussian state space models
is challenging because the likelihood is usually an intractable integral over the
latent states.  In a seminal contribution, \citet{Andrieu:2010} propose two batch particle
Markov chain Monte Carlo (PMCMC) methods for state space models. The
first method is the particle marginal Metropolis-Hastings (PMMH), where
the parameters are generated with the unobserved latent states \lq \lq integrated
out\rq \rq; this technically means that the likelihood is replaced by its unbiased estimate. The second method is the particle Gibbs (PG), 
which generates the parameters
conditional on the states and conversely. 
Both methods run a particle filter algorithm
within an MCMC sampling scheme at each iteration.
\citeauthor{Andrieu:2010} show that for any finite number
of particles, the augmented target density used by these two algorithms
has the joint posterior distribution of the parameters and states as its
marginal distribution. A number of papers, including \citet{Lindsten2013}, \citet{Lindsten:2014}, \citet{Olsson2011},
\citet{Mendes2020}, and \citet{Gunawan2020PHS}, extend their work.

Density tempering sequential Monte Carlo (SMC) is an alternative to MCMC as a computational approach for Bayesian inference.
It is a sequential importance sampling approach \citep{Neal:2001},
where samples are first drawn from an
easily-generated distribution, e.g., the prior,
 and then moved towards the target distribution
through a sequence of intermediate stages, with each stage having its own
target distribution. \citet{DelMoral:2006} outline
the sequential Monte Carlo (SMC) method, which consists of resampling, reweighting and Markov move steps at each stage of the SMC.  
\comment{The Markov move step effectively runs
a few MCMC iterations for each combination of the parameters and latent
variables to help diversify them  so that they better approximate  the
tempered target density at that temperature level.}

It is generally recognised  \citep{DelMoral:2006}
that SMC has the following advantages compared
to MCMC, and in particular, PMCMC approaches:
(a)~the  Markov chain generated by an PMCMC sampler often 
gets trapped in local modes;  it is also difficult to assess whether the chain mixes adequately and converges to its invariant distribution. SMC explores the parameter space
more efficiently when the target distribution is multimodal; assessing convergence is also easier.
(b)~It is usually difficult to use PMCMC to estimate the marginal likelihood, which  is often used to for model choice~\citep{Kass:1995,Chib2001}. Using SMC to estimate the marginal likelihood is more straightforward. Section \ref{subsec:diagnostic} also shows the usefulness of the sequential SMC approach to estimate the sequence of independent standard normal random variables used to form the goodness of fit statistics to test for the model adequacy of a general time series state space model as in \citet{Smith1985}, \citet{Shephard1994}, and \citet{Gerlach1998}. Using PMCMC  to compute the goodness of fit statistics can be very time-consuming  as it is necessary  to repeatedly construct the full Markov chain for each time period $t$; see, for example, \citet{Gerlach1998}.
(c)~PMCMC algorithms are not parallelisable in general, whereas SMC algorithms can be parallelised (subject to some caveats); see, for example, \citet{Neal:2001}, \citet{DelMoral:2006}, and \citet{Duan2015}. Section \ref{subsec:Annealing-Approach-for state space models} discusses this further. 


\citet{Duan2015} extend the annealing/density tempering method of \citet{Neal:2001},
and \citet{DelMoral:2006} to time series state space models having intractable likelihoods. Their approach processes the time series as a batch.
\citet{Fulop2013} and \citet{Chopin2013} propose full sequential Bayesian inference on the parameters and call their algorithm Marginalised Resample-Move (MRM) and $\textrm{SMC}^{2}$, respectively.

Our paper builds on this literature by proposing flexible SMC approaches that use Markov
move steps that are more efficient than the density tempered SMC approach
of \citet{Duan2015}, MRM of \citet{Fulop2013}, and $\textrm{SMC}^{2}$ of \citet{Chopin2013}. 
The first Markov move approach is
based on the PG algorithm of \citet{Andrieu:2010}; we call it
sequential Monte Carlo
with particle Gibbs (SMC-PG). The second
is based on the Hamiltonian Monte Carlo (HMC) algorithm \citep{Duane1987,Neal:2011,Girolami:2011,Hoffman:2014,Betancourt2017}; we call it
sequential Monte Carlo
with Hamiltonian Monte Carlo (SMC-HMC). The third is based on the particle hybrid sampler of \citet{Gunawan2020PHS}, which combines a step similar to the correlated PMMH of \citet{Deligiannidis2018} and the particle Gibbs of \citet{Andrieu:2010}; we call it sequential Monte Carlo with the particle hybrid sampler (SMC-PHS). The Markov move step based on PHS generates the parameters that are highly correlated with the states in PMMH steps. All other parameters are generated by particle Gibbs steps conditioning on the latent state variables. 


The Markov moves based on PG, PHS and HMC are important for two reasons. First,
they permit applications to much higher dimensional parameter spaces
and much better proposals than random walk proposals; in particular, in some
cases we are able to sample from the full conditional distribution using Gibbs steps. Second, \secref{subsec:Multivariate-Factor-Stochastic examples-1} shows that using particle Gibbs Markov moves can substantially
reduce the number of particles required because it is then unnecessary
for the variance of the log of the likelihood estimate to be in the range of 1 to 3. In addition, if we suspect that some parameters converge slowly because they are highly correlated with the latent states in the model, then they can be   generated using the PMMH step of the PHS.

\citet{dufays2016evolutionary} robustifies data tempering by adding batch tempering at each data point. His approach is suitable for models with a tractable likelihood enabling the Markov move steps to be carried out by MCMC. 
The target densities in \citeauthor{dufays2016evolutionary}
depend only on the parameters and the method is applied  to standard GARCH models and GARCH models with change-point, and both only have a few parameters. Our article extends \citeauthor{dufays2016evolutionary}'s
approach to handle state space models  with intractable likelihoods and with the latent states also appearing in the target densities;  this makes it  possible to obtain joint inference on both the parameters and the latent states. 

Section~\ref{SVoutliers} and Section~\ref{SVbreaks} of the Supplement show that the proposed SMC is more stable than the methods of \citet{Fulop2013} and \citet{Chopin2013} in datasets with outliers and structural breaks because it  incorporates information from the data more gradually.

Our  proposed sequential approach with tempering is different to previous approaches. It provides joint sequential inference on latent states and parameters up to time $t$, $p(\bs{x}_{1:t},\bs\theta|\bs{y}_{1:t})$, whereas the $\textrm{SMC}^{2}$ provides sequential inference on the parameters only. \citet{Johansen2015} does not propose density tempering or data tempering or a combination of both for making inference on the parameters, states, or both.  \citeauthor{Johansen2015} proposes a block tempered particle filter to estimate the likelihood unbiasedly and provides the filtered state distribution $p(\bs{x}_{1:t}|\bs{y}_{1:t},\bs\theta)$. 
\citeauthor{Johansen2015} mentions that the block tempered particle filter can be used within the PMCMC for making inference on the parameters of the state space models, although he does not show it. His particle filter algorithm can also be used within our SMC algorithm, although we do not pursue this idea. 

The Markov move step in SMC  helps to diversify the particles  to provide a better approximation to the
tempered target distribution at that temperature level.
\citet{Duan2015}, \citet{Fulop2013} and \citet{Chopin2013} use a particle marginal Metropolis-Hastings (PMMH) approach with the likelihood estimated unbiasedly
by the particle filter in the Markov move step. There are two drawbacks
to a  Markov move based on PMMH. First, it is computationally
costly  to follow the guidelines
of \citet{Pitt:2012} for high dimensional state space models and set the optimal number of particles in the particle
filter such that the variance of the log of the estimated likelihood is in the range of
$1$ to $3$;
see the empirical example in \secref{subsec:Multivariate-Factor-Stochastic examples-1}.
It is possible to implement the correlated PMMH of \citet{Deligiannidis2018},
which is more efficient than standard PMMH for low dimensional state
space models. However, the correlated PMMH is ineffective
for a model with a high dimensional state, e.g. the factor SV model in
\secref{subsec:Multivariate-Factor-Stochastic examples-1}. Second, it is hard
to implement the PMMH Markov move efficiently
for a high dimensional parameter  as it is difficult
to obtain a good proposal for the parameter
because the first and second
derivatives with
respect to the parameters can only be estimated. \citet{Nemeth2016} and \citet{Mendes2020} find that the behaviour of the particle Metropolis adjusted Langevin algorithm (MALA) depends critically on how accurately we can estimate the gradient of the log-posterior. If the variance of the estimate  of the 
gradient of the log-posterior is too large, then there is no advantage in using particle MALA over the random walk proposal. The random walk
proposal is easy to implement, but is very inefficient in high dimensions.

We illustrate the proposed SMC methods
empirically using a sample of daily US stock returns to
estimate the univariate stochastic volatility (SV) models with outliers and structural breaks
and the multivariate high dimensional factor SV model.
The factor SV model is  popular because it parsimoniously models a vector of returns. Most  applications of multivariate factor stochastic volatility models are in financial econometrics \citep{Nardari2007, Aguilar:2000, Zhou:2014}.
Current popular MCMC estimation approaches for the factor SV model by
\citet{Chib2006} and \citet{Kastner:2017} are neither exact nor flexible; neither paper
corrects
for the approximation errors. They follow the approach proposed by \citet{Kim1998} to approximate the distribution of log outcome innovations by a carefully constructed seven component mixture of normal distributions. In the empirical application in Section \ref{subsec:Multivariate-Factor-Stochastic examples-1}, we show that the MCMC sampler of \citet{Kastner:2017} can give very different results to the PHS for some parameters of the factor SV model. The accuracy of the proposed SMC approaches are also compared to the PHS. 
Section~\ref{subsec:Multivariate-Factor-Stochastic examples-1} shows that the proposed SMC samplers give accurate estimates of the posterior densities of the parameters and predictive densities. We also consider a factor SV with the log-volatilities of the idiosyncratic errors to follow continuous time Ornstein-Uhlenbeck (OU) processes \citep{Stein1991}. Although the continuous
time OU diffusion model has a closed form transition density \citep{Brix2018}, the SMC methods are applied to the Euler approximation of the OU process and hence can handle the diffusion processes that do not admit closed form transition densities \citep{Ignatieva2015}. Section~\ref{OUprocess} gives further details.

The rest of the article is organised as follows.  \secref{sec:Flexible-Annealing-Approach}
outlines the  state space model and its estimation by SMC.
 \secref{MarkovMovesSec}
discusses the proposed flexible Markov moves. Section \ref{univariateexample} presents empirical results for the univariate example. \secref{sec:Multivariate-Factor-Stochastic} introduces
 and presents empirical results for the factor SV model.
The online Supplement to the paper contains some
further technical and empirical results.The following notation is used in both the main paper and the online Supplement.
Eq.~(1), Section~1, Algorithm~1 and Sampling Scheme~1, etc. refer
to the main article, while Eq.~(S1), Section~S1, Algorithm~S1, and Sampling Scheme~S1, etc. refer to the Supplement.

%
%

\section{Flexible Density Tempering \label{sec:Flexible-Annealing-Approach}}

This section introduces the state space model and discusses
the SMC approaches.


\subsection{The State Space Model \label{subsec:State-Space-Model}}
{We use the colon notation for collections of variables, i.e.
$\boldsymbol{a}_{t}^{1:N}=\left(a_{t}^{1},...,a_{t}^{N}\right)$,
$\boldsymbol{a}_{1:t}^{1:N}=\left(\boldsymbol{a}_{1}^{1:N},...,\boldsymbol{a}_{t}^{1:N}\right)$,
     and $\boldsymbol{a}_{1:M,1:t}=\left(a_{1,1},...,a_{M,1},a_{1,2},...,a_{M,2},...,a_{1,t},...,a_{M,t}\right)$.}


We consider a state space model where the latent states $\bs X_{t}$ determine
the evolution of the system. The density of $\bs X_{1}$ is $p\left(\bs x_{1}|\bs\theta\right)$
and the transition density of $\bs X_{t}$ given $\boldsymbol{X}_{1:t-1} = \bs {x}_{1:t-1}$
is $p\left(\bs x_{t}|\bs x_{t-1}, \boldsymbol{\theta}\right)$ for $t\geq2$.
The observations $\bs Y_{t}= \bs y_{t}$ for $t=1,...,T$ are linked to the
latent states through the observation equation $p\left(\bs y_{t}|\bs x_{t}, \boldsymbol{\theta}\right)$.
Denote the $d$-dimensional Euclidean space by $\mathbb{R}^d$. We suppose that
(i) the state vector $\bs x_t \in \mathcal{X}$ and the observation $ \bs y_t \in \mathcal{Y}$, where $\mathcal{X} \subset \mathbb{R}^{d_x}$ and $\mathcal{Y} \subset \mathbb{R}^{d_y}$; (ii)~the parameter vector $\boldsymbol{\theta}\in\boldsymbol{\Theta}$,
where $\boldsymbol{\Theta}$ is a subset of $\mathbb{R}^{d_{\theta}}$.

Our aim is to perform  Bayesian inference over the
latent states and the parameters conditional on the observations $\boldsymbol{y}_{1:T}$.
By using Bayes rule, the joint posterior of the latent states and
the parameters is
\begin{equation*}\label{eq: joint posterior}  
\begin{aligned}
p\left(\boldsymbol{\theta},\boldsymbol{x}_{1:T}|\boldsymbol{y}_{1:T}\right)
 & =\frac{p\left(\boldsymbol{x}_{1:T},\boldsymbol{y}_{1:T}|\boldsymbol{\theta}\right)p\left(\boldsymbol{\theta}\right)}{p\left(\boldsymbol{y}_{1:T}\right)},\\
\text{where}\quad  p\left(\boldsymbol{x}_{1:T},\boldsymbol{y}_{1:T}|\boldsymbol{\theta}\right)& =p\left(\bs x_{1}| \bs \theta\right)\prod_{t=2}^{T}
p\left(\bs x_{t}|\bs x_{t-1}, \bs \theta\right)\prod_{t=1}^{T}p\left(\bs y_{t}|\bs x_{t},\bs\theta\right);
\end{aligned}
\end{equation*}
$p\left(\boldsymbol{\theta}\right)$ is the prior density of $\boldsymbol{\theta}$
and $p\left(\boldsymbol{y}_{1:T}\right)$ is the marginal likelihood
of $\boldsymbol{y}_{1:T}$. For notational simplicity, we often write $\bs x = \bs x_{1:T}$ and $\bs y = \bs y_{1:T}$.

\subsection*{Examples}
This section illustrates the state space model by the univariate stochastic volatility model,
\begin{equation}\label{eq: simple SV model}
\begin{aligned}
y_{t} & =  \exp\left(x_{t}/2\right)\epsilon_{t}, \quad
x_{t+1}  =  \mu+\phi\left(x_{t}-\mu\right)+\tau\eta_{t+1}, \quad
x_{1}  \sim  N\left(\mu,\frac{\tau^{2}}{1-\phi^{2}}\right),
\end{aligned}
\end{equation}
with $\epsilon_{t}\sim N\left(0,1\right)$ and $\eta_{t}\sim N\left(0,1\right)$ and independent.
We call $\boldsymbol{x}_{1:T}$ the latent log volatility process.
The vector of unknown parameters is $\boldsymbol{\theta}=(\mu,\phi,\tau^{2})$. To ensure that the model is stationary, we restrict the persistence parameter $\phi$ so that $|\phi|<1$.


\subsection{The SMC Approach for State Space Models \label{subsec:Annealing-Approach-for state space models}}
\subsubsection{Batch Estimation Method \label{Batch Estimation}}
This section discusses the proposed SMC method for batch estimation problems. The main
idea is to begin with an easy-to-sample distribution
and propagate a particle cloud $\left\{ \boldsymbol{\theta}_{1:M}^{\left(p\right)},\boldsymbol{x}_{1:M}^{\left(p\right)},\bs W_{1:M}^{\left(p\right)}\right\} $
through a sequence of tempered target densities $\xi_{a_{p}}\left(\boldsymbol{\theta},\boldsymbol{x}\right)$,
for $p=0,...,P$, to the posterior density of interest which is much harder to sample
from directly. The tempered densities are defined as
\begin{equation*}
\begin{aligned}\label{eq: tempered density} 
\xi_{a_{p}}\left(\boldsymbol{\theta},\boldsymbol{x}\right) &:=\eta_{a_{p}}\left(\boldsymbol{\theta},\boldsymbol{x}\right)/Z_{a_{p}},\quad
\text{where} \quad Z_{a_{p}}  :=\int\eta_{a_{p}}\left(\boldsymbol{\theta},\boldsymbol{x}\right)d\boldsymbol{\theta}d\boldsymbol{x}\\
\text{and}\quad
\eta_{a_{p}}\left(\boldsymbol{\theta},\boldsymbol{x}\right)& :=\left(\pi_{0}\left(\boldsymbol{\theta},\boldsymbol{x}\right)\right)^{1-a_{p}}
\left(p\left(\boldsymbol{y}|\boldsymbol{\theta},\boldsymbol{x}\right)p\left(\boldsymbol{x}|\boldsymbol{\theta}\right)p\left(\boldsymbol{\theta}\right)\right)^{a_{p}}.
\end{aligned}
\end{equation*}

The tempering sequence $\boldsymbol{a}_{0:P}$ is
such that $a_{0}=0<a_{1}<...<a_{P}=1$. If it is both easy to generate
from the densities $p\left(\boldsymbol{\theta}\right)$ and $p\left(\boldsymbol{x}|\boldsymbol{\theta}\right)$,
and evaluate them, then we take $\pi_{0}\left(\boldsymbol{x},\boldsymbol{\theta}\right)=p\left(\boldsymbol{\theta}\right)p\left(\boldsymbol{x}|\boldsymbol{\theta}\right)$,
and hence
\begin{equation}\label{eq: simplified tempered density} 
\eta_{a_{p}}\left(\boldsymbol{\theta},\boldsymbol{x}\right)=p\left(\boldsymbol{y}|\boldsymbol{\theta},\boldsymbol{x}\right)^{a_{p}}
p\left(\boldsymbol{x}|\boldsymbol{\theta}\right)p\left(\boldsymbol{\theta}\right).
\end{equation}

Algorithm \ref{alg:Generic-AISIL-Algorithm}  summarizes the general SMC approach, which we
 now discuss.
The initial particle
cloud $\left\{ \boldsymbol{\theta}_{1:M}^{\left(0\right)},\boldsymbol{x}_{1:M}^{\left(0\right)},\bs W_{1:M}^{\left(0\right)}\right\} $
is obtained by generating the $\left\{ \boldsymbol{\theta}_{1:M}^{\left(0\right)},\boldsymbol{x}_{1:M}^{\left(0\right)}\right\} $
from $\pi_{0}\left(\boldsymbol{x},\boldsymbol{\theta}\right)$, and
giving the particles equal weight, i.e., $\bs W_{1:M}^{\left(0\right)}=1/M$.
 The
weighted particles (particle cloud)
$\left\{ \boldsymbol{\theta}_{1:M}^{\left(p-1\right)},\boldsymbol{x}_{1:M}^{\left(p-1\right)},\bs W_{1:M}^{\left(p-1\right)}\right\} $
at the $\left(p-1\right)$st level (or stage) of the
annealing is an estimate of $\xi_{a_{p-1}}\left(\boldsymbol{\theta},\boldsymbol{x}\right)$.
Based on this estimate of $\xi_{a_{p-1}}\left(\boldsymbol{\theta},\boldsymbol{x}\right)$,
the SMC algorithm goes through the following steps to obtain an
estimate of $\xi_{a_{p}}\left(\boldsymbol{\theta},\boldsymbol{x}\right)$.
The move from $\xi_{a_{p-1}}\left(\boldsymbol{\theta},\boldsymbol{x}\right)$
to $\xi_{a_{p}}\left(\boldsymbol{\theta},\boldsymbol{x}\right)$ is
implemented by reweighting the particles by the ratio of the two
unnormalized densities $\eta_{a_p}/\eta_{a_{p-1}}$, yielding the following weights, $\bs W_{1:M}^{\left(p\right)}=\bs w^{(p)}_{1:M}/ \sum_{j=1}^{M} w^{(p)}_{j}$, where
\begin{equation}
\begin{aligned}
w^{(p)}_{i}=W_{i}^{\left(p-1\right)}\frac{\eta_{a_{p}}\left(\boldsymbol{\theta}^{(p-1)}_{i},\boldsymbol{x}^{(p-1)}_{i}\right)}{\eta_{a_{p-1}}\left(\boldsymbol{\theta}^{(p-1)}_{i},\boldsymbol{x}^{(p-1)}_{i}\right)}=W_{i}^{\left(p-1\right)}p\left(\boldsymbol{y}|\boldsymbol{\theta}^{(p-1)}_{i},\boldsymbol{x}^{(p-1)}_{i}\right)^{a_{p}-a_{p-1}},\label{weightsmc}
\end{aligned}
\end{equation}
where $\eta_{a_{p}}$ is defined in \eqref{eq: simplified tempered density}.  We follow \citet{DelMoral2012} and choose the tempering sequence
adaptively to ensure sufficient  particle diversity by selecting
the next value of $a_{p}$ automatically such that the effective sample
size (ESS) stays close to some target value $\textrm{ESS}_\textrm{target}$ chosen by the user. The  ESS  is used to measure the variability in the $W_i^{(p)}$, and is defined
\def\sumweights{\left (  \sum_{i=1}^{M}\left(W_{i}^{(p)}\right)^{2}\right )}
as $\textrm{ESS} := {\sumweights}^{-1}$.
The ESS varies between 1 and $M$, with a low value of ESS indicating that
the weights are concentrated only on a few particles and a value of $M$ means that the particles are equally weighted.
An ESS close to the target value is achieved by evaluating the ESS over a grid $\bs a_{1:G,p}$ of
potential $a_{p}$ values and selecting as $a_{p}$ the value of $a_{j,p}$
whose ESS is the closest to $\textrm{ESS}_\textrm{target}$. Other approaches to select the tempering sequence, such as that in \citet{DelMoral2012}, may be used instead. 
The particles $\left\{ \boldsymbol{\theta}_{1:M}^{\left(p\right)},\boldsymbol{x}_{1:M}^{\left(p\right)}\right\} $ are then resampled
proportionately to their weights $\bs W_{1:M}^{\left(p\right)}$, which means that the resampled particles then
have equal weight $\bs W_{1:M}^{\left(p\right)}=1/M$. This has the
effect of eliminating particles with negligible weights and replicating
the particles with large weights, so the ESS is now $M$. We use multinomial resampling for all the examples in the paper.

Repeatedly reweighting and resampling can seriously reduce the diversity
of the particles, and lead to particle depletion. The
particle cloud
 $\left\{ \boldsymbol{\theta}_{1:M}^{\left(p\right)},\boldsymbol{x}_{1:M}^{\left(p\right)},\bs W_{1:M}^{\left(p\right)}\right\} $
may then be a poor representation of $\xi_{a_{p}}\left(\boldsymbol{\theta},\boldsymbol{x}\right)$.
To improve the approximation of the particle cloud to $\xi_{a_{p}}\left(\boldsymbol{\theta},\boldsymbol{x}\right)$,
we carry out $R$ Markov move steps for each particle using a Markov
kernel $K_{a_{p}}$ which retains $\xi_{a_{p}}\left(\boldsymbol{\theta},\boldsymbol{x}\right)$
as its invariant density. Thus, at each annealing level, we run
a short MCMC scheme for each of the $M$ particles.

Steps~$1$ and 2a-2d of Algorithm~\ref{alg:Generic-AISIL-Algorithm}
are standard and apply to any model, with only
slight model-specific modifications; hence, it is only necessary to discuss the Markov
move step in detail for each model. We propose three Markov move algorithms
that leave the target density $\xi_{a_{p}}\left(\boldsymbol{\theta},\boldsymbol{x}\right)$
invariant. The first algorithm is based on particle Gibbs \citep{Andrieu:2010}, and is denoted
as SMC-PG. The second algorithm is based on Hamiltonian
Monte Carlo \citep{Neal:2011}, and is denoted as SMC-HMC. The third algorithm is based on the particle hybrid sampler (PHS) of \citet{Gunawan2020PHS}, and is denoted as SMC-PHS.

By \citet{DelMoral:2006}, the SMC algorithm
provides consistent inference for the target density $p\left(\boldsymbol{\theta},\boldsymbol{x}_{1:T}|\boldsymbol{y}_{1:T}\right)$
as $M$ goes to infinity. See also \citet{Beskos2016} for
recent consistency results for the adaptive algorithm employed here
and the associated central limit theorem.


Section \ref{sec:Introduction} states that the SMC algorithm can be parallelised. However, parallelising SMC is not straightforward due to the resampling in step (2d) of \algref{alg:Generic-AISIL-Algorithm}. In our applications given in Sections \ref{univariateexample} and \ref{subsec:Multivariate-Factor-Stochastic examples-1}, the optimal number of annealing levels, $P$, obtained from the adaptive approach described above is moderate and therefore resampling has little effect on the efficiency of the SMC algorithm. The reweighting in step (2c) and the Markov moves in step (2e) are easily parallelised for each SMC sample because the computations required for each sample are independent ofthe other SMC samples.

\begin{algorithm}[h]
\caption{Generic SMC Algorithm \label{alg:Generic-AISIL-Algorithm}}  
\begin{enumerate}
\item Set $p=0$ and initialise
the particle cloud $\left\{ \boldsymbol{\theta}_{1:M}^{\left(0\right)},\boldsymbol{x}_{1:M}^{\left(0\right)},\bs W_{1:M}^{\left(0\right)}\right\} $,
by generating the $\left\{ \boldsymbol{\theta}_{1:M}^{\left(0\right)},\boldsymbol{x}_{1:M}^{\left(0\right)}\right\} $
from $\pi_{0}\left(\boldsymbol{x},\boldsymbol{\theta}\right)$, and
giving them equal weight, i.e., $W_{i}^{\left(0\right)}=1/M$, for
$i=1,...,M$
\item While the tempering sequence $a_{p}<1$ do
\begin{enumerate}
\item Set $p\leftarrow p+1$
\item Find $a_{p}$ adaptively by searching across a grid of $a_{p}$ values to
maintain ESS around some constant $\textrm{ESS}_\textrm{target}$.
\item Compute new weights,
\begin{equation}
\bs W_{1:M}^{\left(p\right)}=\frac{\bs w^{(p)}_{1:M}}{\sum_{j=1}^{M}w^{(p)}_{j}}\;\;\textrm{where}\;\;w^{(p)}_{i}=W_{i}^{\left(p-1\right)}\frac{\eta_{a_{p}}\left(\boldsymbol{\theta_{i}}^{(p-1)},\boldsymbol{x_{i}}^{(p-1)}\right)}{\eta_{a_{p-1}}\left(\boldsymbol{\theta_{i}}^{(p-1)},\boldsymbol{x_{i}}^{(p-1)}\right)}.\label{eq:updated weights}
\end{equation}
\item Resample $\left(\boldsymbol{\theta}_{i}^{\left(p-1\right)},\boldsymbol{x}_{i}^{\left(p-1\right)}\right)$
using the weights $\bs W_{1:M}^{\left(p\right)}$ to obtain $\left\{ \boldsymbol{\theta}_{1:M}^{\left(p\right)},\boldsymbol{x}_{1:M}^{\left(p\right)},\bs W_{1:M}^{\left(p\right)}=1/M\right\} $.
\item Markov moves

\begin{enumerate}
\item Let $K_{a_{p}}\left(\left(\boldsymbol{\theta},\boldsymbol{x}\right),\cdot\right)$
be a Markov kernel having invariant density $\xi_{a_{p}}\left(\boldsymbol{\theta},\boldsymbol{x}\right)$.
For $i=1,...,M$, move each $\left(\boldsymbol{\theta}_{i}^{\left(p\right)},\boldsymbol{x}_{i}^{\left(p\right)}\right)$
$R$ times using the Markov kernel $K_{a_{p}}$ to obtain $\left\{ \widetilde{\boldsymbol{\theta}}_{i},\widetilde{\boldsymbol{x}}_{i}\right\} $.
\item Set $\left(\boldsymbol{\theta}_{1:M}^{\left(p\right)},\boldsymbol{x}_{1:M}^{\left(p\right)}\right)\leftarrow
    \left(\widetilde{\boldsymbol{\theta}}_{1:M},\widetilde{\boldsymbol{x}}_{1:M}\right)$.
\end{enumerate}
\end{enumerate}
\end{enumerate}
\end{algorithm}

\subsubsection{Sequential Estimation Method \label{Sequential Estimation}}
We now discuss the extension of the density tempered SMC approach in Section
\ref{Batch Estimation} to the situation where we are interested in sequential estimation
of both the states and parameters as new observations arrive. This approach
has very useful applications, especially in finance and economics.
For example, in finance, investors need to update their volatility
and return forecasts whenever new data become available. As another
example, economists need to revise their forecasts as news on the
state of the economy changes. Here, PMCMC techniques are
expensive and time-consuming to use as they need to repeatedly restart the MCMC
 as new observations arrive.

The main challenge is how to jointly sample and propagate the states
and parameters as new observations arrive. Our framework deals with this by extending the SMC sequence
of target distributions as the number of observations increases.
The main idea is to propagate a particle cloud $\left\{ \boldsymbol{\theta}_{1:M}^{\left(p\right)},
\boldsymbol{x}_{1:t-1,1:M}^{\left(p\right)},\bs W_{1:M}^{\left(p\right)}\right\} $
that reflects the joint posterior of states $\boldsymbol{x}_{1:t-1}$
and parameters $\boldsymbol{\theta}$ at time $t-1$ through a sequence
of tempered target densities $\xi_{a_{p}}\left(\boldsymbol{\theta},\boldsymbol{x}\right)$,
for $p=0,...,P$, to the joint posterior density of states $\boldsymbol{x}_{1:t}$
and parameters $\boldsymbol{\theta}$ at time $t$. The tempered densities
are defined as
\begin{multline*}
\xi_{a_{p}}\left(\boldsymbol{\theta},\boldsymbol{x}_{1:t}\right):=\eta_{a_{p}}\left(\boldsymbol{\theta},\boldsymbol{x}_{1:t}\right)/Z_{a_{p}},\;\;\textrm{where}\;\;Z_{a_{p}}:=\int\eta_{a_{p}}\left(\boldsymbol{\theta},\boldsymbol{x}_{1:t}\right)d\boldsymbol{\theta}d\boldsymbol{x}_{1:t},\\
\textrm{and}\;\eta_{a_{p}}\left(\boldsymbol{\theta},\boldsymbol{x}\right):=p\left(\bs y_{t}|\bs x_{t},\boldsymbol{\theta}\right)^{a_{p}}p\left(\bs x_{t}|\bs x_{t-1},\boldsymbol{\theta}\right)p\left(\boldsymbol{\theta},\boldsymbol{x}_{1:t-1}|\bs y_{1:t-1}\right).
\end{multline*}
The move from $\xi_{a_{p-1}}\left(\boldsymbol{\theta},\boldsymbol{x}_{1:t-1}\right)$
to $\xi_{a_{p}}\left(\boldsymbol{\theta},\boldsymbol{x}_{1:t}\right)$ is
implemented by reweighting the particles by the ratio of the two unnormalised
densities $\eta_{a_{p}}/\eta_{a_{p-1}}$, yielding the weights
\[
w_{i}^{\left(p\right)}:=W_{i}^{\left(p-1\right)}\frac{\eta_{a_{p}}\left(\boldsymbol{\theta}_{i}^{\left(p-1\right)},\boldsymbol{x}_{i,1:t}^{\left(p-1\right)}\right)}{\eta_{a_{p-1}}\left(\boldsymbol{\theta}_{i}^{\left(p-1\right)},\boldsymbol{x}_{i,1:t}^{\left(p-1\right)}\right)}=W_{i}^{\left(p-1\right)}p\left(\bs y_{t}|\bs x^{(p-1)}_{i,t},\boldsymbol{\theta}^{(p-1)}_{i}\right)^{a_{p}-a_{p-1}}.
\]






Section \ref{SVoutliers} and  Supplement \ref{SVbreaks}  show that our sequential approach is robust to outliers and structural breaks; this happens because tempering incorporates the information in new observations gradually. It also means that it is easy to maintain the ESS close to its target $\textrm{ESS}_\textrm{target}$.

The $\textrm{SMC}^{2}$ sequential approach in \citet{Chopin2013} only targets the posterior density $p(\bs\theta|\bs{y}_{1:t})$ of the parameters $\bs \theta$  by \lq \lq integrating out\rq\rq{} the latent states, $\bs x_{1:t}$, up to time $t$. Our sequential tempering approach  targets the joint posterior density $p(\bs\theta,\bs{x}_{1:t}|\bs{y}_{1:t})$ of the parameters $\bs \theta$ and latent states, $\bs x_{1:t}$, up to time $t$. 




\section{Flexible Markov Moves \label{MarkovMovesSec}}
This section discusses the Markov move steps in step (2e) of Algorithm \ref{alg:Generic-AISIL-Algorithm}. The first is based on the Hamiltonian Monte Carlo (HMC) algorithm
\citep{Neal:2011}.
The second 
is based on the particle Gibbs (PG) algorithm of  \citet{Andrieu:2010}. The third is based on the particle marginal Metropolis-Hastings (PMMH) algorithm of \citet{Andrieu:2010}. The fourth is based on the particle hybrid sampler of \citet{Gunawan2020PHS}. 

\subsection{Hamiltonian Monte Carlo \label{HMCMarkovMove}}
The Markov move has two steps.
The first step updates the latent states $\boldsymbol{x}_{1:T}$ conditional on the parameters $\boldsymbol{\theta}$ using Hamiltonian Monte Carlo (HMC). The second step updates the parameters $\boldsymbol{\theta}$ conditional on the states $\boldsymbol{x}_{1:T}$. 

Sampling the high dimensional latent state vector $\boldsymbol{x}_{1:T}$ is discussed first. 
Suppose we want to sample from a $T$-dimensional distribution with a density
proportional to $\exp\left(\mathcal{L}\left(\boldsymbol{x}\right)\right)$,
where $\mathcal{L}\left(\boldsymbol{x}\right)=\log p\left(\boldsymbol{x}|\boldsymbol{\theta},\boldsymbol{y}\right)$.
In Hamiltonian Monte Carlo \citep{Neal:2011}, we augment $\bs x$ with an auxiliary
momentum vector $\boldsymbol{r}$, having the same dimension as the
latent state vector $\boldsymbol{x}$, with the density $p\left(\boldsymbol{r}\right)=N\left(\boldsymbol{r}|0,\widetilde{M}\right)$,
where $\widetilde{M}$ is a mass matrix. We define the joint conditional
density of $\left(\boldsymbol{x},\boldsymbol{r}\right)$ as
\begin{align*}\label{eq:jointHamiltonian}
p\left(\boldsymbol{x},\boldsymbol{r}|\textrm{rest}\right) & \propto \exp\left(-H\left(\boldsymbol{x},\boldsymbol{r}\right)\right),
\end{align*}
where
$H\left(\boldsymbol{x},\boldsymbol{r}\right)=-\mathcal{L}\left(\boldsymbol{x}\right)+\frac{1}{2}\boldsymbol{r}^{T}\widetilde{M}^{-1}\boldsymbol{r}$
is called the Hamiltonian. In an idealised Hamiltonian step, the state
vector $\boldsymbol{x}$ and the momentum variables $\boldsymbol{r}$
move continuously according to the differential equations
\begin{equation*}\label{eq: hmc diff eqns}
\begin{aligned}
\frac{d\boldsymbol{x}}{dt}=\frac{\partial H}{\partial\boldsymbol{r}}=\widetilde{M}^{-1}\boldsymbol{r},
\frac{d\boldsymbol{r}}{dt}=-\frac{\partial H}{\partial\boldsymbol{x}}=\nabla_{\boldsymbol{x}}\mathcal{L}\left(\boldsymbol{x}\right),
\end{aligned}
\end{equation*}
where $\nabla_{\boldsymbol{x}}$ denotes the gradient with respect
to $\boldsymbol{x}$. In practice, this continuous time Hamiltonian
dynamics needs to be approximated by discretizing time, using a step
size $\epsilon$. We can then simulate the evolution over time of
$\left(\boldsymbol{x},\boldsymbol{r}\right)$ via the \lq\lq leapfrog\rq\rq{}
integrator using \algref{alg:Hamiltonian-Monte-Carlo}. 


Sections \ref{subsec:Sampling-the-Latent volatility} and \ref{SS: steps ii to iv} of the Supplement give the details  for the univariate SV model. Sections \ref{SS: Markov moves for HMC} and \ref{SS: samplingparamSV} of the Supplement give the details  for the factor SV model. 

\begin{algorithm}[H]
\caption{The Hamiltonian Monte Carlo part of a single Markov move step\label{alg:Hamiltonian-Monte-Carlo} \citep{Hoffman:2013}}

Given initial values of $\boldsymbol{x}$, $\epsilon$, $L$, where
$L$ is the number of leapfrog updates

Sample $\boldsymbol{r}\sim N\left(0,\wt{M}\right)$

For $l=1$ to $L$

\hspace{20pt}Set $\left(\boldsymbol{x}^{*},\boldsymbol{r}^{*}\right)\leftarrow\textrm{Leapfrog}\left(\boldsymbol{x},\boldsymbol{r},\epsilon\right)$

end for

With probability
\begin{align*}
\alpha &=\min\left(1,\frac{\exp\left(\mathcal{L}\left(\boldsymbol{x}^{*}\right)-\frac{1}{2}\boldsymbol{r}^{*T}\wt{M}^{-1}\boldsymbol{r}^{*}\right)}
{\exp\left(\mathcal{L}\left(\boldsymbol{x}\right)-\frac{1}{2}\boldsymbol{r}^{T}\wt{M}^{-1}\boldsymbol{r}\right)}\right),
\end{align*}
set $\boldsymbol{x}=\boldsymbol{x}^{*}$, $\boldsymbol{r}^{*}=-\boldsymbol{r}$, else retain current $\bs x$.

\vspace{5mm}

function Leapfrog($\boldsymbol{x}$,$\boldsymbol{r}$,$\epsilon$) \\
Set $\boldsymbol{r}+\epsilon\nabla_{\boldsymbol{x}}\mathcal{L}\left(\boldsymbol{x}\right)/2$\\
Set $\boldsymbol{x}+\epsilon\boldsymbol{\wt{M}}^{-1}\boldsymbol{r}$,\\
Set $\boldsymbol{r}+\epsilon\nabla_{\boldsymbol{x}}\mathcal{L}\left(\boldsymbol{x}\right)/2$,\\
return $\boldsymbol{x}^{*}$, $\boldsymbol{r}^{*}$

\end{algorithm}


\subsection{Particle Gibbs} 
The Markov moves based on PG, PMMH, and PHS run a particle filter algorithm at each iteration.
\secref{subsec:PF}  discusses the particle filter. \secref{subsec:AISIL-PG-Augmented-Target} then discusses particle Gibbs based Markov move steps.

\subsubsection{Particle Filter\label{subsec:PF}}
This section describes the particle filter methods used to obtain sequential
approximations to the densities $p\left(\boldsymbol{x}_{1:t}|\boldsymbol{y}_{1:t},\boldsymbol{\theta}\right)$
for $t=1,...,T$. The particle filter algorithm recursively
produces a set of weighted particles $\left\{ \boldsymbol{x}_{1:t}^{\left(j\right)},\widetilde{W}_{t}^{\left(j\right)}\right\} _{j=1}^{N}$
such that the intermediate densities $p\left(\boldsymbol{x}_{1:t}|\boldsymbol{y}_{1:t},\boldsymbol{\theta}\right)$
are approximated by
\[
\widehat{p}\left(\boldsymbol{x}_{1:t}|\boldsymbol{y}_{1:t},\boldsymbol{\theta}\right)
=\sum_{j=1}^{N}\widetilde{W}_{t}^{\left(j\right)}\delta_{\boldsymbol{x}_{1:t}^{(j)}}\left(d\boldsymbol{x}_{1:t}\right),
\]
where $\delta_{a}\left(dx\right)$ is the Dirac delta distribution
located at $a$. In more detail, given $N$ samples $\left\{ \boldsymbol{x}_{t-1}^{\left(j\right)},\widetilde{W}_{t-1}^{\left(j\right)}\right\}_{j=1}^N $
representing the filtering density $p\left(\boldsymbol{x}_{t-1}|\boldsymbol{y}_{1:t-1},\boldsymbol{\theta}\right)$
at time $t-1$, we use
\[
p\left({x}_{t}|\boldsymbol{y}_{1:t},\boldsymbol{\theta}\right)\propto\int p\left({y}_{t}|{x}_{t},\boldsymbol{\theta}\right)p\left({x}_{t}|{x}_{t-1},\boldsymbol{\theta}\right)p\left({x}_{t-1}|\boldsymbol{y}_{1:t-1},\boldsymbol{\theta}\right)d{x}_{t-1},
\]
to obtain the particles $\left\{{x}_{t}^{\left(j\right)},\widetilde{W}_{t}^{\left(j\right)}\right\}_{j=1}^N $  representing
$p({x}_t|\bs y_{1:t}, \bs \theta) $, by first drawing from a known and easily sampled proposal density
function $m\left({x}_{t}|{x}_{t-1},{y}_{t},\boldsymbol{\theta}\right)$
and then computing the unnormalised weights $\widetilde{w}_{t}^{(j)}$ to
account for the difference between the target posterior density and
the proposal, where
\[
\widetilde{w}_{t}^{\left(j\right)}:=\wt W_{t-1}^{(j)} \frac{p\left({y}_{t}|
{x}_{t}^{\left(j\right)},\boldsymbol{\theta}\right)p\left({x}_{t}^{\left(j\right)}|{x}_{t-1}^{\left(j\right)},
\boldsymbol{\theta}\right)}{m_{t}\left({x}_{t}^{\left(j\right)}|{x}_{t-1}^{\left(j\right)},{y}_{t},
\boldsymbol{\theta}\right)},\;j=1,...,N;
\]
these are then normalized as $\widetilde{W}_{t}^{\left(j\right)}:=\widetilde{w}_{t}^{\left(j\right)}/\sum_{r=1}^{N}\widetilde{w}_{t}^{\left(r\right)}$.

\citet{Chopin2004} shows that as the number of time periods $t$
increases, the normalised weights of the particle system become concentrated
only on a few particles and eventually the normalised weight of a single
particle converges to one. This is known as the \lq weight degeneracy\rq{}
problem. One way to reduce the impact of weight degeneracy is to include
resampling steps in a particle filter algorithm. A resampling scheme
is defined as $\M\left(\widetilde{\boldsymbol{b}}_{t-1}^{1:N}|\widetilde{\bs W}_{t-1}^{1:N}\right)$,
where $\widetilde{b}_{t-1}^{j}$ indexes a particle in $\left\{{x}_{t-1}^{\left(j\right)},\widetilde{W}_{t-1}^{\left(j\right)}\right\} _{j=1}^{N}$
that is chosen with probability $\widetilde{W}_{t-1}^{\left(j\right)}$.
\secref{sec:Assumptions} of the Supplement states some assumptions on the proposal density $m_t\left({x}_{t}|{x}_{t-1},{y}_{t},\boldsymbol{\theta}\right)$
and resampling scheme.
In our empirical applications, we use
the bootstrap filter with $p\left({x}_{t}|{x}_{t-1},\boldsymbol{\theta}\right)$
as a proposal density and multinomial resampling.

The particle filter produces the unbiased estimate of the likelihood
\begin{align*}
\widehat{p}\left(\boldsymbol{y}_{1:T}|\boldsymbol{\theta}\right) & =  \widehat{p}\left({y}_{1}|\boldsymbol{\theta}\right)\prod_{t=2}^{T}\widehat{p}\left({y}_{t}|{y}_{t-1},\boldsymbol{\theta}\right)
  =  \prod_{t=1}^{T}\left\{ \frac{1}{N}\sum_{j=1}^{N}\widetilde{w}_{t}^{(j)}\right\},
\end{align*}
i.e. $\E\left(\widehat{p}\left(\boldsymbol{y}_{1:T}|\boldsymbol{\theta}\right)\right)=p\left(\boldsymbol{y}_{1:T}|\boldsymbol{\theta}\right)$ \citep{DelMoral:2004,Pitt:2012}.


\subsubsection{Particle Gibbs (PG)  Algorithm\label{subsec:AISIL-PG-Augmented-Target}}

This section discusses the Markov move based on the particle Gibbs algorithm. Unless stated otherwise, we write PG to denote both the
particle Gibbs (PG) and particle Metropolis within Gibbs Markov moves.
Let $\boldsymbol{U}_{1:T}^{1:N}=\left(\boldsymbol{x}_{1:T}^{1:N},\widetilde{\boldsymbol{b}}_{1:T-1}^{1:N}\right)$
denote the collection of particles in the particle filter algorithm, together with their associated indices $\widetilde{\boldsymbol{b}}_{1:T-1}^{1:N}$ defined in Section \ref{subsec:PF}.
The SMC-PG constructs a sequence of tempered densities
$\xi_{a_{p}}\left(\boldsymbol{\theta},\boldsymbol{x}_{1:T}\right)$,
$p=1,...,P$, based on the augmented tempered target density that includes all the particles generated by the particle filter.  \secref{targetdensityPG} of the Supplement gives further details of the augmented tempered target density.


    
    





Algorithm \ref{alg:Particle-Gibbs-Markov1} gives the Markov move based on the particle Gibbs algorithm.
Let $\boldsymbol{\theta}\coloneqq\left(\boldsymbol{\theta}_{1},...,\boldsymbol{\theta}_{H}\right)$
be a partition of the parameter vector into $H$ components, where each component may be a vector.

\begin{algorithm}[H]
\caption{Markov move based on the Particle Gibbs algorithm \label{alg:Particle-Gibbs-Markov1}}

For $i=1,...,M$ SMC samples, 
\begin{enumerate}
\item For $h=1,...,H$, sample $\boldsymbol{\theta}_{ih}^{*}$ from the
proposal density\newline  $q_{h}\left(\cdot|\bs j_{i,1:T},\boldsymbol{x}_{i,1:T}^{\bs j_{i,1:T}},\boldsymbol{\theta}_{-ih},\boldsymbol{\theta}_{ih}\right)$.
\item Accept the proposed values $\boldsymbol{\theta}_{ih}^{*}$ with probability
\[
\min\left(1,\frac{\xi_{a_{p}}\left(\boldsymbol{\theta}_{ih}^{*}|\boldsymbol{x}_{i,1:T}^{\bs j_{i,1:T}},\bs j_{i,1:T},\boldsymbol{\theta}_{-ih}\right)}{\xi_{a_{p}}\left(\boldsymbol{\theta}_{ih}|\boldsymbol{x}_{i,1:T}^{\bs j_{i,1:T}},\bs j_{i,1:T},\boldsymbol{\theta}_{-ih}\right)}\frac{q_{h}\left(\boldsymbol{\theta}_{ih}|\boldsymbol{\theta}_{-ih},\boldsymbol{\theta}_{ih}^{*},\boldsymbol{x}_{i,1:T}^{\bs j_{i,1:T}},\bs j_{i,1:T}\right)}{q_{h}\left(\boldsymbol{\theta}_{ih}^{*}|\boldsymbol{\theta}_{-ih},\boldsymbol{\theta}_{ih},\boldsymbol{x}_{i,1:T}^{\bs j_{i,1:T}},\bs j_{i,1:T}\right)}\right).
\]
\item Sample $\bs U_{i,1:T}^{\left(-\bs j_{i,1:T}\right)}$ using conditional sequential Monte Carlo given in Algorithm \ref{alg:Conditional-Sequential-Monte carlo} in \secref{sec:Particle-Filter-and CPF} of the Supplement. 
\item Sample $J_{i,t}=j_{i,t}\sim\widetilde{\xi}_{a_{p}}\left(\bs j_{i,1:T}|\boldsymbol{x}_{i,1:t}^{1:N},\wt{\boldsymbol{b}}_{i,1:t-1}^{1:N},\boldsymbol{\theta}_{i},\boldsymbol{x}_{i,t+1:T}^{\bs j_{i,t+1:T}},\bs  j_{i,t+1:T}\right)$ for \\$t=T-1,...,1$ and  $J_{i,T}=j_{i,T}\sim\widetilde{\xi}_{a_{p}}\left(j_{i,T}|\boldsymbol{\theta}_{i},\boldsymbol{x}_{i,1:T}^{1:N},\wt{\boldsymbol{b}}_{i,1:T-1}^{1:N}\right)$
using the backward simulation algorithm given in Algorithm \ref{alg:The-backward simulation algorithm} of \secref{sec:Particle-Filter-and CPF} of the supplement.
\end{enumerate}
\end{algorithm}





Parts 1 and 2 of Algorithm \ref{alg:Particle-Gibbs-Markov1} update each parameter block of $\bs\theta$ conditional on a  trajectory $\boldsymbol{x}_{1:T}^{\bs j_{1:T}}=\left(x_{1}^{j_{1}},...,x_{T}^{j_{T}}\right)$. 
Part 3 is the conditional sequential Monte Carlo of \citet{Andrieu:2010} that is the
key part of the Markov move steps. It is a particle filter algorithm in which a particle $\boldsymbol{x}_{1:T}^{\bs j_{1:T}}=\left(\boldsymbol{x}_{1}^{j_{1}},...,\boldsymbol{x}_{T}^{j_{T}}\right)$,
and the associated sequence of ancestral indices are kept unchanged
with all the other particles and indices resampled and updated. The collection of all random variables except the chosen particle
trajectory is denoted by $\boldsymbol{U}_{1:T}^{- \bs j_{1:T}}$. 
Part 4 of Algorithm \ref{alg:Particle-Gibbs-Markov1} samples the new indices $J_{T},J_{T-1},...,J_{1}$ sequentially using backward simulation given in Algorithm \ref{alg:The-backward simulation algorithm} of \secref{sec:Particle-Filter-and CPF} of the Supplement, obtains the new particle trajectory $\boldsymbol{x}_{1:T}^{\bs j_{1:T}}=\left(x_{1}^{j_{1}},...,x_{T}^{j_{T}}\right)$, and discards the rest of the particles.
Following \citet{Whiteley2010} and \citet{Lindsten2013}, the conditional density in Part 4 is 
\[
\widetilde{\xi}_{a_{p}}\left(j_{t}|\boldsymbol{\theta},{x}_{1:t}^{1:N},\widetilde{\boldsymbol{b}}_{1:t-1}^{1:N},\boldsymbol{x}_{t+1:T}^{\bs j_{t+1:T}},\bs j_{t+1:T}\right)\propto\widetilde{w}_{t}^{j_{t}}p\left(\bs{x}_{t+1}^{j_{t+1}}|\bs\theta,\bs{x}_{t}^{j_{t}}\right),
\]
for $t=T-1,...,1$ and $\widetilde{\xi}_{a_{p}}\left(j_{T}|\boldsymbol{\theta},\boldsymbol{x}_{1:T}^{1:N},\widetilde{\boldsymbol{b}}_{1:T-1}^{1:N}\right)\propto\widetilde{w}_{T}^{j_{T}}$. The Markov move based on the PG algorithm is useful in generating the vector of high-dimensional parameters,  except when some  of the parameters are highly correlated with the latent states.

\subsection{Particle Marginal Metropolis-Hastings \label{PMMHstep}}
Algorithm \ref{PMMH_Markov} describes the Markov move based on the particle marginal Metropolis-Hastings (PMMH) approach 
\citep{Andrieu:2010}; this is useful for generating parameters that are highly correlated with the states. The PMMH algorithm requires an unbiased estimate of the likelihood, which is obtained from the particle filter.

\begin{algorithm}[H]
\caption{Particle Marginal Metropolis-Hastings (PMMH) Algorithm \label{PMMH_Markov}}

Given initial values for $\boldsymbol{U}_{i,1:T}$, $\boldsymbol{\theta}_{i}$,
and $\bs{j}_{i,1:T}$, for $i=1,...,M$ SMC samples. One iteration of PMMH involves
the following steps:
\begin{enumerate}
\item Sample $\boldsymbol{\theta}_{i}^{*}$ from the proposal density $q\left(\boldsymbol{\theta}_{i}^{*}|\boldsymbol{\theta}_{i},\boldsymbol{U}_{i,1:T}\right)$.
\item Sample $\boldsymbol{U}_{i,1:T}^{*}\sim\psi\left(\boldsymbol{U}_{i,1:T}^{*}|\boldsymbol{\theta}_{i}^{*}\right)$
by running the particle filter algorithm in Algorithm \ref{alg:Sequential Monte-Carlo-Algorithm-1-1} of Section~ \ref{sec:Particle-Filter-and CPF} of the Supplement 
 and compute a tempered estimate of the likelihood $\widehat{p}\left(\boldsymbol{y}_{1:T}|\boldsymbol{\theta}_{i}^{*}\right)^{a_{p}}$. 
\item Sample $J_{i,1:T}^{*}=j_{i,1:T}^{*}$ using the backward simulation
algorithm given in Algorithm \ref{alg:The-backward simulation algorithm} of Section~ \ref{sec:Particle-Filter-and CPF} of the Supplement.
\item Accept the proposed values of $\boldsymbol{\theta}_{i}^{*}$, $\boldsymbol{U}_{i,1:T}^{*}$,
and $\bs{j}^{*}_{i,1:T}$ with probability
\begin{equation}\label{acceptancePMMH}
\min\left\{ 1,\frac{\widehat{p}\left(\boldsymbol{y}_{1:T}|\boldsymbol{\theta}_{i}^{*}\right)^{a_{p}}p\left(\boldsymbol{\theta}_{i}^{*}\right)}{\widehat{p}\left(\boldsymbol{y}_{1:T}|\boldsymbol{\theta}_{i}\right)^{a_{p}}p\left(\boldsymbol{\theta}_{i}\right)}\frac{q\left(\boldsymbol{\theta}_{i}|\boldsymbol{\theta}_{i}^{*}\right)}{q\left(\boldsymbol{\theta}_{i}^{*}|\boldsymbol{\theta}_{i}\right)}\right\}.
\end{equation}
\end{enumerate}
\end{algorithm}

Parts 1 to 3 of the Algorithm \ref{PMMH_Markov} propose the values of $\boldsymbol{\theta}_{i}^{*}$, $\boldsymbol{U}_{i,1:T}^{*}\sim\psi\left(\boldsymbol{U}_{i,1:T}^{*}|\boldsymbol{\theta}_{i}^{*}\right)$, and $\bs{j}_{i,1:T}^{*}$, which 
are then accepted with the probability in \eqref{acceptancePMMH}. Algorithm \ref{PMMH_Markov} targets the joint posterior density of the latent states $\bs{x}_{1:T}$ and the parameters $\bs\theta$. \citet{Duan2015} use a PMMH method, with the likelihood estimated unbiasedly by the particle filter, in the Markov move component. However, they only target the posterior distribution of the parameters $\bs\theta$. We call their method SMC-PMMH-DF.



\subsection{Particle Hybrid Sampler \label{SMC_PHS}}
\citet{Mendes2020} propose a new particle MCMC sampler that combines the particle Gibbs and PMMH algorithms of \citet{Andrieu:2010}. The sampler generates parameters that are highly correlated with the states using PMMH, with the rest generated by PG. \citet{Deligiannidis2018} propose the correlated PMMH method, which correlates the random numbers used in constructing the  likelihood estimates at the current and proposed values of the parameters. They show that the correlated PMMH can scale up with the number of observations compared to standard PMMH, as long as the state dimension is not too large. \citet{Gunawan2020PHS} propose a novel PMCMC which involves a non-trivial combination of the correlated PMMH and the PG algorithms;  \citeauthor{Gunawan2020PHS} call it the particle hybrid sampler (PHS). Unlike \citet{Andrieu:2010}, its augmented target density is expressed in terms of basic uniform and standard normal random numbers. The parameters that are efficiently generated by conditioning on the latent states are generated in a particle Gibbs step. All other parameters are drawn with PMMH steps by conditioning on the basic uniform and standard normal variables used in the particle filter algorithm. The PHS sampler is scalable in the number of observations and the number of parameters. See \citeauthor{Gunawan2020PHS} for further details.

\subsection*{The PHS Sampling Scheme \label{sub:Flexible-Correlated-PMMH+PG sampling scheme}}
Algorithm \ref{alg:Sampling-Scheme:-The correlated PMMH+PG} outlines the Markov move steps based on the particle hybrid sampler. 
For simplicity, let $\bs\theta\coloneqq\left(\bs{\theta}_{1},\bs{\theta}_{2}\right)$ partition the parameter vector into 2 components.  
 It generates $\bs{\theta}_{1}$ using a PMMH step and the vector  parameter $\bs{\theta}_{2}$  using  a PG step. The elements  in $\bs{\theta}_{1}$ and $\bs{\theta}_{2}$ are sampled separately in the  PMMH step and  PG step, respectively. 




\begin{algorithm}[H]
\caption{ The Particle Hybrid Sampler (PHS).\label{alg:Sampling-Scheme:-The correlated PMMH+PG}}

Given initial values for $\bs{v}_{x,i,1:T}^{1:N}$, $\bs{v}_{B,i,1:T-1}$, $\bs{j}_{i,1:T}$,
and $\bs\theta_i$. For $i=1,...,M$ (SMC samples), one iteration of the PHS involves the following steps:
\begin{enumerate}
\item [Part 1:] PMMH sampling. 
\begin{enumerate}
\item Sample $\bs{\theta}_{i,1}^{*}\sim q_{1}\left(\cdot|\bs{\theta}_{i,2},\bs{\theta}_{i,1}\right)$
\item Run the particle filter with Euclidean sort  in Algorithm \ref{alg:The-correlated particle filter algorithm} of Section \ref{SS: SMC algorithms22} of the Supplement and evaluate $\widehat{p}(\bs{y}_{1:T}|\bs{\theta}^{*}_{i,1},\bs{\theta}_{i,2})^{a_p}$.
\item Accept the proposed values $\bs{\theta}_{i,1}^{*}$ with probability
\begin{align}\label{eq: MH ratio1}
\alpha\left(\bs\theta_{i,1};\bs\theta_{i,1}^{*}|\bs{u}_{x,i,1:T}^{1:N},\bs{u}_{A,i,1:T-1},\bs\theta_{i,2}\right) & =\notag \\
1\land\frac{\widehat{p}(\bs{y}_{1:T}|\bs\theta^{*}_{i,1},\bs\theta_{i,2})^{a_p}    p\left(\bs\theta_{i,1}^{*}|\bs\theta_{i,2}\right)}
{\widehat{p}(\bs{y}_{1:T}|\bs\theta_{i,1},\bs\theta_{i,2})^{a_p} p\left(\bs\theta_{i,1}|\bs\theta_{i,2}\right)} & \times\frac{q_{1}\left(\bs\theta_{i,1}|\bs\theta_{i,2},\bs\theta_{i,1}^{*}\right)}
{q_{1}\left(\bs\theta_{i,1}^{*}|\bs\theta_{i,2},\bs\theta_{i,1}\right)}.
\end{align}

\end{enumerate}
\item [Part 2:] Sample $\bs{J}_{i,1:T}=\bs{j}_{i,1:T}$ using the backward simulation
algorithm given in Algorithm \ref{alg:The-backward simulation algorithm} of Section \ref{sec:Particle-Filter-and CPF} of the Supplement.

\item [Part 3:] Particle Gibbs (PG) sampling. 
\begin{enumerate}
\item Sample $\bs\theta_{i,2}^{*}\sim q_{2}\left(\cdot|\bs\theta_{i,1},\bs\theta_{i,2}\right)$
\item Accept the proposed values $\bs\theta_{i,2}^{*}$ with probability
\begin{align*}
\alpha\left(\bs\theta_{i,2};\bs\theta_{i,2}^{*}|\bs{x}_{i,1:T}^{j_{i,1:T}},\bs{j}_{i,1:T},  \bs\theta_{i,1}\right) & \\
= & 1\land \frac{p\left(\bs{y}_{1:T}|\bs{x}_{i,1:T}^{j_{i,1:T}},\bs\theta_{i,2}^{*},\bs\theta_{i,1}\right)^{a_p}
p\left(\bs{x}_{i,1:T}^{j_{i,1:T}}|\bs\theta_{i,2}^{*},\bs\theta_{i,1}\right)p\left(\bs\theta_{i,2}^{*}|\bs\theta_{i,1}\right)}{p\left(\bs{y}_{1:T}|\bs{x}_{i,1:T}^{j_{i,1:T}},\bs\theta_{i,2},\bs\theta_{i,1}\right)^{a_p}
p\left(\bs{x}_{i,1:T}^{j_{i,1:T}}|\bs\theta_{i,2},\bs\theta_{i,1}\right)p\left(\bs\theta_{i,2}|\bs\theta_{i,1}\right)} \\
& \times\frac{q_{2}\left(\bs\theta_{i,2}|\bs\theta_{i,1},\bs\theta_{i,2}^{*}\right)}{q_{2}\left(\bs\theta_{i,2}^{*}|\bs\theta_{i,1},\bs\theta_{i,2}\right)}. \label{eq: MH ratio 2}
\end{align*}
\end{enumerate}
\item [Part 4:] Sample $\left(\bs{v}_{i,x,1:T},\bs{v}_{i,A,1:T-1}\right)$ 
using the constrained conditional sequential Monte Carlo algorithm (Algorithm \ref{alg:The-conditional Sequential-Monte carlo algorithm} of Section  \ref{sub:Conditional-Sequential-Monte Carlo constrained}) of the Supplement
and evaluate $\widehat{p}(\bs{y}_{1:T}|\bs\theta_{i,1},\bs\theta_{i,2})^{a_p}$.
\end{enumerate}
\end{algorithm}

Part 1 generates $\bs{\theta_1}^{*}$ using PMMH, conditioning on the random numbers $\bs{v}_{x,1:T}^{1:N}$ and $\bs{v}_{B,1:T-1}$ obtained from Part~4, the parameters $\bs\theta_2$, and the current parameters $\bs{\theta}_1$. Part 2 samples the new indices $\bs{j}_{1:T}$ using the backward simulation algorithm, obtains the new particle trajectory  $\boldsymbol{x}_{1:T}^{\bs j_{1:T}}=\left(x_{1}^{j_{1}},...,x_{T}^{j_{T}}\right)$, and discards the rest of the particles. Part~3 generates the parameters $\bs{\theta}^{*}_{2}$ using PG, conditioning on the selected particle trajectory $\boldsymbol{x}_{1:T}^{\bs j_{1:T}}$, the parameter $\bs\theta_1$ and the current parameter $\bs\theta_2$. Part~4 updates the basic random numbers using constrained conditional sequential Monte Carlo \citep{Gunawan2020PHS}; see Section~ \ref{sub:Conditional-Sequential-Monte Carlo constrained} of the Supplement.

\subsection{Marginal Likelihood Estimation\label{subsec:Estimating-Marginal-Likelihood}}

The marginal likelihood $p\left(\boldsymbol{y}_{1:T}\right)$ is often
used in the Bayesian literature to compare models \citep{Chib2001}.
An advantage of the SMC method is that it offers a natural way to
estimate the marginal likelihood. We note that $p\left(\boldsymbol{y}_{1:T}\right)=Z_{a_{P}}$,
$Z_{a_{0}}=1$, so that
\[
p\left(\boldsymbol{y}_{1:T}\right)=\prod_{p=1}^{P}\frac{Z_{a_{p}}}{Z_{a_{p-1}}}\;\;\textrm{with}\;\;\frac{Z_{a_{p}}}{Z_{a_{p-1}}}
=\int\left(\frac{\eta_{a_{p}}\left(\boldsymbol{\theta},\boldsymbol{x}\right)}{\eta_{a_{p-1}}\left(\boldsymbol{\theta},
\boldsymbol{x}\right)}\right)\xi_{a_{p-1}}\left(\boldsymbol{\theta},\boldsymbol{x}\right)d\boldsymbol{\theta}d\boldsymbol{x}.
\]
Because the particle cloud $\left\{ \boldsymbol{\theta}_{1:M}^{\left(p-1\right)},\boldsymbol{x}_{1:M}^{\left(p-1\right)},\bs W_{1:M}^{\left(p-1\right)}\right\} $
obtained after iteration $p-1$ approximates $\xi_{a_{p-1}}\left(\boldsymbol{\theta},\boldsymbol{x}\right)$,
the ratio above is estimated by
\[
\widehat{\frac{Z_{a_{p}}}{Z_{a_{p-1}}}}=\sum_{i=1}^{M}W_{i}^{\left(p-1\right)}\frac{\eta_{a_{p}}\left(\boldsymbol{\theta}_{i}^{\left(p-1\right)},\boldsymbol{x}_{i}^{\left(p-1\right)}\right)}{\eta_{a_{p-1}}\left(\boldsymbol{\theta}_{i}^{\left(p-1\right)},\boldsymbol{x}_{i}^{\left(p-1\right)}\right)},
\]
giving the marginal likelihood estimate
\[
\widehat{p\left(\boldsymbol{y}_{1:T}\right)}=\prod_{p=1}^{P}\widehat{\frac{Z_{a_{p}}}{Z_{a_{p-1}}}}.
\]

\section{Univariate Examples \label{univariateexample}}

\subsection{Univariate Stochastic Volatility Model \label{subsec:Univariate-Stochastic-Volatility examples-1}}
This section illustrates the proposed SMC-PG and SMC-HMC methods for batch estimation problems by applying
them to the univariate stochastic volatility (SV) model and compares their performance to the exact particle hybrid sampler (PHS) of \citet{Gunawan2020PHS} and the SMC-PMMH-DF of \citet{Duan2015}. Posterior distributions obtained from the PHS method are treated as the \lq ground truth\rq{}  for comparing the accuracy of the posterior density approximations for two reasons. The first is that we have extensively applied PHS to stochastic volatility models, and in our experience it has always worked well. Second, PHS belongs to the class of PMCMC methods of \citet{Andrieu:2010} and their  key property is that they are an \lq exact approximation\rq{} to the idealised MCMC algorithms targeting the joint posterior density of the latent states and parameters $p(\bs{x}_{1:T},\bs\theta|y_{1:T})$ \citep{Andrieu:2010}. We follow \citet{South2019} and \citet{Gunawan2020Subsampling} by comparing the posterior densities of the parameters estimated using different SMC algorithms and exact MCMC methods.




The vector of unknown parameters of the SV model in Section \ref{subsec:State-Space-Model} is $\boldsymbol{\theta}=(\mu,\phi,\tau^{2})$. The parameters have the following priors, and are assumed to be independent apriori:
(a) $p\left(\mu\right)\propto I\left(-10<\mu<10\right)$. (b)~Following \citet{Jensen:2010}, the prior for $\tau^{2}$ is inverse
Gamma $\textrm{IG}\left(v_{0}/2,s_{0}/2\right)$ with $v_{0}=10$
and $s_{0}=0.5$. (c)~To ensure stationarity, the persistence parameter is restricted to $\mid \phi\mid <1$; we follow \citet{Kim1998} and choose
the prior for $\phi$ as $\left(\phi+1\right)/2\sim\textrm{Beta}\left(a_{0},b_{0}\right)$,
with $a_{0}=100$ and $b_{0}=1.5$. We apply our methods to a sample of daily US food industry stock returns
obtained from the Kenneth French website, using a sample from
December 11th, 2001 to the 11th November 2013, a total of $3001$
observations.

The performance of SMC-HMC depends on choosing suitable values for the three tuning parameters i)~the mass matrix $\widetilde{M}$, ii)~the step size $\epsilon$, and iii)~the number of leapfrog steps $L$.  The step size $\epsilon$ determines how well the leapfrog integration
approximates the Hamiltonian dynamics. If it is  too large, then
a low acceptance rate may result, but if it is too small, then
it becomes computationally expensive to obtain distant proposals.
Similarly, if $L$ is too small, then the proposal will be close to
the current value of the latent state vectors, resulting in undesirable
random walk behaviour. If $L$ is too large, then HMC will generate
proposals that retrace their steps.
The precision matrix $\Sigma^{-1}$ of the AR(1) process of the latent states is
a sparse tridiagonal matrix
whose diagonal elements are  $0.5a_{p}+\left(1+\phi^{2}\right)/\tau^{2}$, 
except for the first and last diagonal elements which are
 $0.5a_{p}+1/\tau^{2}$; the super- and sub-diagonal
elements  are $-{\phi}/{\tau^{2}}$. We set $\widetilde{M}=\Sigma^{-1}$. Our article uses two adaptive approaches.
The first follows an adaptive
method based on \citet{Garthwaite:2015} to select an $\epsilon$ that
yields a specified average acceptance probability across all $M$
particles and $L$ is set to some fixed value;  we denote this method as SMC-HMC-I. The second adaptive approach is to select both $\epsilon$ and $L$ based on the adaptive method by \citet{Buchholz2020} and  denote it as SMC-HMC-II.

\tabref{tab:Univariate-SV-model-results} summarises the estimation results
for the univariate SV model estimated using the PHS,
SMC-HMC, SMC-PG, and SMC-PMMH-DF methods. The PHS chain consists of 5000 iterates for burnin and another 50000 iterates
used for inference. All the SMC estimates are
obtained using 10 independent runs, each  with $M=560$ samples, to generate
a total of $5600$ samples for each algorithm.  We set the constant $\textrm{ESS}_{T}=0.8M$.
The computations are done using Matlab on a single desktop computer with 6-CPU cores.

\begin{figure}[H]
\caption{The kernel density estimates of the marginal posterior densities of the univariate
SV parameters $\tau^2$ estimated using 
SMC-PMMH-DF (left), SMC-PG (middle), and SMC-HMC (right).   Each method is compared with PHS (PMCMC). \label{fig:The-Kernel-DensitySV} }
\begin{centering}
\psfrag{tausq}{$\tau^2$}
\includegraphics[width=15cm,height=9cm]{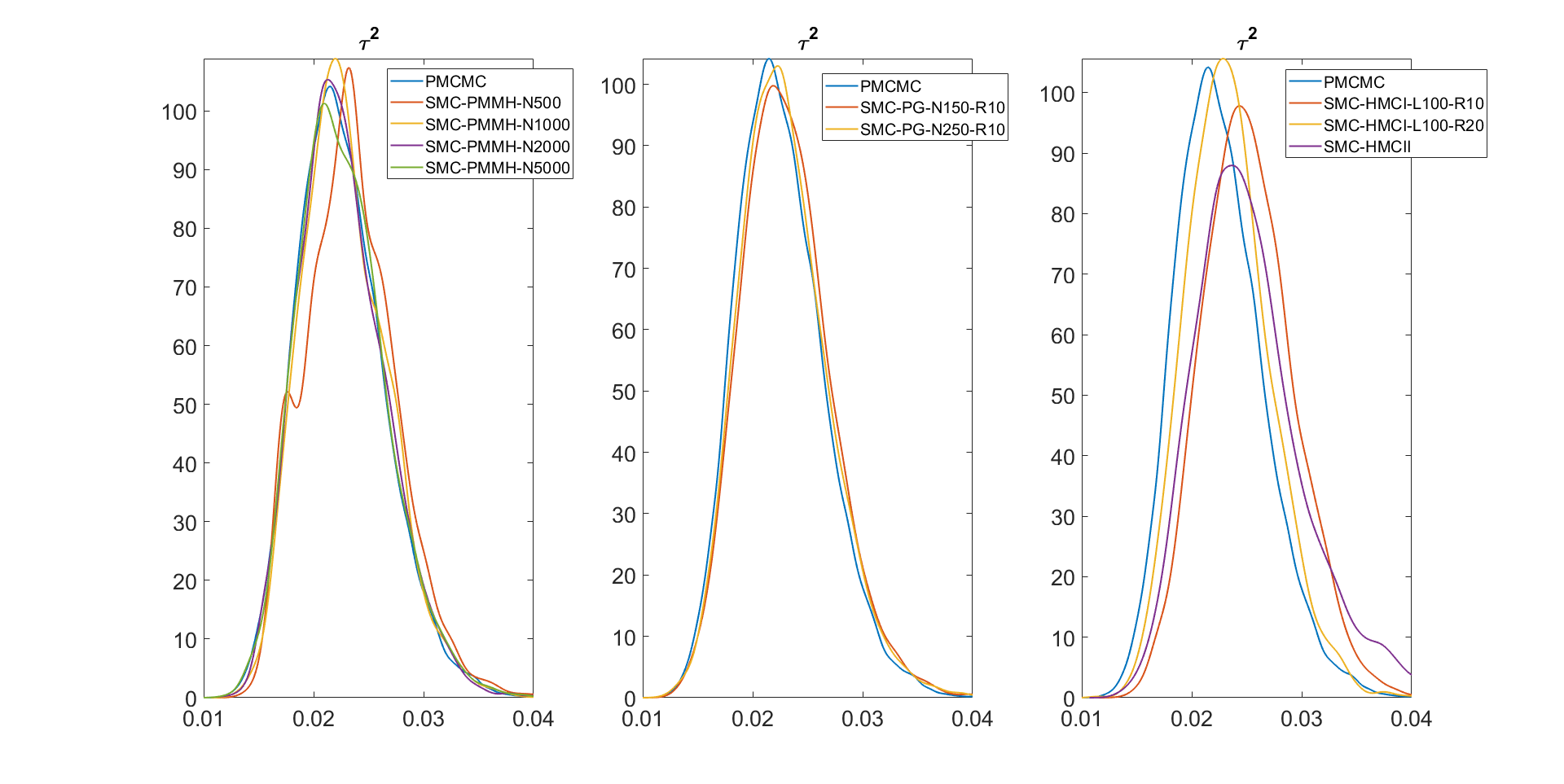}
\par\end{centering}
\end{figure}

\tabref{tab:Univariate-SV-model-results} shows  that all the SMC-PG 
estimates are very close to the PHS estimates for all
parameters even with as few as $150$ particles and $R=10$ Markov
move steps.
The middle panel of \figref{fig:The-Kernel-DensitySV} shows the
kernel density estimates of the marginal posteriors of the univariate
stochastic volatility parameters $\tau^2$ 
estimated using the PHS and the SMC-PG methods
with different numbers of particles. The density
estimates from the SMC-PG methods are very close to the density estimates
of the PHS sampler.

\tabref{tab:Univariate-SV-model-results} shows that the estimates of $\tau^{2}$ and $\phi$ estimated using the SMC-HMC-I with $R=20$ is the closest to the
estimates of the PHS compared to SMC-HMC-I with $R=10$ and SMC-HMC-II. But in general, they are very close to each other.
The right panel of \figref{fig:The-Kernel-DensitySV} shows the kernel density estimates
of the marginal posteriors of the univariate SV parameters $\tau^2$  estimated using the PHS and the SMC-HMC methods. The figure confirms that the densities estimates of $\tau^{2}$ 
estimated from SMC-HMC-I with $R=20$ are the closest to the estimates from the PHS, but still sligthly inaccurate compared to the SMC-PG estimates. In general, SMC-PG is more accurate than the SMC-HMC, but its CPU time is slightly larger. \tabref{tab:Univariate-SV-model-results} also shows that the optimal number of Markov move steps obtained from the adaptive approach described in Section \ref{subsec:Annealing-Approach-for state space models} are similar for SMC-PG and SMC-HMC methods.





The left panel of \figref{fig:The-Kernel-DensitySV} shows the kernel density estimates of the marginal posteriors of the univariate SV parameters $\tau^2$  estimated using the PHS and the SMC-PMMH-DF methods. The density estimates from SMC-PMMH-DF with $N=2000$ and $5000$ particles are close to the PHS estimates. The number of Markov move steps of SMC-PMMH-DF are smaller compared to the SMC-PG and SMC-HMC methods because it only targets the posterior density of SV parameters and not the latent log-volatilities. The SMC-PG with $N=250$ particles is still slightly faster than SMC-PMMH-DF with $N=2000$ particles.

\begin{figure}[H]
\caption{Standard univariate stochastic volatility example. The plots of the log-volatility estimates using the PHS (with $N=100$), SMC-PG ($N=250$), SMC-HMC-I ( $R=20$), and SMC-PMMH-DF ($N=2000$). 
 \label{fig:The-logvolatilitySMCPG} }
\begin{centering}
\psfrag{tausq}{$\tau^2$}
\includegraphics[width=15cm,height=9cm]{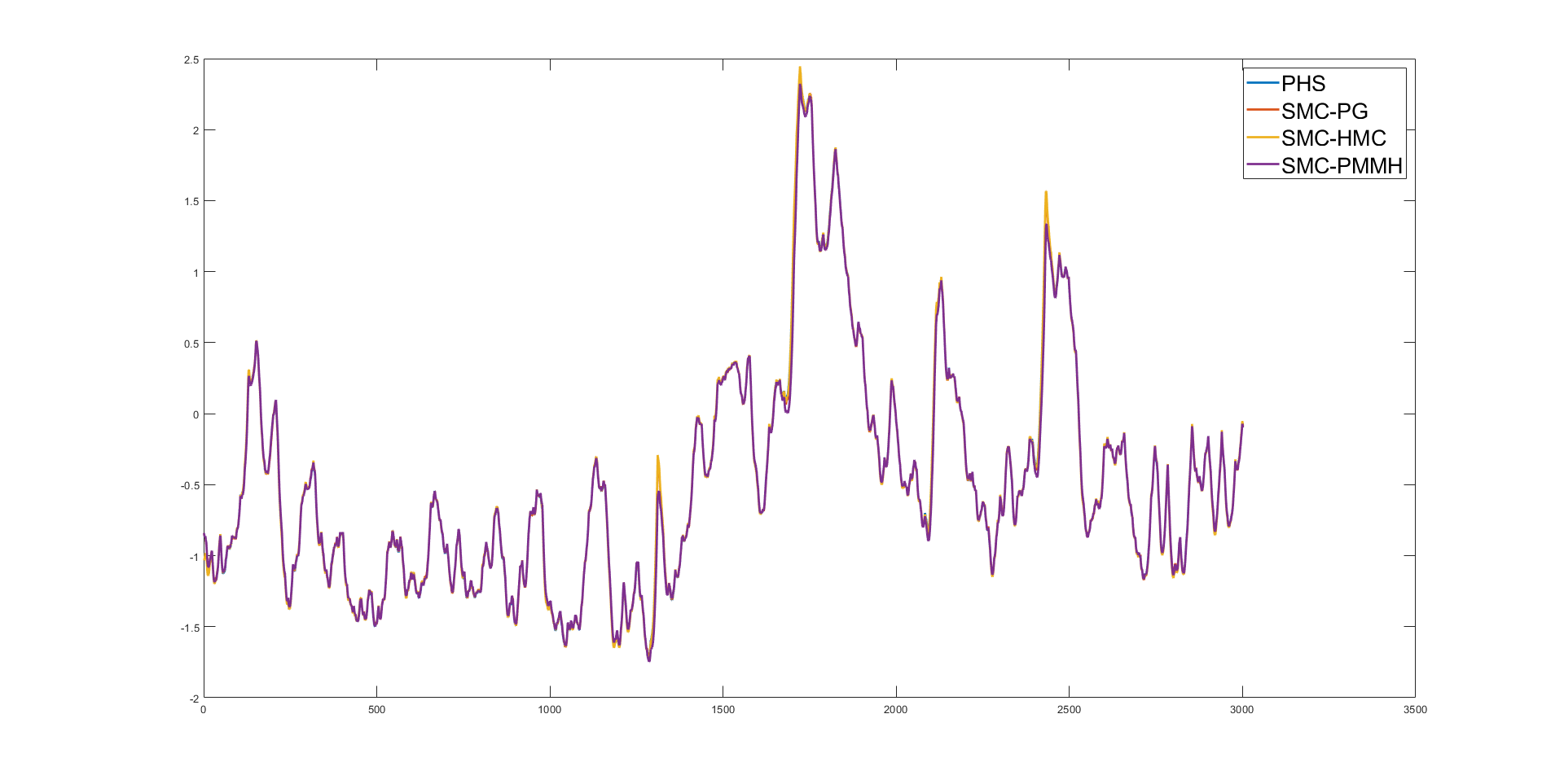}
\par\end{centering}
\end{figure}

\figref{fig:The-logvolatilitySMCPG} shows the posterior mean estimates of the latent log-volatilities estimated using PHS with $N=100$, SMC-PG with $N=250$, SMC-HMC-I with $R=20$, and SMC-PMMH-DF with $N=2000$. The SMC-PG, SMC-PMMH-DF, and PHS estimates are indistinguishable. The estimates from SMC-HMC-I are slightly less accurate at some time points. This study suggets that (a) The SMC-HMC methods are less accurate compared to SMC-PG and SMC-PMMH-DF methods for estimating the standard univariate SV models; (b) The SMC-PG with $N=250$ is as accurate as the SMC-PMMH-DF with $N=2000$, and faster than the SMC-PMMH-DF for this example; (c) The SMC-PMMH-DF provides inference only on the parameters, and not the latent states. Two steps are needed to obtain the posterior density of the latent states. First, the SMC draws of the parameters are obtained from running the SMC-PMMH-DF algorithm. Second, the particle filter and backward simulation algorithms are run for each parameter draw to give the posterior density of the latent states.  

\begin{table}[H]
\caption{Results for the univariate SV model estimated using the PHS, SMC-HMC, SMC-PG, and SMC-PMMH-DF samplers for the US food stock returns data with $T=3001$.
The SMC results are obtained using 10 independent runs of each
algorithm. Time is the time in minutes for one run of the algorithm. The posterior standard deviations of the SV parameters are in brackets.
\label{tab:Univariate-SV-model-results}}
\centering{}%
\begin{tabular}{ccccccccc}\\
\hline
Method & $N$ & $L$ & $R$ & $\mu$ & $\phi$ & $\tau^{2}$ & $P$ & Time\tabularnewline
\hline
PHS & 100 & - & - & $\underset{\left(0.2054\right)}{-0.4886}$ & $\underset{\left(0.0040\right)}{0.9853}$ & $\underset{\left(0.0040\right)}{0.0226}$ & - & 513\tabularnewline
HMC-I & - & 100 & 10 & $\underset{\left(0.1992\right)}{-0.4934}$ & $\underset{\left(0.0043\right)}{0.9839}$ & $\underset{\left(0.0041\right)}{0.0252}$ & 57  & 62.87 \tabularnewline
HMC-I & - & 100 & 20 & $\underset{\left(0.2007\right)}{-0.4888}$ & $\underset{\left(0.0039\right)}{0.9849}$ & $\underset{\left(0.0039\right)}{0.0234}$ & 57  & 123.68 \tabularnewline
HMC-II & - & 100 & 20 & $\underset{\left(0.2006\right)}{-0.4959}$ & $\underset{\left(0.0045\right)}{0.9839}$ & $\underset{\left(0.0051\right)}{0.0253}$ & 57 & 137.23 \tabularnewline
PG & 150 & - & 10 & $\underset{\left(0.2096\right)}{-0.4996}$ & $\underset{\left(0.0040\right)}{0.9850}$ & $\underset{\left(0.0040\right)}{0.0231}$ & 59  & 109.27 \tabularnewline
PG & 250 & - & 10 & $\underset{\left(0.2106\right)}{-0.4986}$ & $\underset{\left(0.0041\right)}{0.9851}$ & $\underset{\left(0.0040\right)}{0.0229}$ & 57  & 156.18 \tabularnewline
PMMH-DF & 500 & - & 10 & $\underset{\left(0.2185\right)}{-0.4590}$ & $\underset{\left(0.0040\right)}{0.9852}$ & $\underset{\left(0.0041\right)}{0.0234}$ & 15 & 58.03\tabularnewline
PMMH-DF & 1000 & - & 10 & $\underset{\left(0.2052\right)}{-0.4802}$ & $\underset{\left(0.0040\right)}{0.9855}$ & $\underset{\left(0.0039\right)}{0.0229}$ & 14 & 105.63 \tabularnewline
PMMH-DF & 2000 & - & 10 & $\underset{\left(0.2059\right)}{-0.4797}$ & $\underset{\left(0.0041\right)}{0.9850}$ & $\underset{\left(0.0041\right)}{0.0229}$ & 14 & 226.18 \tabularnewline
PMMH-DF & 5000 & - & 10 & $\underset{\left(0.2038\right)}{-0.4878}$ & $\underset{\left(0.0040\right)}{0.9853}$ & $\underset{\left(0.0040\right)}{0.0228}$ & 13 & 671.94\tabularnewline

\hline
\end{tabular}
\end{table}

The above analyses suggest that for this example the SMC-PG and SMC-PMMH-DF are the most accurate SMC methods. The next example compares the performance of the SMC-PG and SMC-PMMH-DF for estimating the univariate SV model with a larger
number of parameters. The measurement equation is now  
\begin{equation}
y_{t}=\bs{l}_{t}^{\top}\bs{\beta}+\exp\left(x_{t}/2\right)\epsilon_{t},\;\textrm{where}\;\;\epsilon_{t}\sim\N\left(0,1\right),
\end{equation}
where $\bs{\beta}=(\beta_1,...,\beta_{50})^{\top}$ and $l_t=(l_{1,t},...,l_{50,t})^{\top}$ is a $(50\times1)$ vector of covariates at time $t$. We compare the performance of the following samplers: 
(I) the PHS with $N=100$, (II) SMC-PG with $N=250$, (III) SMC-PMMH-DF with $N=1000$ and $2500$. We apply the methods to  simulated data with $T=6000$ observations, $\phi=0.99$, $\tau^2=0.5$, $\rho=-0.2$, $\mu=0.5$. The parameters $\beta_k$ are generated independently of each other from normal distributions with mean zero and standard deviation $1$, for all $k=1,...,50$. The covariates are generated randomly from a multivariate normal $\N(\bs{0},I)$. The prior for $\beta_k$ is $\N(0,1)$ for $k=1,...,50$. 

\begin{figure}[H]
\caption{The kernel density estimates of some marginal posterior densities of the univariate
SV parameters with the covariate coefficients  estimated using the PHS (PMCMC), SMC-PMMH-DF and SMC-PG methods.
 \label{fig:SV_covariates} }
\begin{centering}
\psfrag{tausq}{$\tau^2$}
\includegraphics[width=15cm,height=9cm]{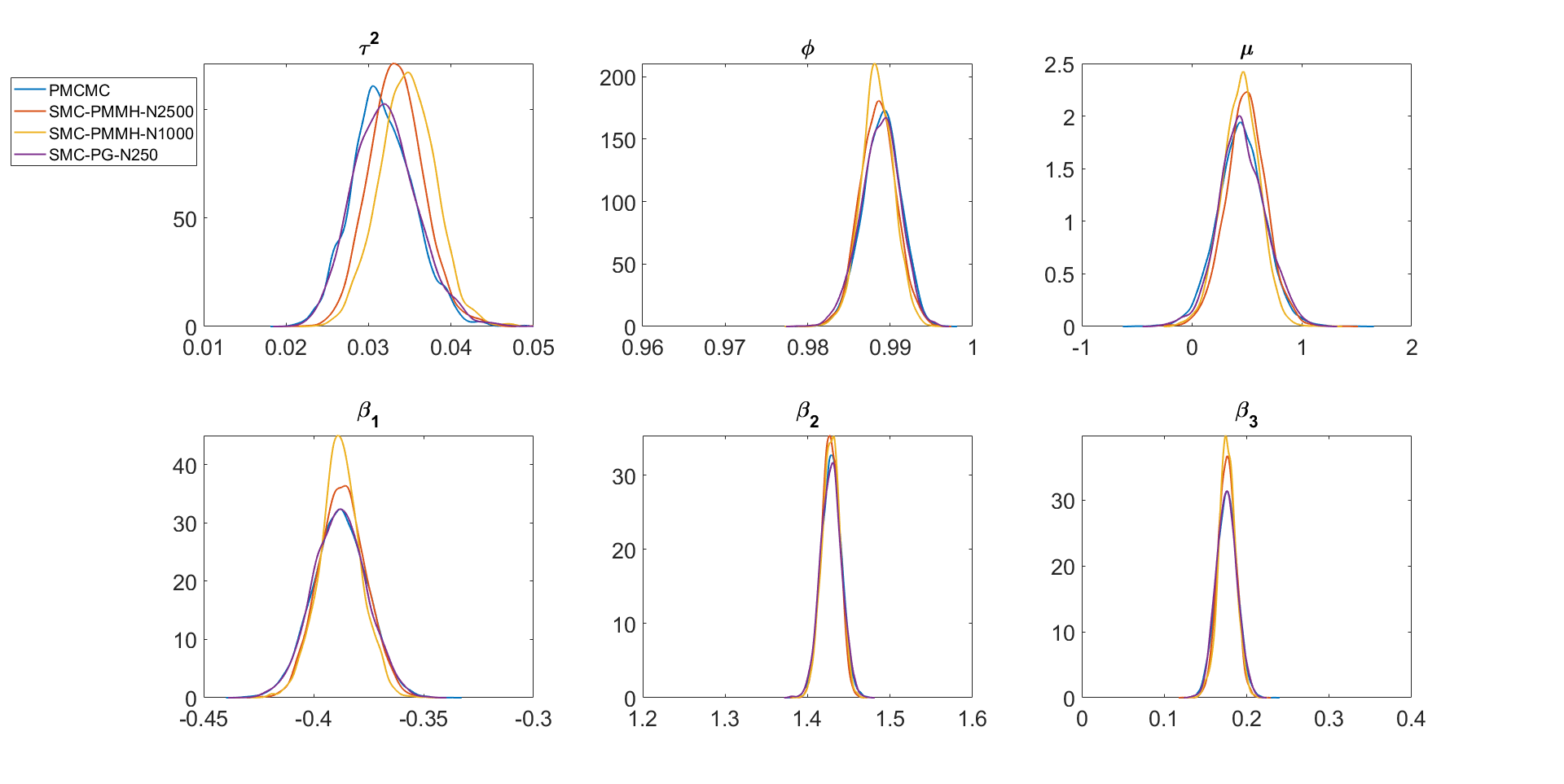}
\par\end{centering}
\end{figure}

\figref{fig:SV_covariates} shows the kernel density estimates of some marginal posterior densities of the univariate SV parameters with the covariate coefficients estimated using the PHS, the SMC-PMMH-DF and the SMC-PG methods. The figure shows that the SMC-PG estimates are very close to the PHS estimates for all SV parameters and the three covariate coefficients $(\beta_1,\beta_2,\beta_3)$. The estimates from the SMC-PMMH-DF are different to the PHS even with $N=2500$ particles. Similar conclusions hold for other $\beta$ coefficients given in Figure \ref{fig:betaunivsv} of Supplement \ref{univSVmodeladditional}.  \tabref{Table:SVcovariates} summarizes the estimation results of the SV model with covariates. 
Interestingly, the number of Markov move steps for the SMC-PMMH-DF is similar to the SMC-PG method for this example. The SMC-PG method is more accurate and 4.76 times faster than the SMC-PMMH-DF with $N=2500$ particles. 

This example suggests that (a) The SMC-PG performs better than the SMC-PMMH-DF for estimating the univariate SV parameters with covariate coefficients. The vector of parameters $\beta$ are high dimensional and not highly correlated with the states, so it is efficient to generate them in a PG step. (b) SMC-PMMH-DF gives different estimates to the PHS even with $N=2500$ particles. SMC-PMMH-DF uses a PMMH approach with a random walk proposal for the parameters in the Markov move component. The random walk is easy to implement, but is not efficient for high-dimensional parameters. However, SMC-PMMH-DF is a more general algorithm than SMC-PG. It only needs an efficient estimate of the likelihood. Section \ref{OUprocess} discusses an example that uses the SMC-PHS method where it is useful to generate parameters that are highly correlated with the states using PMMH steps and the other parameters are generated using PG steps by conditioning on the states.

\begin{table}[H]
\caption{Simulated data with $T=6000$ for the univariate SV with covariates
estimated using the SMC-PG and SMC-PMMH-DF samplers. Time is in minutes for one run of the algorithm.\label{Table:SVcovariates}}

\centering{}%
\begin{tabular}{ccccc}
\hline 
Method & N & R & P & Time\tabularnewline
\hline 
PG & 250 & 10 & 89 & 754.79\tabularnewline
PMMH-DF & 1000 & 10 & 72 & 1081.29\tabularnewline
PMMH-DF & 2500 & 10 & 72 & 3590.80\tabularnewline
\hline 
\end{tabular}
\end{table}




\subsection{Time Series Diagnostics \label{subsec:diagnostic}}

The sequential SMC approach in Section \ref{Sequential Estimation} allows us to efficiently estimate the
sequence of standard normal random variables used to form the goodness
of fit statistics to test for the adequacy of a general time series
state space model \citep{Gerlach1998}. Let $y_{1},...,y_{T}$ be
a sequence of time series observations generated by the continuous random variables $Y_{1},...,Y_{T}$. If the correct model is fitted to the data, then the sequence
\begin{equation}
u_{t}=p\left(Y_{t}\leq y_{t}|\bs y_{1:t-1}\right),\;t=1,...,T, \label{eq:uniformequation}
\end{equation}
is a realization of independent uniform $U\left(0,1\right)$ random
variables. A sequence of independent standard normal random variables
$N\left(0,1\right)$ is obtained by setting $v_{t}=\Phi^{-1}\left(u_{t}\right)$,
where $\Phi$ is the standard normal cumulative distribution function.
To obtain $u_{t}$, it is necessary to integrate out the parameters
and the latent states by evaluating
\begin{equation}
u_{t}=\int p\left(Y_{t}\leq y_{t}|\bs y_{1:t-1},\bs \theta,x_{t}\right)p\left(x_{t}|x_{t-1},\bs \theta\right)p\left(\bs \theta,\bs x_{1:t-1}|\bs y_{1:t-1}\right)d\bs \theta \, d{\bs x}_{1:t}.\label{eq:uniformintegraleqn}
\end{equation}

\begin{figure}[H]
\caption{Simulated data, left panel  -- the autocorrelation plot of the $\widehat{v}_{t}$ series;
right panel -- the QQ-plot of the $\widehat{v}_{t}$ series versus the standard normal
\label{fig:Simulated-data diagnostic sim} }

\centering{}\includegraphics[width=15cm,height=6cm]{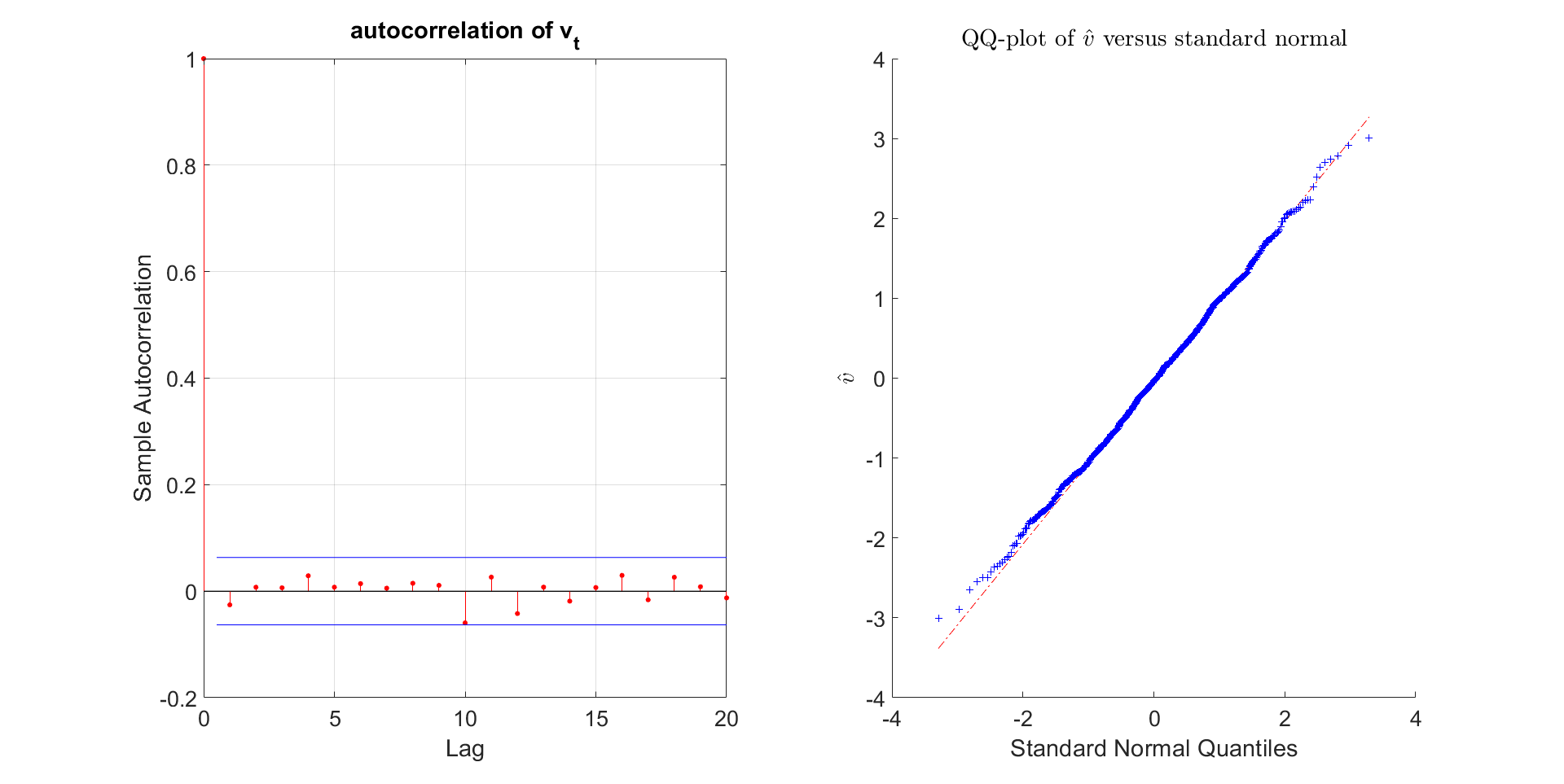}
\end{figure}

\begin{figure}[H]
\caption{US Food stock returns data, left panel  -- the autocorrelation plot of the $\widehat{v}_{t}$ series;
right panel -- the QQ-plot of the $\widehat{v}_{t}$ series versus the standard normal
\label{fig:real-data diagnostic real-1}}

\centering{}\includegraphics[width=15cm,height=6cm]{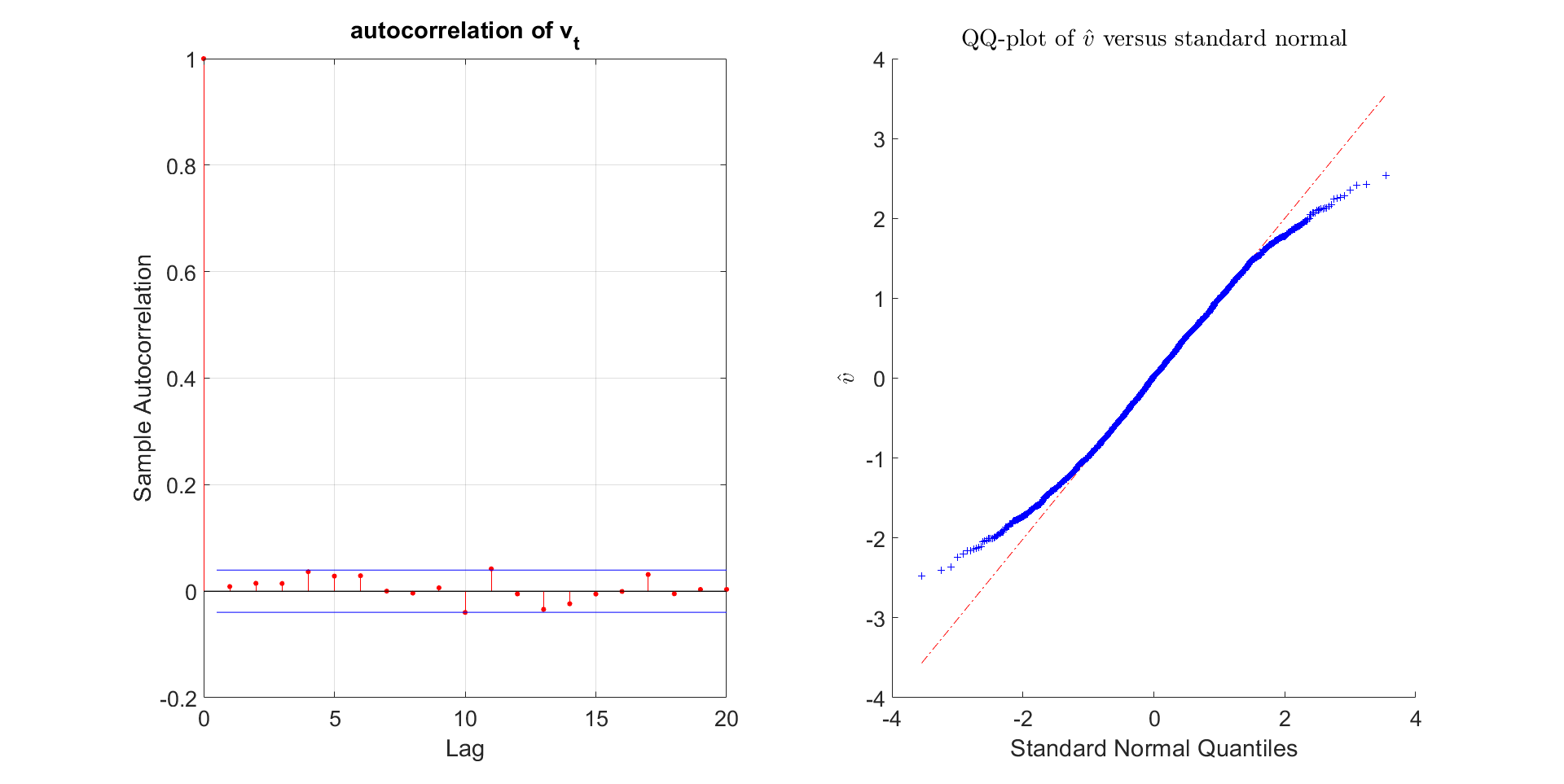}
\end{figure}

For many statistical models, such as the stochastic volatility model, the
integral in \eqref{eq:uniformintegraleqn} cannot be evaluated
analytically. One way to estimate $u_{t}$ is to use
MCMC or particle MCMC methods. Given a sample of MCMC draws, $\bs \theta_{m}$
and $x_{m,t}$ ($m=1,...,M$), from $p\left(\bs \theta,\bs x_{1:t}|\bs y_{1:t-1}\right)$, then
\[
\widehat{u}_{t}=\frac{1}{M}\sum_{m=1}^{M}p\left(Y_{t}\leq y_{t}|\bs y_{1:t-1},\bs \theta_{m},x_{m,t}\right).
\]
is an estimate of $u_{t}$. 
The sequence $\widehat{v}_{t}$ is obtained by setting $\widehat{v}_{t}=\Phi^{-1}\left(\widehat{u}_{t}\right)$
for $t=1,...,T$. However, MCMC  can be very time-consuming
as it is necessary to repeatedly construct the full Markov chain for each
time period $t$; see, e.g., \citep{Gerlach1998}. The sequential approach, denoted by SMC-PG-seq, can be used to
estimate $u_{t}$ efficiently. If the model is correct,
the sequence of $\{u_{t}\}$ is uniform and independent and the sequence of $\{v_{t}\}$ is standard normal and independent. We apply the approach to test for model adequacy
using simulated and real data. The simulated data uses the parameter values
$\phi=0.98$, $\mu=-0.48$, $\tau^{2}=0.02$, and $T=1000$ observations. A single run of the sequential approach, SMC-PG-seq,
with $M=560$ samples is used to obtain the estimate of $\widehat{u}_{t}$
and $\widehat{v}_{t}$, for $t=1,...,T$. Fig. \ref{fig:Simulated-data diagnostic sim}
shows the diagnostic plots for the series $\widehat{v}_{t}$
for $t=1,...,T$. The autocorrelation plot in the
figure suggests  that the  $v_{t}$
series is uncorrelated. 
The QQ-plot suggests  the $\widehat{v}_{t}$ series is normal.
The Anderson-Darling test \citep{Stephens1974} can also be used to test the null hypothesis that
the series $v_{t}$ is normally distributed;
the null hypothesis is not rejected at the 5\% level of significance (p-value=0.65).
This is expected because the dataset is simulated from the SV model.
We now apply the methods to a sample of daily US food industry stock
returns as in the previous section. Fig.
\ref{fig:real-data diagnostic real-1} shows the diagnostic plots
for the series $\widehat{v}_{t}$ for the real data. The autocorrelation plot in the figure suggests that the $v_t$ series
is uncorrelated; (2) the
quantiles of $\widehat{v}_{t}$ are similar to the quantiles of the standard
normal distribution, except in both tails. The Anderson-Darling test for normality of the
$v_t$ series is rejected at the 5\% level of significance (p-value=0.00). We  conclude from this result that the standard univariate SV model is inadequate for this series. 

This shows the usefulness of the SMC-PG-seq to estimate efficiently the sequence $u_t$  used to form the goodness of fit statistics to test for the adequacy of a general time series state space model. Another important advantage of sequential SMC over MCMC is that it provides sequential one-step ahead predictive densities of the log returns, and hence prices; this is particularly useful for financial applications. See Section \ref{univSVmodeladditional} of the Supplement. 



\subsection{SV Model with Outliers \label{SVoutliers}}
This section uses data simulated from a standard SV model contaminated by outliers to study the performance of the proposed sequential approach
SMC-PG-seq with tempering (denoted by SMC-PG-seqT) in Section \ref{Sequential Estimation} and compares it to
the standard sequential approach without tempering (denoted by SMC-PG-seqNT). 
In each scenario, a dataset is generated with $T=1000$ observations using parameter
values $\phi=0.98$, $\mu=-0.48$, and $\tau^{2}=0.02$. The
generated observations are then contaminated with noise $B_{t}\times\beta_{t}$,
where $B_{t}$ is a Bernoulli indicator with parameter $0.01$ and
$\beta_{t}$ is normally distributed with mean zero and standard deviation
$\kappa$. The parameter $\kappa$ controls the degree of contamination
and is set to $\left(5,10,15,25\right)$. A single dataset is simulated
in each case. \citet{Duan2015} find that the density tempered SMC method is more robust to outliers than the standard sequential approach because it incorporates the information from the data gradually. We use the estimates of the SMC-PG method as the
``gold standard'' to assess the accuracy of the sequential SMC-PG-seqT and SMC-PG-seqNT
approaches. All the SMC estimates are obtained using 10 independent
runs, each with $M=560$ samples. For the SMC-PG and SMC-PG-seqT, $\textrm{ESS}_\textrm{target}$ is set to $0.8M$.



Table \ref{tab:Simulation-results-for SV model with contamination}
summarises the simulation results for the SV models with different degrees
of data contamination $\kappa=\left(5,10,15,25\right)$. The first
three rows report the standard error of the posterior mean estimates
of the SV model parameters over the 10 runs of the algorithms;
the fourth and fifth rows report the mean and standard error of the
log of the marginal likelihood estimates. The SMC-PG-seqT and SMC-PG
provide stable results for all values of $\kappa$, but the performance
of SMC-PG-seqNT deteriorates as $\kappa$ increases. When
$\kappa=25$, the standard error of $\tau^{2}$ from SMC-PG-seqNT is $2.84$
and $2.09$ times larger than from SMC-PG-seqT and SMC-PG, respectively.
The Monte Carlo error of the SMC-PG-seqNT increases as outliers become
larger in magnitude. The outliers lead to highly variable weights
in the reweighting step of the SMC-PG-seqNT algorithm and very low effective
sample size (ESS) of the particles. More importantly, the Monte Carlo
error of the log of the marginal likelihood estimates seem to deteriorate
much faster than the parameter estimates as $\kappa$ increases. When $\kappa=25$, the standard
error of the log of the marginal likelihood estimate from SMC-PG-seqNT
is 28.05 and 21.49 times larger than from SMC-PG-seqT and SMC-PG, respectively.
The log of the marginal likelihood estimates from the PG-seqNT are
substantially different to the estimates from SMC-PG-seqT and SMC-PG
for $\kappa=10,15,25$. Section \ref{AddOutlier} of the Supplement gives further empirical results. Section \ref{SVbreaks} of the Supplement discusses the flexibility of the SMC-PG-seqT to estimate a two state Markov switching SV model.

\begin{sidewaystable}
\caption{Simulation results for the SV models with different degrees of data
contamination $\kappa=\left(5,10,15,25\right)$. The results are computed
from 10 independent runs of each algorithm. The first three rows report
the standard error of the posterior mean estimates of the SV parameters
$\left(\mu,\phi,\tau^{2}\right)$ over the 10 independent runs. The
fourth and fifth rows report the mean and standard error of the log
of the marginal likelihood estimates over 10 runs. \label{tab:Simulation-results-for SV model with contamination}}

\begin{centering}
{\footnotesize{}}%
\begin{tabular}{ccccccccccccc}
\hline
 & \multicolumn{4}{c}{{\footnotesize{}SMC-PG-seqT}} & \multicolumn{4}{c}{{\footnotesize{}SMC-PG-seqNT}} & \multicolumn{4}{c}{{\footnotesize{}SMC-PG}}\tabularnewline
\hline
{\footnotesize{}$\kappa$} & {\footnotesize{}5} & {\footnotesize{}10} & {\footnotesize{}15} & {\footnotesize{}25} & {\footnotesize{}5} & {\footnotesize{}10} & {\footnotesize{}15} & {\footnotesize{}25} & {\footnotesize{}5} & {\footnotesize{}10} & {\footnotesize{}15} & {\footnotesize{}25}\tabularnewline
\hline
{\footnotesize{}$\mu$ std err.} & {\footnotesize{}0.0100} & {\footnotesize{}0.0062} & {\footnotesize{}0.0058} & {\footnotesize{}0.0048 } & {\footnotesize{}0.0094} & {\footnotesize{}0.0051} & {\footnotesize{}0.0081} & {\footnotesize{}0.0047} & {\footnotesize{}0.0040} & {\footnotesize{}0.0026} & {\footnotesize{}0.0025} & {\footnotesize{}0.0020}\tabularnewline
{\footnotesize{}$\phi$ std err.} & {\footnotesize{}0.0006} & {\footnotesize{}0.0023} & {\footnotesize{}0.0014} & {\footnotesize{}0.0021} & {\footnotesize{}0.0007} & {\footnotesize{}0.0022} & {\footnotesize{}0.0018} & {\footnotesize{}0.0039} & {\footnotesize{}0.0003} & {\footnotesize{}0.0040} & {\footnotesize{}0.0010} & {\footnotesize{}0.0025}\tabularnewline
{\footnotesize{}$\tau^{2}$ std err.} & {\footnotesize{}0.0011} & {\footnotesize{}0.0058} & {\footnotesize{}0.0027} & {\footnotesize{}0.0069} & {\footnotesize{}0.0009} & {\footnotesize{}0.0063} & {\footnotesize{}0.0034} & {\footnotesize{}0.0196} & {\footnotesize{}0.0004} & {\footnotesize{}0.0087} & {\footnotesize{}0.0028} & {\footnotesize{}0.0094}\tabularnewline
{\footnotesize{}$\log\widehat{p}\left(\bs y\right)$ mean} & {\footnotesize{}-1381.15} & {\footnotesize{}-1431.94} & {\footnotesize{}-1276.53} & {\footnotesize{}-1348.71} & {\footnotesize{}-1381.83} & {\footnotesize{}-1457.13} & {\footnotesize{}-1289.08} & {\footnotesize{}-1459.16} & {\footnotesize{}-1382.53} & {\footnotesize{}-1432.79} & {\footnotesize{}-1278.11} & {\footnotesize{}-1349.54}\tabularnewline
{\footnotesize{}$\log\widehat{p}\left(\bs y\right)$ std err.} & {\footnotesize{}0.3812} & {\footnotesize{}0.5907} & {\footnotesize{}0.3311 } & {\footnotesize{}0.5871} & {\footnotesize{}0.5808} & {\footnotesize{}13.1455} & {\footnotesize{}7.2123} & {\footnotesize{}16.4654} & {\footnotesize{}0.1548} & {\footnotesize{}0.3069} & {\footnotesize{}0.3321} & {\footnotesize{}0.7661}\tabularnewline
\hline
\end{tabular}{\footnotesize\par}
\par\end{centering}
\end{sidewaystable}

\section{The Multivariate Factor Stochastic Volatility Model \label{sec:Multivariate-Factor-Stochastic}}

\subsection{Model \label{subsec:Model}}
The factor SV model is a parsimonious multivariate stochastic volatility model that is often used to
model a vector of stock returns; see, for example, \citet{Chib2006}
and \citet{Kastner:2017}. It is a high dimensional state space model having a large number of parameters and a large number of latent states.

Suppose that $\boldsymbol{P}_{t}$ is a
$S\times1$ vector of daily stock prices and define $\boldsymbol{y}_{t}\coloneqq\log\boldsymbol{P}_{t}-\log\boldsymbol{P}_{t-1}$
as the vector of stock returns. We model $\boldsymbol{y}_{t}$ as
a factor SV model
\begin{equation}\label{eq:factor model}
\boldsymbol{y}_{t}={\beta}\boldsymbol{f}_{t}+{V}_{t}^{\frac{1}{2}}\boldsymbol{\epsilon}_{t},\:\left(t=1,...,T\right),
\end{equation}
where $\boldsymbol{f}_{t}$ is a $K\times 1$ vector of latent factors
(with $K \ll S$), and ${\beta}$ is a $S\times K$ factor
loading matrix of unknown parameters. The model for the latent factor
is $\boldsymbol{f}_{t}\sim N\left(0,{D}_{t}\right)$ with
$\boldsymbol{\epsilon}_{t}\sim N\left(0,I\right)$. The time varying
variance matrices ${V}_{t}$ and ${D}_{t}$
 depend on the unobserved random variables $\boldsymbol{h}_{t}=\left(h_{1t},...,h_{St}\right)$
and $\boldsymbol{\lambda}_{t}=\left(\lambda_{1t},...,\lambda_{Kt}\right)$
such that
\[
{V}_{t}\coloneqq\textrm{diag}\left\{ \exp\left(h_{1t}\right),...,\exp\left(h_{St}\right)\right\} ,\;{D}_{t}:=\textrm{diag}\left\{ \exp\left(\lambda_{1t}\right),...,\exp\left(\lambda_{Kt}\right)\right\} .
\]
Each of the  log-volatilites $\lambda_{kt}$ and $h_{st}$ is assumed to follow  an independent first order autoregressive
process, with
\begin{equation}
h_{st}-\mu_{\epsilon s}=\phi_{\epsilon s}\left(h_{st-1}-\mu_{\epsilon s}\right)+\eta_{\epsilon st},\quad \eta_{\epsilon st}\sim N\left(0,\tau_{\epsilon s}^{2}\right),\quad s=1,...,S\label{eq:statetransitionidiosyncratic}
\end{equation}
and
\begin{equation}
\lambda_{kt}=\phi_{fk}\lambda_{kt-1}+\eta_{fkt},\quad \eta_{fkt}\sim N\left(0,\tau_{fk}^{2}\right), \quad k=1,...,K.\label{eq:statetransitionfactor}
\end{equation}
For $s=1,...,S$ and $k=1,...,K$, we choose the priors for the persistence
parameters $\phi_{\epsilon s}$ and $\phi_{fk}$, the priors $\tau_{\epsilon s}^{2}$,
$\tau_{fk}^{2}$, and $\mu_{\epsilon s}$ as in \secref{univariateexample}.
For every unrestricted element of the factor loadings matrix ${\beta}$,
we follow \citet{Kastner:2017} and choose independent Gaussian distributions
$N\left(0,1\right)$. These priors cover most possible values
in practice.

We also consider the factor stochastic volatility models, where
the log-volatilities $h_{s,t}$ follow a continuous time Ornstein-Uhlenbeck
(OU) process $\left\{ h_{s,t}\right\} _{t\geq1}$, introduced by \citet{Stein1991}.
The process satisfies, 
\begin{equation}
dh_{s,t}=\left\{ \alpha_{\epsilon,s}\left(\mu_{\epsilon,s}-h_{s,t}\right)\right\} dt+\tau_{\epsilon,s}dW_{s,t},
\end{equation}
where $W_{s,t}$ are independent Wiener processes. Although the continuous
time OU diffusion model has a closed form transition density \citep{Brix2018}, we
investigate the performance of the proposed SMC samplers
to estimate state space models using an approximation such as the
Euler discretisation.

The Euler scheme can be used as an approximation by placing $M-1$
evenly spaced points between times $t$ and $t+1$. We denote the
intermediate volatility components by $h_{s,t,1},...,h_{s,t,M-1}$
and set $h_{s,t,0}=h_{s,t}$ and $h_{s,t,M}=h_{s,t+1}$. The equation
for the Euler evolution, starting at $h_{s,t,0}$ is 
\begin{equation}
h_{s,t,j+1}|h_{s,t,j}\sim N\left(h_{s,t,j}+\left\{ \alpha_{\epsilon,s}\left(\mu_{\epsilon,s}-h_{s,t,j}\right)\right\} \delta,\tau_{\epsilon,s}^{2}\delta\right),
\end{equation}
for $j=0,...,M-1$, where $\delta=1/M$. 
We use the following priors for the OU parameters  and  assume the parameters are independent apriori:
(a) $p\left(\mu_{\epsilon,s}\right)\propto I\left(-10<\mu_{\epsilon,s}<10\right)$, (b) the priors for $\tau^{2}_{\epsilon,s}$ and $\alpha_{\epsilon,s}$ are inverse
Gamma $\textrm{IG}\left(v_{0}/2,s_{0}/2\right)$ with $v_{0}=10$
and $s_{0}=1$, for $s=1,...,S$. \secref{sec:Sampling-factor-loading} of the Supplement discusses parameterisation and identification issues regarding the factor loading matrix ${\beta}$ and the latent factors $\bs f_t$.

\subsubsection*{Conditional Independence in the factor SV model}
The key to making the estimation of the factor SV model tractable
is that given the values of $\left(\boldsymbol{y}_{1:T},\boldsymbol{f}_{1:T},{\beta}\right)$,
the factor model in \eqref{eq:factor model} separates into
$S+K$ independent components consisting of $K$ univariate SV models
for the latent factors with $f_{kt}$ the $t$th \lq observation\rq{} of the $k$th
factor univariate SV model and $S$ univariate SV models for the idiosyncratic
errors with $\epsilon_{st}$ the $t$th \lq observation\rq{} on the $s$th idiosyncratic error
SV model.  \secref{augmentedtargetdensityfactorSVmodel} of the Supplement
discusses the SMC-HMC, SMC-PG, and SMC-PHS methods for the factor SV model.


\subsection{Examples \label{subsec:Multivariate-Factor-Stochastic examples-1}}

This section investigates the performance of the SMC samplers to estimate the 
multivariate factor SV model discussed in
\secref{subsec:Model} using one factor. 
A sample of  daily returns for $S=26$ value weighted industry portfolios is used,
from December 11th, 2001 to 29th November 2005, a total of $1000$
observations. The data is obtained from the Kenneth French website, with the industry portfolios used
listed in
\secref{sec:List-of-Industry} of the Supplement. As in \secref{subsec:Univariate-Stochastic-Volatility examples-1}, the PHS is regarded as the \lq\lq gold
standard\rq\rq{}; it is run for $50000$
iterates, with another $5000$ iterates used as burn-in.


Our first study discusses how well  SMC-PMMH-DF \citep{Duan2015} estimates the factor SV model.
SMC-PMMH-DF uses the PMMH Markov
steps and follows the  \citet{Pitt:2012} guidelines to set the optimal
number of particles in the particle filter to ensure that the variance
of the log of the estimated likelihood is around $1$. The PMMH Markov step
generates the parameters of the latent factors, with the factor and idiosyncratic
log-volatilities \lq\lq integrated out\rq\rq{},   resulting in a $\left(S+K\right)$
dimensional state vector. The tempered measurement density at the $p$th stage is
\begin{equation*}\label{eq:PMMH observation density}
\bigg \{ N\left(\bs y_t; 0,\Sigma_{t}={\beta}D_{t}{\beta}^{'}+{V}_{t}\right)\bigg \}^{a_{p}},
\end{equation*}
 \eqref{eq:statetransitionidiosyncratic} gives the state transition densities for the
 idiosyncratic log-volatilities ($s=1, \dots, S$) and \eqref{eq:statetransitionfactor}
gives the state transition equations for the factor log-volatilities ($k=1, \dots, K$).

Table \ref{tab:The-Variance-of log-likelihood}
shows the variance of log of the estimated likelihood for different numbers
of particles evaluated at posterior means of the parameters obtained
using the PHS of \citet{Gunawan2020PHS} with $a_{p}=1$. It
shows that even with $5000$ particles, the PMMH Markov step would
get stuck. \citet{Deligiannidis2018} proposed the correlated
PMMH method and it is possible to implement
it in the Markov move step instead of the standard PMMH
method of \citet{Andrieu:2010}. 

\begin{table}[H]
\caption{The variance of the log of estimated likelihood for the PMMH step for different numbers
of particles for the US stock returns dataset; $T=1000$, $S=26$,
and $K=1$ with the tempered sequence $a_{p}$  set to 1, evaluated
at the posterior means of the parameters obtained from the PHS of \citet{Gunawan2020PHS}. Time is time in seconds  to
compute the log of the estimated likelihood once.\label{tab:The-Variance-of log-likelihood}}
\centering{}%
\begin{tabular}{ccc}
\\
\hline
Number of Particles & Variance of log of estimated likelihood & Time\tabularnewline
\hline
\hline
250 & 2198.72 & 4.86\tabularnewline
500 & 1164.51 & 9.88\tabularnewline
1000 & 813.53 & 20.13\tabularnewline
2500 & 439.05 & 50.43\tabularnewline
5000 & 345.85 & 99.24\tabularnewline
\hline
\end{tabular}
\end{table}

The correlated PMMH correlates the random
numbers used in constructing the estimators of the likelihood at current
and proposed values of the parameters and sets the correlation very
close to 1 to reduce the variance of the difference in the logs of
estimated likelihoods at the current and proposed values of the parameters
appearing in the Metropolis-Hastings (MH) acceptance ratio.
\citet{Deligiannidis2018} show that the correlated PMMH can be much more efficient
and can significantly
reduce the number of particles required by the standard PMMH approach
when the dimension of the latent states is small. However, the current
 example considers
a high dimensional latent state vector in the one factor-Factor SV model. \citet{Mendes2020} found that
it is very challenging to preserve the correlation between the
logs of the estimated likelihoods for such a high dimensional state space model.
The Markov move based on the correlated PMMH approach will also  get stuck
at lower temperatures unless enough particles are used to ensure the variance of the log
of the estimated likelihood is around 1.

A second drawback of the PMMH
Markov move step as in \citet{Duan2015} is that the dimension of the parameter
space in the factor SV model is large making it very hard to implement the
PMMH Markov step efficiently. Section \ref{subsec:Univariate-Stochastic-Volatility examples-1} shows that SMC-PG is much more efficient than SMC-PMMH-DF for estimating univariate SV model with $50$ covariates. It is difficult to obtain good proposals for the high-dimensional parameters because the first and second derivatives of log of the estimated likelihood with
respect to the parameters are unavailable analytically and can only
be estimated. \citet{Sherlock2015} note that in general it is
even more difficult to obtain accurate estimate of the gradient of the
log of the estimated likelihood  than it is to obtain accurate estimates of the log of the estimated likelihood.
 \citet{Nemeth2016} and \citet{Mendes2020} found that the behaviour of particle Metropolis adjusted Langevin Algorithm depends critically on how accurately we can estimate the gradient of the log-posterior. If the variance of the gradient of the log-posterior is insufficiently small, then there is no advantage in using particle MALA over the random walk proposal. Table \ref{tab:The-Variance-of log-likelihood} shows that the variance of the log of the estimated likelihood is still very large, even with $5000$ particles. Therefore, there is no advantage in using particle MALA. The random walk proposal is easy to implement, but it is very inefficient
in high dimensions.

The second study compares PHS to the approximate MCMC sampler of \citet{Kastner:2017} for estimating the factor SV model.   \citeauthor{Kastner:2017} use the approach proposed by \citet{Kim1998} to approximate the distribution of innovations in the log outcomes by a mixture of normals. \citeauthor{Kim1998} correct their approximation by importance sampling, which gives a simulation consistent estimation for the univariate SV model. However, the \citeauthor{Kastner:2017} estimator is not simulation consistent because it 
does not correct for these approximations for the factor SV model.  The PHS is simulation consistent in the sense that as the number of samples of the parameters $\boldsymbol{\theta}$ and latent states $\boldsymbol{x}_{1:T}$ tends to infinity, the PHS estimates converge to their true posterior distributions.
We implement the MCMC of \citeauthor{Kastner:2017}  using the R package factorstochvol \citep{Hosszejni2019}, using 
the default priors in the R package factorstochvol
in the comparison. The priors for $\tau^{2}_{\epsilon,s}$ for $s=1,...,S$ and $\tau^{2}_{f,k}$ for $k=1,...,K$ are $G(0.5,1)$ gamma density with shape $0.5$ and scale $1$. The prior for $\mu_{\epsilon,s}$ for $s=1,...,S$ is $N(0,10^{14})$.
We use the priors given in Section \ref{subsec:Model} for the other parameters.

\begin{figure}[H]
\caption{The marginal posterior density plots of the parameters $\mu_{\epsilon,16}$, $\phi_{\epsilon,16}$, and $\tau_{\epsilon,16}$ of the factor SV model estimated using PMCMC (PHS) and the \citeauthor{Kastner:2017} MCMC. \label{muphitauFactorSV26Kastner}}

\centering{}\includegraphics[width=15cm,height=8cm]{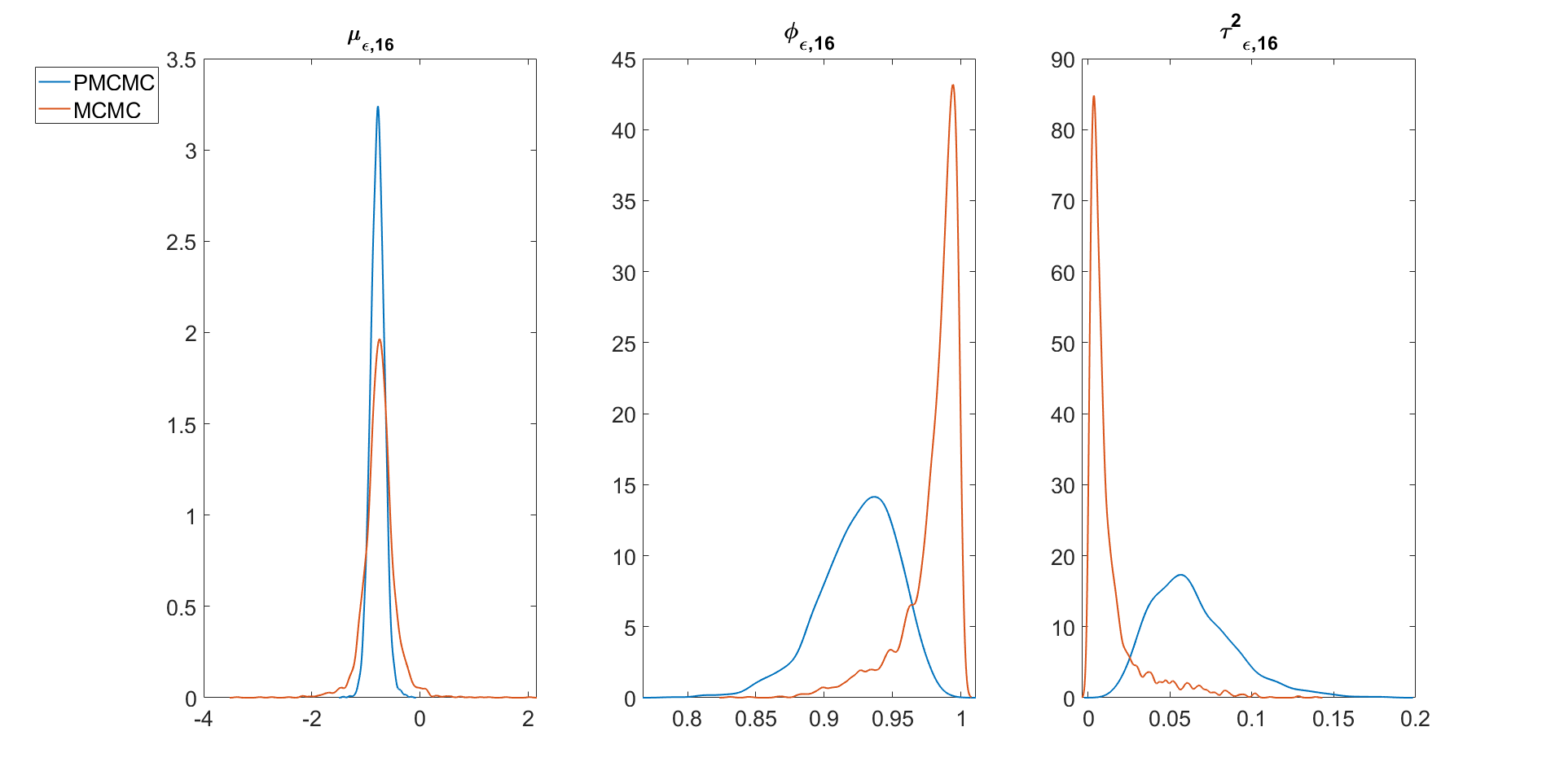}
\end{figure}

\figref{muphitauFactorSV26Kastner} shows the marginal posterior density estimates of the parameters $\mu_{\epsilon,16}$, $\phi_{\epsilon,16}$, and $\tau_{\epsilon,16}$ of the factor SV model using PMCMC (PHS) and the MCMC sampler of \citeauthor{Kastner:2017}. The figure shows that the MCMC sampler gives different estimates to PHS. Figures~\ref{tau2FactorSVall} to \ref{B1FactorSVall} of Section  \ref{additionalresultfactorsvmodel} of the Supplement give results for the other parameters in the factor SV model. The figures show that although some of the parameters estimated using the MCMC \citeauthor{Kastner:2017} sampler are close to the PHS,  some others are quite different. In addition, the Gibbs type MCMC sampler as in \citeauthor{Kastner:2017} cannot handle the factor SV with the log-volatility following diffusion processes discussed in Section \ref{OUprocess}. It is well known that the Gibbs sampler is inefficient for generating parameters for a diffusion model, in particular the variance parameter \citep{Stramer2011}. 


The third study investigates the performance of the SMC-PG and SMC-PHS methods and compares them to the PHS. 
The SMC estimates are obtained using 10 independent runs with $M=280$
samples to generate a total of $2800$ samples for each algorithm.
For the SMC methods, $\textrm{ESS}_\textrm{target}=0.8M$. The computation is done using a Matlab implementation of the algorithms and 28 CPU-cores of a high performance computer cluster.

We use the following notation:
$\textrm{SMC-PG}\left(\theta\right)$ means sampling
$\theta$ using the SMC-PG method;
$\textrm{SMC-PHS}\left(\theta_{1};\theta_{2}\right)$ means sampling $\theta_{1}$ in the PMMH step in Part 1 of Algorithm \ref{alg:Sampling-Scheme:-The correlated PMMH+PG}
and $\theta_{2}$ in the PG step. The strategy for the SMC-PHS method is to generate parameters that are subject to slow convergence using a PMMH step and the rest of the parameters are generated conditional on the states using PG steps. We first run the  particle Gibbs (PMCMC) sampling algorithm for all the parameters to identify which parameters have convergence issues. We then generate these parameters in the PMMH step in SMC-PHS algorithm.  \citet{Gunawan2020PHS} find that the particle Gibbs sampler usually generates   the variance parameters in the state transition equation inefficiently, i.e., $\tau^{2}_{\epsilon,s}$ for $s=1,...,S$ and $\tau^{2}_{f,k}$ for $k=1,...,K$. We compare the following methods: (I) the SMC-PHS $\left(\bs{\tau}_{\epsilon}^{2},\bs{\tau}_{f}^{2};\bs{\mu}_{\epsilon},\bs{\phi}_{\epsilon},\bs{\phi}_{f},\bs{f}_{1:T},\bs\beta\right)$,
(II) the $\textrm{SMC-PG}\left(\bs{\tau}_{\epsilon}^{2},\bs{\tau}_{f}^{2},\bs{\mu}_{\epsilon},\bs{\phi}_{\epsilon},\bs{\phi}_{f},f_{1:T},\bs{\beta}\right)$, (III) PHS $\left(\bs{\tau}_{\epsilon}^{2},\bs{\tau}_{f}^{2};\bs{\phi}_{f},\bs{f}_{1:T},\bs{\beta},\bs{\mu}_{\epsilon},\bs{\phi}_{\epsilon}\right)$.



\begin{figure}[H]
\caption{The marginal posterior density plots of the parameters $\mu_{\epsilon,16}$, $\phi_{\epsilon,16}$, and $\tau_{\epsilon,16}$ of the factor SV model estimated using PMCMC (PHS), SMC-PG, and SMC-PHS. \label{muphitauFactorSV26}}

\centering{}\includegraphics[width=15cm,height=8cm]{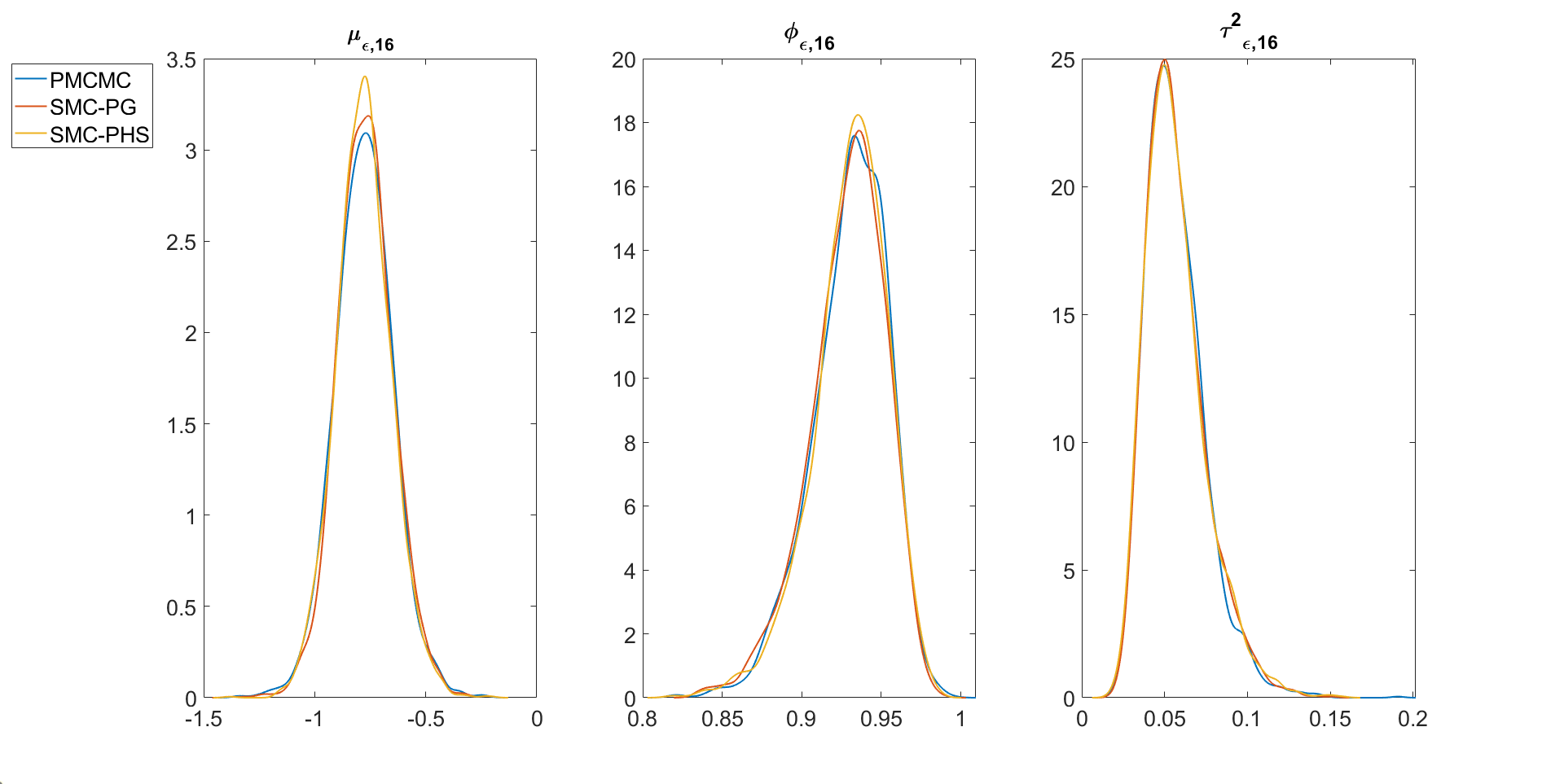}
\end{figure}

\figref{muphitauFactorSV26} plots the marginal posterior density estimates of the parameters $\mu_{\epsilon,16}$, $\phi_{\epsilon,16}$, and $\tau_{\epsilon,16}$ of the factor SV model estimated using PMCMC (PHS), SMC-PG and SMC-PHS. The figure shows that the SMC-PG and SMC-PHS methods give the same estimates as PHS, with SMC-PG  slightly more accurate at estimating $\mu_{\epsilon, 16}$. Figures~\ref{tau2FactorSVallSMCPG} to \ref{B1FactorSVallSMCPG} in the Supplement show all the other parameters of the factor SV model. In general, the SMC-PG and SMC-PHS estimates are very close to the PHS estimates for all parameters. Table \ref{FactorSVTable} shows the estimates
of the log of the marginal likelihood for the one factor model estimated using the SMC-PHS and SMC-PG methods. The table shows  that the estimated standard errors of the estimates of the log of the marginal likelihood
estimated using SMC-PHS are 1.87 times bigger than for the SMC-PG methods. The standard errors can be reduced by increasing the number of SMC samples $M$ and setting the $\textrm{ESS}_\textrm{target}>0.8M$. The table also shows that the number of annealing steps of SMC-PG and SMC-PHS methods are comparable.  The Markov move based on the PHS is  computationally more expensive than the Markov move based on PG only because it is necessary to run the particle filter twice at each iteration.

We now compare the predictive performance of the SMC-PG and SMC-PHS methods
with the (exact) PHS. The minimum variance portfolio implied by the $h$ step-ahead time-varying covariance
matrix $\Sigma_{T+h}$, is considered;  it can
be used to uniquely define the optimal portfolio weights  \citep{Bodnar2017},
\[
w_{port,T+h} :=\frac{\Sigma_{T+h}^{-1}\mathbb{1}}{\mathbb{1}^{\top}\Sigma_{T+h}^{-1}\mathbb{1}},
\]
where $\mathbb{1}$ denotes an S-variate vector of ones. The optimal portfolio weights give the lowest possible risk for a given expected portfolio return.

\begin{figure}[H]
\caption{Plot of the predictive densities $\widehat{p}(y_{T+h}|y_{1:T})$ of an optimally weighted portfolio, for $h=1,...,10$, estimated using PMCMC (PHS), SMC-PG, and SMC-PHS. \label{PredDensFactorSV}}

\centering{}\includegraphics[width=15cm,height=8cm]{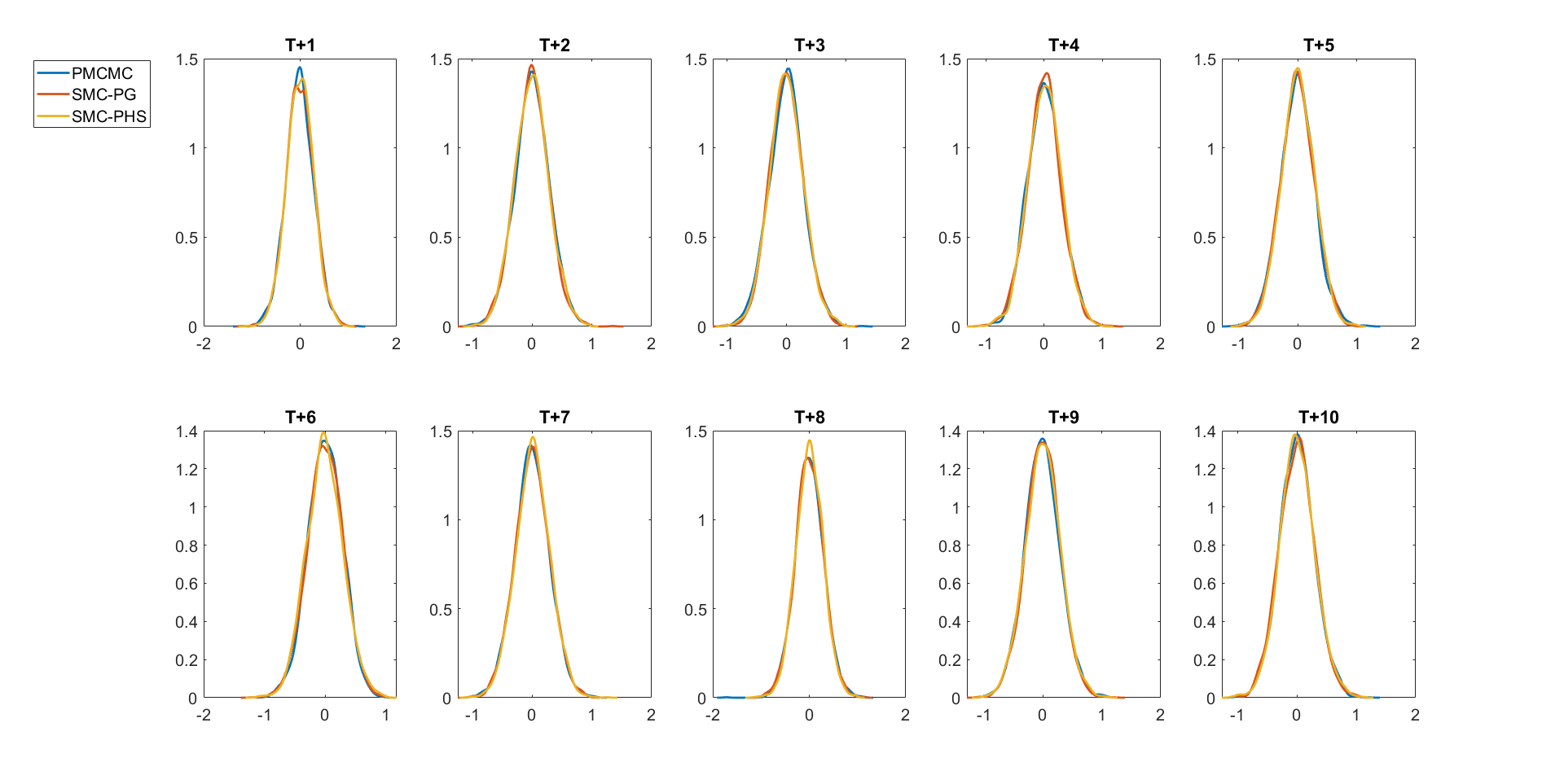}
\end{figure}

Figure~\ref{PredDensFactorSV}
shows the multiple-step ahead predictive densities $\widehat{p}\left({y}_{T+h}|\bs{y}_{1:T}\right)$
of an optimally weighted combination of all series, for $h=1,...,10$ obtained
using SMC-PG, SMC-PHS, and PHS. The figure shows that the
predictive densities obtained using SMC-PG method are very close to the exact predictive densities
obtained from the PHS. The predictive densities obtained using SMC-PHS are slightly less accurate. \secref{additionalresultfactorsvmodel} of the Supplement give additional empirical results for the factor SV model.

This example suggests  that: (a)~the PMMH Markov move is expensive for the factor SV model,  as it requires the number of particles to be greater than $5000$; (b)~the PMMH Markov move is unsuitable for the factor SV model because the dimension of the parameter space in the factor SV model is large; (c)~the approximate MCMC method of \citeauthor{Kastner:2017} can give unreliable estimates for some parameters in the factor SV model; (d)~for the standard factor SV model, SMC-PG is faster and slightly more accurate than SMC-PHS in estimating the posterior densities of the factor SV parameters and the predictive densities.

\begin{table}[H]

\caption{Factor SV model estimated using PHS, SMC-PG, and SMC-PHS 
for the 
US returns data with $T=1000$, $S=26$, and $K=1$. Time is  in minutes for one run of the algorithm. The table gives: (i)
the estimates of the log of the marginal likelihood $\textrm{log}\widehat{p}\left(\boldsymbol{y}_{1:T}\right)$
based on the average of the 10 runs, and the standard error of the
estimate (in brackets). (ii) the average value of the number of annealing
steps $P$ averaged over the 10 runs \label{FactorSVTable}}

\vspace{2mm}

\centering{}%
\begin{tabular}{cccccc}
\hline 
Method & N & R & P & Time & $\textrm{log}\widehat{p}\left(\boldsymbol{y}_{1:T}\right)$\tabularnewline
\hline 
SMC-PG & 250 & 10 & 289 & 2307.38 & $\underset{\left(30.29\right)}{-26149.66}$\tabularnewline
SMC-PHS & 250 & 10 & 274 & 6003.34 & $\underset{\left(56.55\right)}{-26176.99}$\tabularnewline
\hline 
\end{tabular}
\end{table}

\subsection{Factor SV Model with the Ornstein-Uhlenbeck Diffusion Processes \label{OUprocess}}
This section considers the factor stochastic volatility model, where the log-volatilities of the factors follow a standard SV model and 
the idiosyncratic log-volatilities  follow a continuous time Ornstein-Uhlenbeck
(OU) process $\left\{ h_{s,t}\right\} _{t\geq1}$  \citep{Stein1991}.
Although the continuous
time OU diffusion model has a closed form transition density \citep{Brix2018}, this section investigates the performance of SMC-PG and SMC-PHS when estimating the state space model using an approximation such as the Euler discretisation. For this example, we use the first $S=10$ daily portfolio returns listed in \secref{sec:List-of-Industry} of the
Supplement, 
from December 11th, 2001 to 29th November 2005, a total of $1000$
observations.

\begin{figure}[H]
\caption{Trace plots of the parameters $\bs\tau_{\epsilon}$, $\bs{\alpha}_{\epsilon}$, and $\bs\mu_\epsilon$ of the factor SV model, where the idiosyncratic errors follow the OU process estimated using particle Gibbs. \label{PGtrace}}

\centering{}\includegraphics[width=15cm,height=8cm]{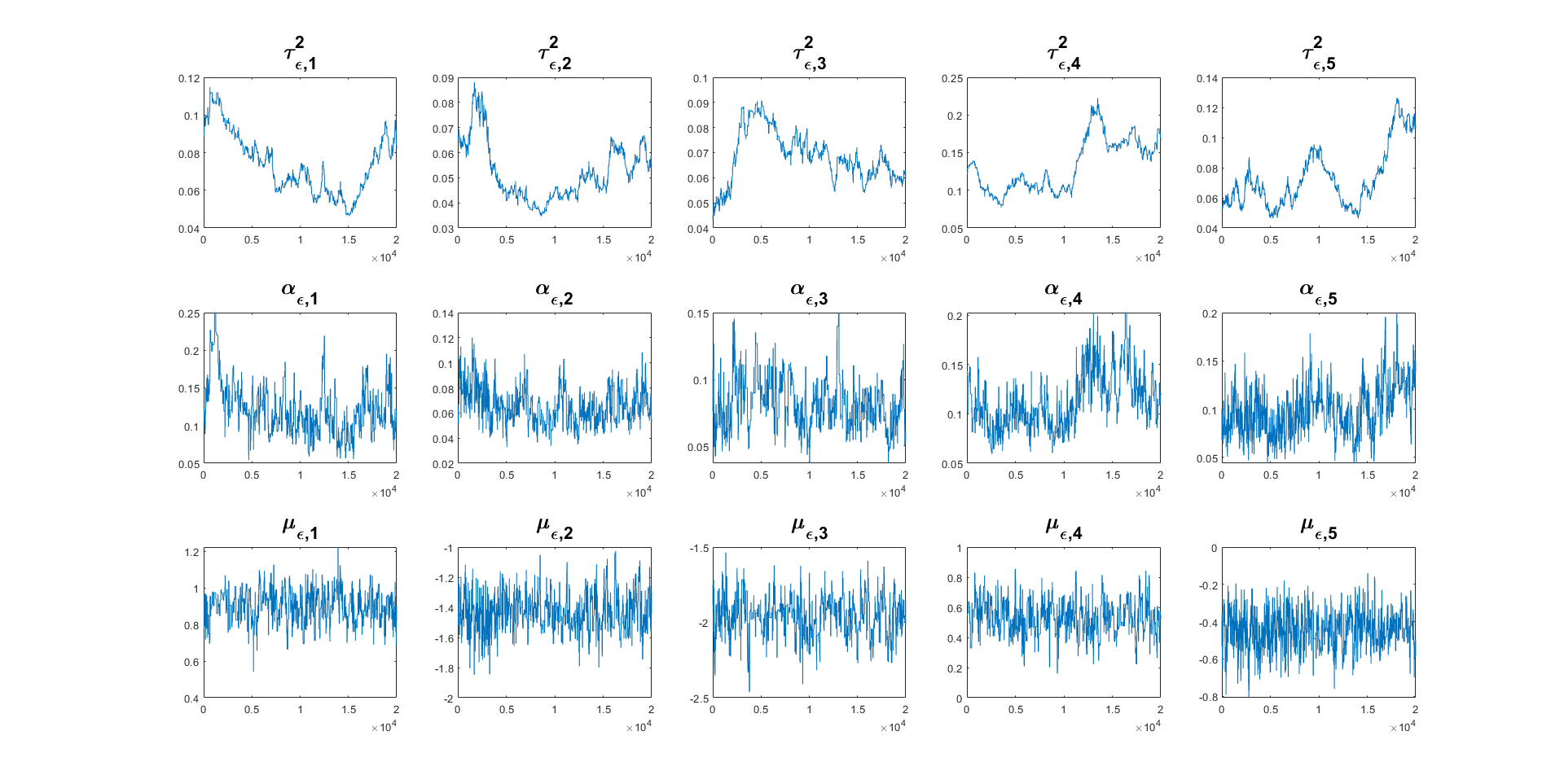}
\end{figure}

\figref{PGtrace} shows the trace plots of the parameters  $\bs\tau_{\epsilon}$, $\bs{\alpha}_{\epsilon}$, and $\bs\mu_\epsilon$ of the factor SV model, where the idiosyncratic log-volatilities follow OU processes and the factor log-volatilities follow the standard SV model estimated using particle Gibbs. The figure shows that the posterior draws of $\bs\tau^{2}_{\epsilon}$ and $\bs{\alpha}_{\epsilon}$ are highly autocorrelated and fail to converge. We estimate all the OU parameters in the PMMH steps in PHS and SMC-PHS. We compare the following samplers: (I) the SMC-PHS $\left(\bs\tau_{\epsilon}^{2},\bs\tau_{f}^{2},\bs\mu_{\epsilon},\bs\alpha_{\epsilon};\bs\phi_{f},\bs{f}_{1:T},\beta\right)$,
(II) the $\textrm{SMC-PG}\left(\bs\tau_{\epsilon}^{2},\bs\tau_{f}^{2},\bs\mu_{\epsilon},\bs\alpha_{\epsilon},\bs\phi_{f},\bs{f}_{1:T},\beta\right)$, (III) PHS $\left(\bs\tau_{\epsilon}^{2},\bs\tau_{f}^{2},\bs\mu_{\epsilon},\bs\alpha_{\epsilon};\bs\phi_{f},\bs{f}_{1:T},\beta\right)$, and (IV) the $\textrm{PG}\left(\bs\tau_{\epsilon}^{2},\bs\tau_{f}^{2},\bs\mu_{\epsilon},\bs\alpha_{\epsilon},\bs\phi_{f},\bs{f}_{1:T},\beta\right)$. The PG and PHS are run for 20000 iterates with another 5000 iterates used as burn-in.



\begin{figure}[H]
\caption{The marginal posterior density estimates of the parameters $\bs\tau_{\epsilon}$ of the factor SV models, where the idiosyncratic errors follow the OU process estimated using PHS, PG, SMC-PG, and SMC-PHS. \label{tauFactordiffusion}}

\centering{}\includegraphics[width=15cm,height=8cm]{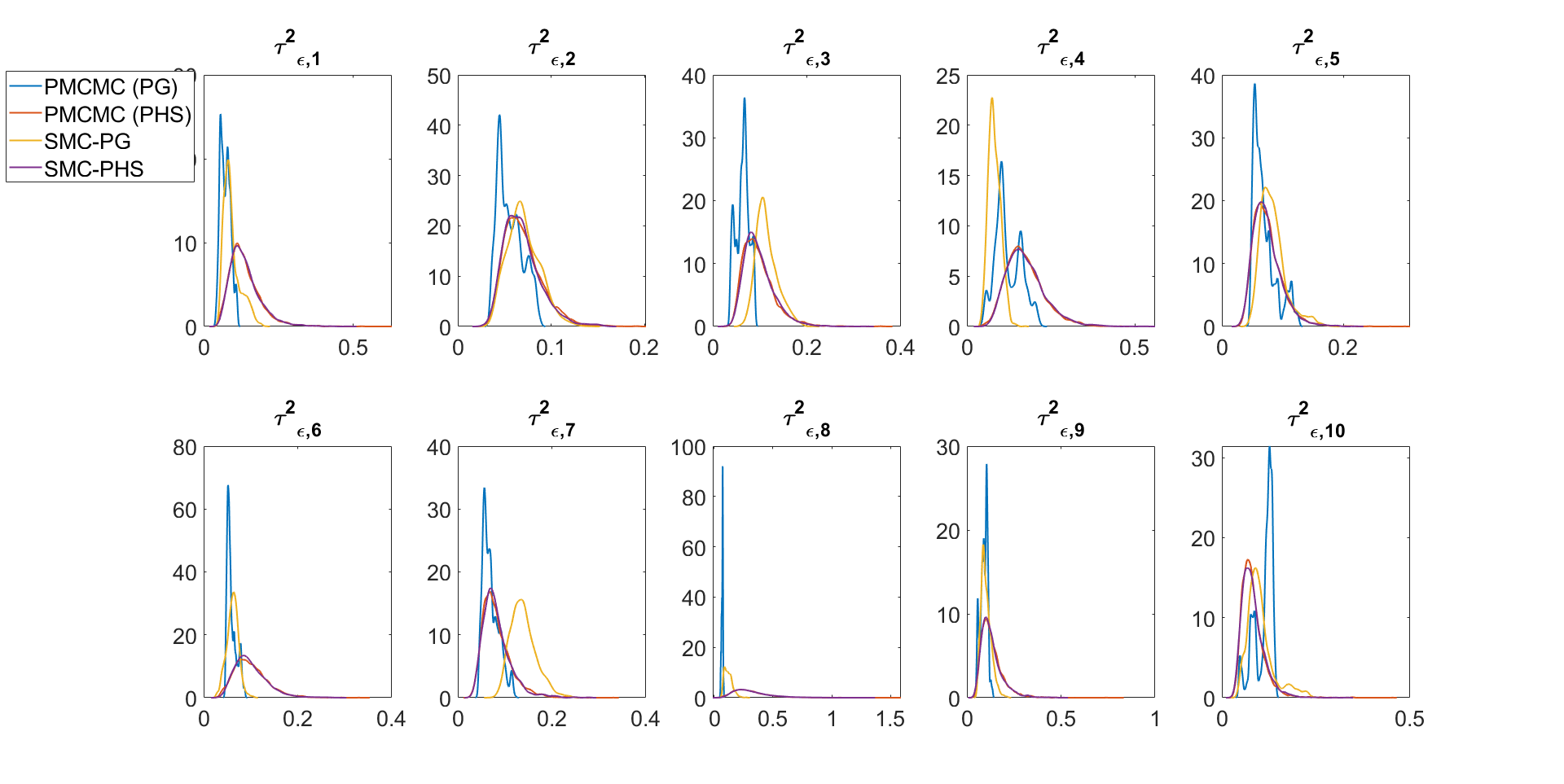}
\end{figure}

\figref{tauFactordiffusion} shows the marginal posterior density plots of the parameters $\tau^{2}_{\epsilon}$ of the factor SV models, where the idiosyncratic errors follow  OU processes estimated using the PHS, PG, SMC-PG, and SMC-PHS methods. The figure shows that the PG and SMC-PG estimates are very different to PHS. SMC-PHS gives the same estimates as PHS. Similar observations hold for the parameters $\bs{\alpha}_{\epsilon}$ in \figref{aFactordiffusion} of Section   \ref{additionalresultfactorsvmodel} of the Supplement. This example suggests  that: (a)  The SMC-PG estimates for the model parameters are unreliable and are highly correlated with the states. (b) The SMC-PHS estimates are  accurate as they generate the parameters that are highly correlated with the states using PMMH steps, and the other parameters are generated using PG steps.

\section{Conclusions \label{sec:Conclusions}}

The paper proposes flexible SMC approaches that can be used for both batch and sequential
state and parameter estimation problems,  and which also lead to accurate one-step and multiple-step ahead predictions. 
The approaches are based on Markov  move steps that are more efficient than in previous SMC approaches in \citet{Duan2015},  \citet{Fulop2013}, and \citet{Chopin2013}.
We show that the methods work well for both simulated and real data and can handle higher-dimensional states and parameters than previously possible by current SMC approaches. 
The data analyses suggest that (a) SMC-PG gives parameter and states estimates close to PHS for both univariate and standard factor SV models. It also accurately estimates the predictive densities of individual stock returns and the minimum variance portfolio. (b) SMC-PG gives inaccurate parameter estimates for $\bs\alpha_{\epsilon}$ and $\bs\tau^{2}_{\epsilon}$ for the factor SV model with the idiosyncratic log-volatilities following OU processes. This is because the OU parameters, in particular, the $\bs\alpha_{\epsilon}$ and $\bs\tau^{2}_{\epsilon}$ are highly correlated with the states.
(c) SMC-PHS provides flexible Markov move steps, where the parameters that are highly correlated with the states are generated by PMMH and all other parameters are generated by PG steps conditioning on the states. \secref{OUprocess} shows that the SMC-PHS gives accurate estimates for the OU parameters. (d) The SMC-PMMH-DF of \citet{Duan2015} cannot handle state space models with a large number of parameters and states. \secref{subsec:Multivariate-Factor-Stochastic examples-1} gives further details. (e) The MCMC sampler of \citet{Kastner:2017} gives similar results to the PHS for most of the factor SV parameters, but some parameter estimates are different. The Gibbs type MCMC sampler as in \citet{Kastner:2017} cannot handle the factor SV with the log-volatility following diffusion processes. (f) The sequential SMC approach allows us to efficiently estimate the sequence of standard normal random variables used to form the goodness of fit statistics to test for the adequacy of a general time series state space model. (g) The sequential approach with tempering provides joint sequential inference for states and parameters and is more robust than the sequential approach without tempering.

\if1\blind
{
\section{Acknowledgement} The research of Robert Kohn and David Gunawan was partially supported by an ARC Center of Excellence grant CE140100049. We thank the AE and the four reviewers for their comments which improved  both the presentation and the technical content of the paper. 
} \fi
\if0\blind
{
\section{Acknowledgement}  We thank the AE and the four reviewers for their comments which improved  both the presentation and the technical content of the paper. 
} \fi

\section{Supplementary Material}
This article has an online supplement that contains additional technical details and  empirical results.





\clearpage
\if1\blind
{
  \title{\bf Online Supplement: Flexible and Robust Particle Tempering for State Space Models}
\author[2,4]{David Gunawan}
\author[1,4]{Robert Kohn}
\author[3,4]{Minh Ngoc Tran}
\affil[1]{School of Economics, UNSW Business School, University of New South Wales}
\affil[2] {School of Mathematics and Applied Statistics, University of Wollongong}
\affil[3]{Discipline of Business Analytics, University of Sydney}
\affil[4]{Australian Center of Excellence for Mathematical and Statistical Frontiers}

\renewcommand\Authands{ and }
\maketitle
} \fi

\if0\blind
{
   \title {\bf Online Supplement: Robust Particle Density Tempering for State Space Models}
\author{}
\maketitle
} \fi
\bigskip

\maketitle


We use the following notation in the supplement. Eq.~(1), Alg.~1,
and Sampling Scheme~1, etc, refer to the main paper, while Eq.~(S1),
Alg.~S1, and Sampling Scheme~S1, etc, refer to the supplement.

\section{Markov move steps for the univariate SV model\label{S: Markov move steps for univariate SV model}}
\subsection{Sampling the latent volatilities using Hamiltonian Monte Carlo \label{subsec:Sampling-the-Latent volatility}}

For the univariate SV model \eqref{eq: simple SV model}, we need the gradient of the log-likelihood
$\nabla_{\boldsymbol{x}}\mathcal{L}\left(\boldsymbol{x}\right)$ with
respect to each of the latent volatilities. The required
gradient for $t=1$ is
\[
\nabla_{x_{1}}\mathcal{L}\left(\boldsymbol{x}\right)=a_{p}\left(-0.5+0.5y_{1}^{2}\exp\left(-x_{1}\right)\right)-
\frac{\left(1-\phi^{2}\right)}{\tau^{2}}\left(x_{1}-\mu\right)+\frac{\phi}{\tau^{2}}\left(x_{2}-\mu-\phi\left(x_{1}-\mu\right)\right);
\]
the gradient for $1<t<T$ is
\begin{align*}
\nabla_{x_{t}}\mathcal{L}\left(\boldsymbol{x}\right) & =a_{p}\left(-0.5+0.5y_{t}^{2}\exp\left(-x_{t}\right)\right)+\frac{\phi}{\tau^{2}}\left(x_{t+1}-\mu-\phi\left(x_{t}-\mu\right)\right)-\\
 & \frac{1}{\tau^{2}}\left(x_{t}-\mu-\phi\left(x_{t-1}-\mu\right)\right);
\end{align*}
and, for $t=T$, the  gradient is
\begin{align*}
\nabla_{x_{T}}\mathcal{L}\left(\boldsymbol{x}\right) & =a_{p}\left(-0.5+0.5y_{T}^{2}\exp\left(-x_{T}\right)\right)-\frac{1}{\tau^{2}}\left(x_{T}-\mu-\phi\left(x_{T-1}-\mu\right)\right).
\end{align*}

\subsection{Sampling the Univariate SV parameters \label{SS: steps ii to iv} }
For $i=1,...,M$, we sample $\mu_{i}|\boldsymbol{x}_{i1:T},\boldsymbol{y}_{1:T},\boldsymbol{\theta}_{-\mu}$
from $N\left(\mu_{\mu},\sigma_{\mu}^{2}\right)$ truncated to 
$\left(-10,10\right)$, where
\[
\sigma_{\mu}^{2}=\frac{\tau_{i}^{2}}{1-\phi_{i}^{2}+\left(T-1\right)\left(1-\phi_{i}\right)^{2}}
\]
and
\[
\mu_{\mu}=\dfrac{\sigma_{\mu}^{2}}{ \tau_{i}^{2}  }
\left ( x_{i1}\left(1-\phi_{i}^{2}\right)+
\sum_{t=2}^{T}x_{it}-\phi_{i}x_{it}+\phi_{i}^{2}x_{it-1}-\phi_{i}x_{it-1}
\right ) .
\]
We sample the persistence parameter $\phi_{i}$ by drawing a proposed
value $\phi_{i}^{*}$ from $N\left(\mu_{\phi},\sigma_{\phi}^{2}\right)$
truncated within $\left(-1,1\right)$, where
\[
\sigma_{\phi}^{2}=\frac{\tau_{i}^{2}}{\sum_{t=2}^{T}\left(x_{it-1}-\mu_{i}\right)^{2}-\left(x_{i1}-\mu_{i}\right)^{2}}
\]
and
\[
\mu_{\phi}=\sigma_{\phi}^{2}\frac{\sum_{t=2}^{T}\left(x_{it}-\mu_{i}\right)\left(x_{it-1}-\mu_{i}\right)}{\tau_{i}^{2}},
\]
and accept with probability
\begin{align*}
\min\left(1,\frac{p\left(\phi_{i}^{*}\right)\sqrt{1-\phi_{i}^{2*}}}{p\left(\phi_{i}\right)\sqrt{1-\phi_{i}^{2}}}\right).
\end{align*}

We sample $\tau_{i}^{2}$ from $\textrm{IG}\left(v_{1}/2,s_{1}/2\right)$,
where $v_{1}=v_{0}+T$ and $s_{1}=s_{0}+\left(1-\phi_{i}^{2}\right)\left(x_{i1}-\mu_{i}\right)^{2}+\sum_{t=2}^{T}\left(x_{it}-\mu_{i}-\phi_{i}\left(x_{it-1}-\mu_{i}\right)\right)^{2}$.

\section{Additional Results for univariate SV models \label{univSVmodeladditional}}


The terms SMC-PG, SMC-PG-seq, and SMC-PG-batch-seq for SMC-PG denote
batch estimation, sequential estimation, and a combination of batch and sequential estimation, respectively.  For the SMC-PG-batch-seq method,  batch estimation is used for the first 80\% of the data and sequential estimation is used for the rest. 

\begin{figure}[H]
\caption{The evolution over time of the posterior mean estimates of the parameters $\mu$, $\phi$, and $\tau^2$ (with 95\% Bayesian credible intervals) for the univariate SV Model estimated using SMC-PG-seq.  \label{fig:Parameter-Learning-in Univariate SV model}}
\begin{centering}
\includegraphics[width=15cm,height=10cm]{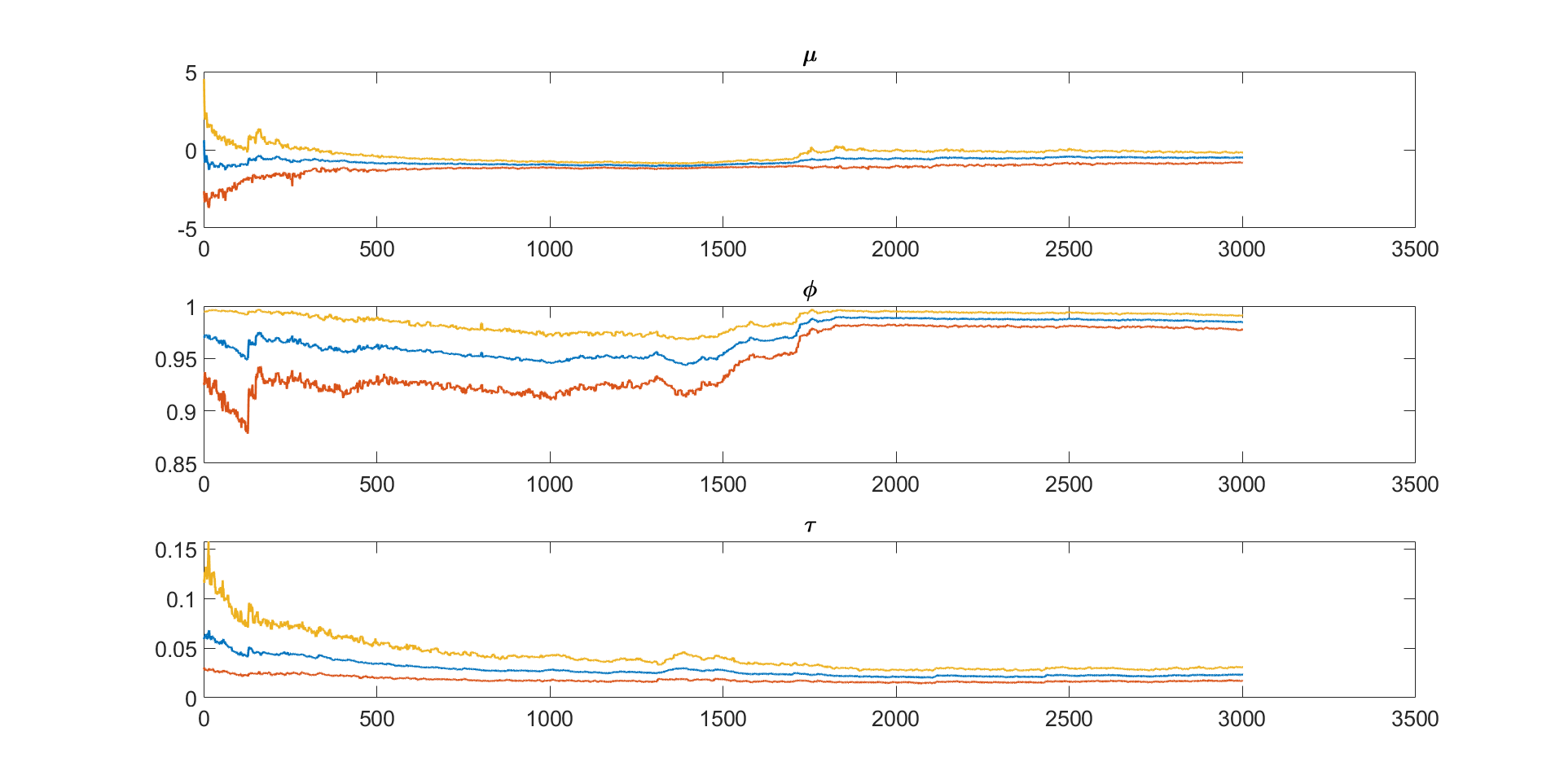}
\par\end{centering}
\end{figure}

\figref{fig:Parameter-Learning-in Univariate SV model} presents the sequential parameter learning in the univariate SV model estimated using SMC-PG-seq. Clearly, all parameter estimates vary a lot at the start when the number of observations is small, then stabilise as the number of observations gets larger. This sequential approach can take into account impacts of parameter and model uncertainties on decision-making over time.

Another important advantage of SMC-PG-batch-seq is that it  provides sequential one-step and multi-step ahead predictive densities of the future observations  $y_{T+h}, h > 0 $.



\begin{figure}[H]
\caption{Sequential one-step ahead predictive density estimates for log return of the
US food industry from 22/06/2011 to 11/11/2013 estimated using the SMC-PG-batch-seq method \label{fig:One-step-ahead-predictive density}}
\begin{centering}
\includegraphics[width=15cm,height=10cm]{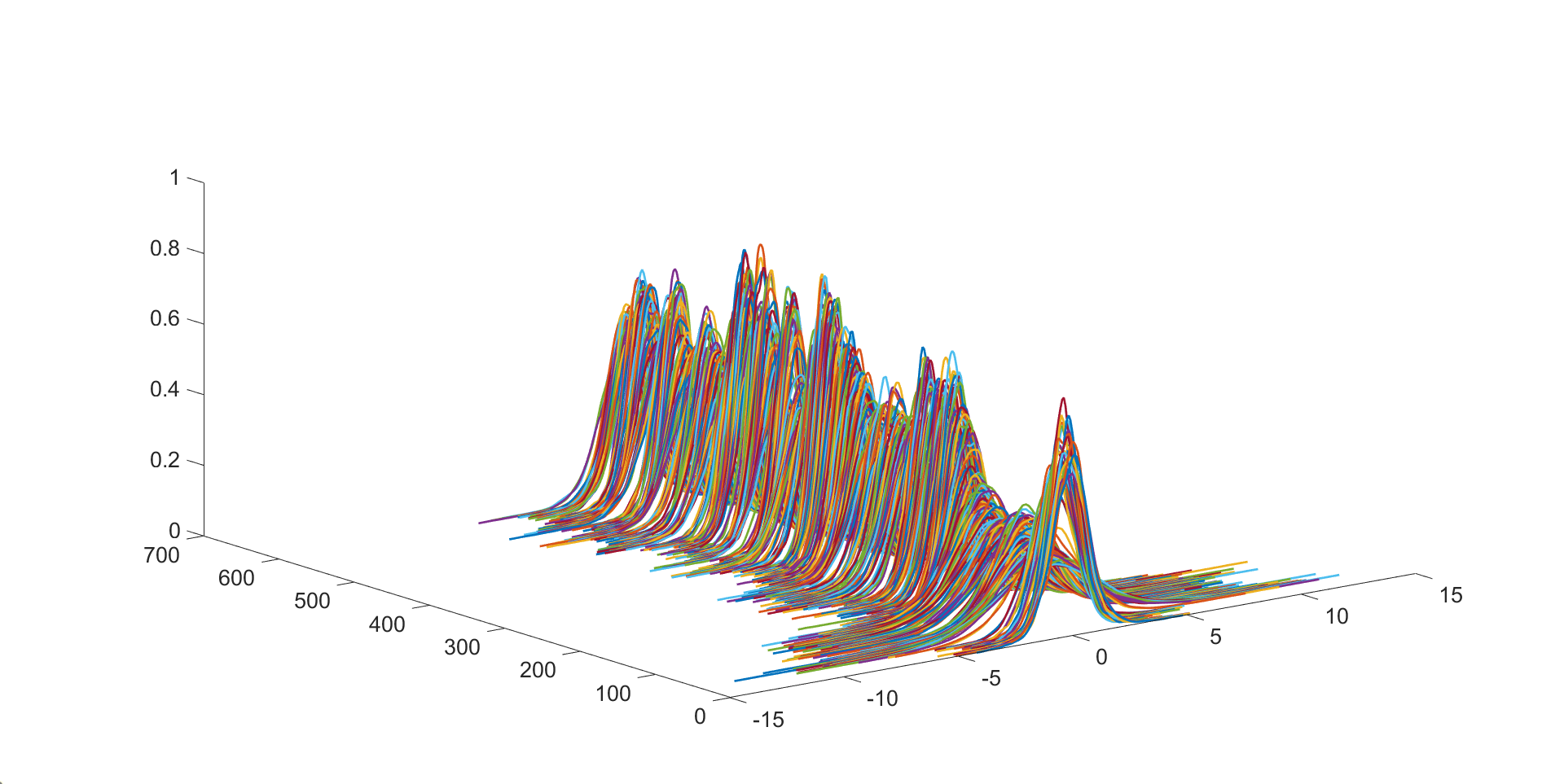}
\par\end{centering}
\end{figure}

\begin{figure}[H]
\caption{Sequential one-step ahead Value at Risk (VaR) estimates for log return of the
US food industry from 22/06/2011 to 11/11/2013 estimated using the SMC-PG-batch-seq method \label{fig:One-step-ahead-Value at Risk}}

\centering{}\includegraphics[width=15cm,height=10cm]{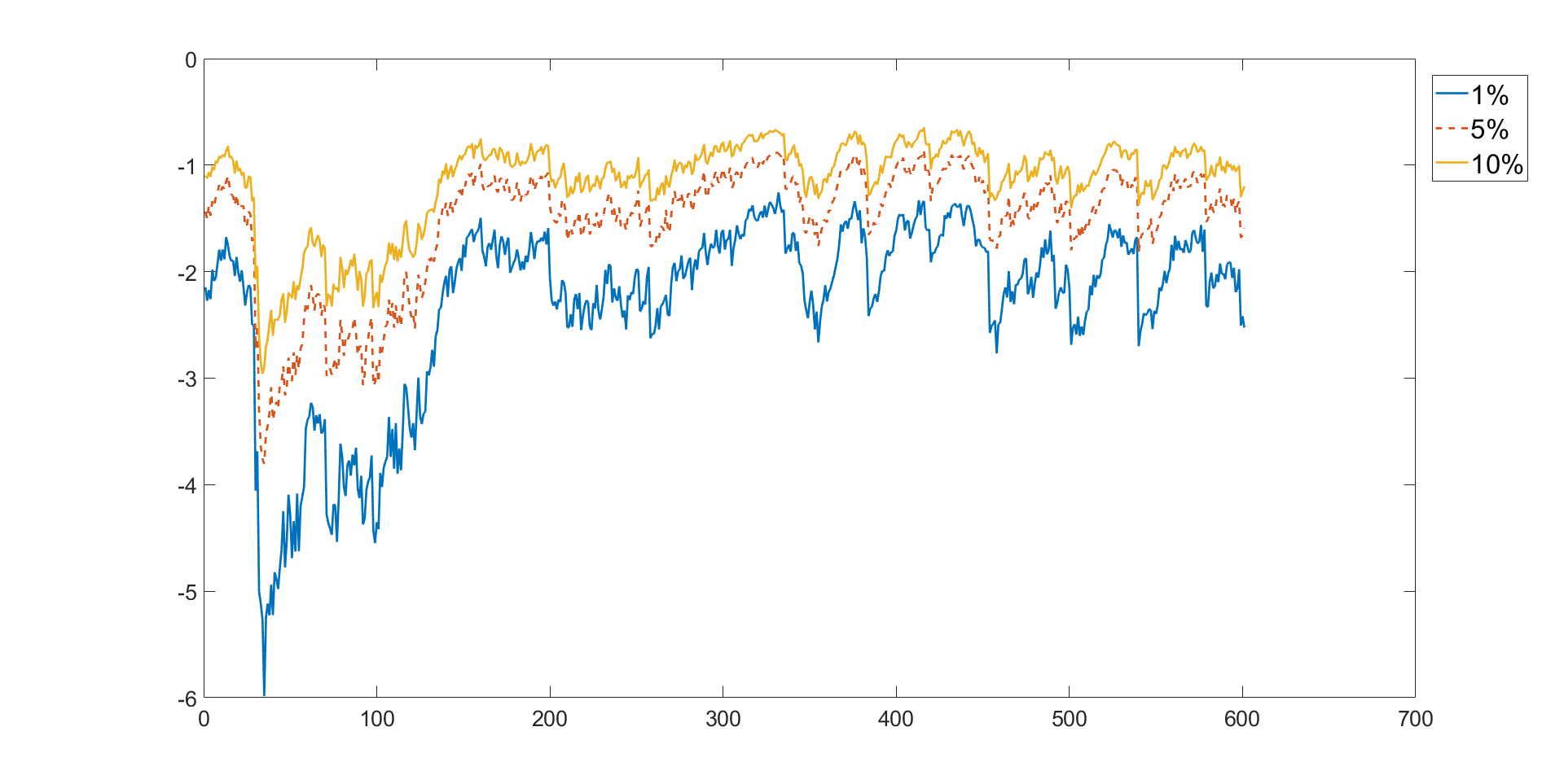}
\end{figure}

\figref{fig:One-step-ahead-predictive density} shows the sequential one-step ahead predictive density for log-return of the US food industry from 22/06/2011 to 11/11/2013 estimated using the SMC-PG-batch-seq method. Financial risk measures such as value at risk (VaR) can be computed from the one step ahead predictive densities. The VaR is the most widely used measure of market risk \citep{Holton2003}. The $t$-period $\pi$-VaR of a return series, denoted as $\textrm{VaR}_{\pi,t}$, is defined by
\[
\pi:=\Pr\left(y_{t}\leq\textrm{VaR}_{\pi,t}|{\cal F}_{t-1}\right),
\]
where ${\cal F}_{t-1}$ is the information available  to time $t-1$. The one-step
ahead $\pi \%$-VaR, $\textrm{VaR}_{\pi,t+1}$,
can be estimated using the $\pi$th-quantile of the posterior
predictive return distribution at time $t+1$ given the information up to time $t$. \figref{fig:One-step-ahead-Value at Risk} shows the sequential one-step ahead VaR estimates at risk levels $0.01$, $0.05$, and $0.10$ for the log return of the US food industry from 22/06/2011 to 11/11/2013. The lowest VaR estimates are between 08/08/2011 to 18/08/2011 obtained using the SMC-PG-batch-seq method.



\begin{figure}[H]
\caption{The kernel density estimates of the marginal posterior densities of the $\beta$ parameters of the univariate SV model with the covariate coefficients estimated using PHS with $N=100$, SMC-PMMH-DF with $N=2500$, and SMC-PG methods with $N=250$ \label{fig:betaunivsv}}

\centering{}\includegraphics[width=15cm,height=10cm]{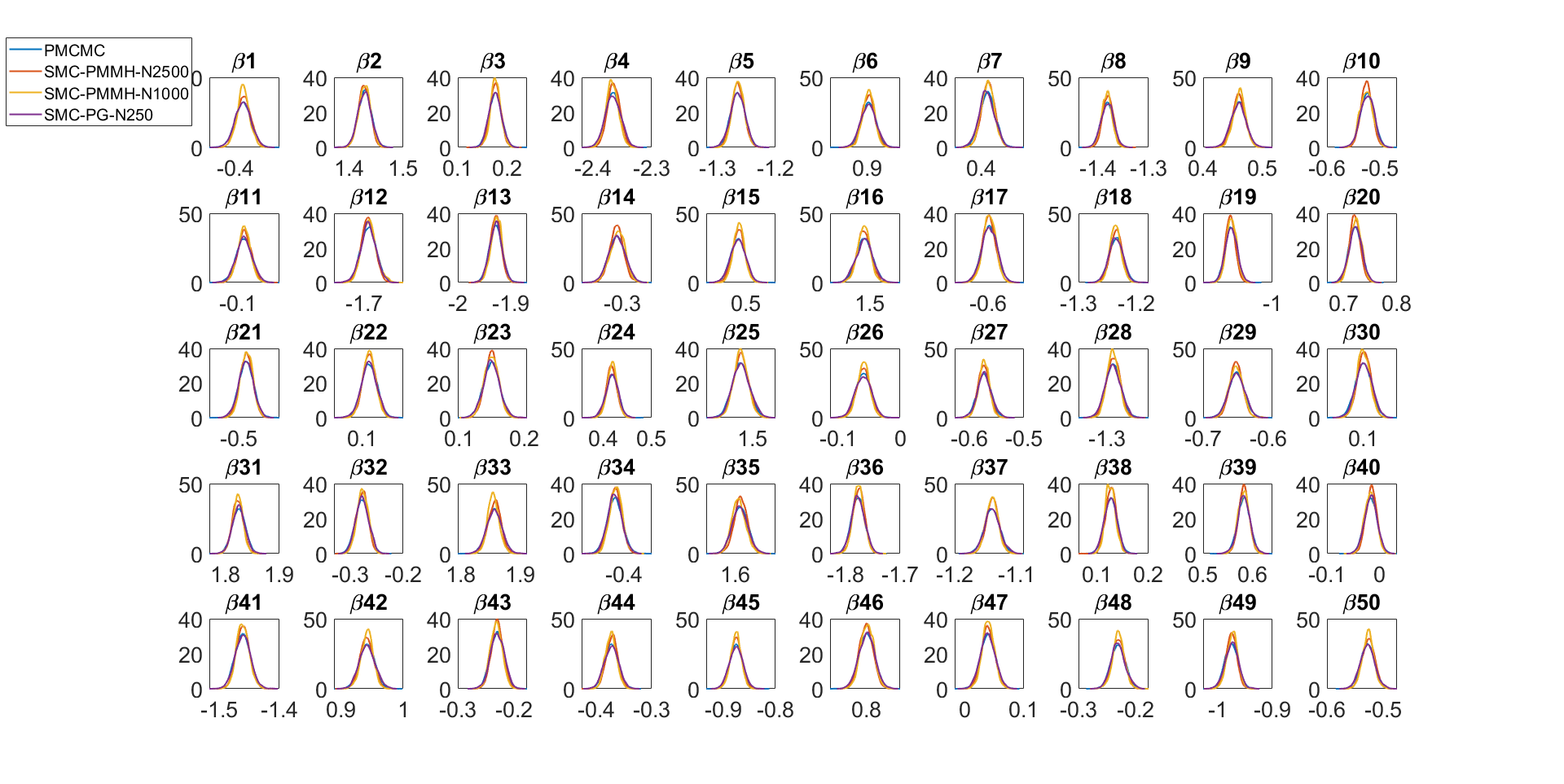}
\end{figure}

Figure \ref{fig:betaunivsv} shows that the SMC-PG estimates are very close to the PHS estimates for all the covariate coefficients. The  SMC-PMMH-DF based estimates differ from the PHS estimates even with $N=2500$ particles. SMC-PMMH-DF uses a PMMH approach with random walk proposal for the parameters in the Markov move component. The random walk is easy to implement, but is not efficient for high-dimensional parameters.



\subsection{Additional results for the SV model with outliers \label{AddOutlier}}

\begin{figure}[H]
\caption{Top panel: The normalised ESS $\left(ESS/M\right)$ over time obtained from
a run of SMC-PG-seqNT algorithm for SV model with $\kappa=25$; Bottom panel:
The series $B_{t}$ for $t=1,...,T; $ $B_{t}=1$ if the observation
is an outlier and $0$, otherwise. \label{fig:Top:-The-normalised ESS} }

\centering{}\includegraphics[width=15cm,height=8cm]{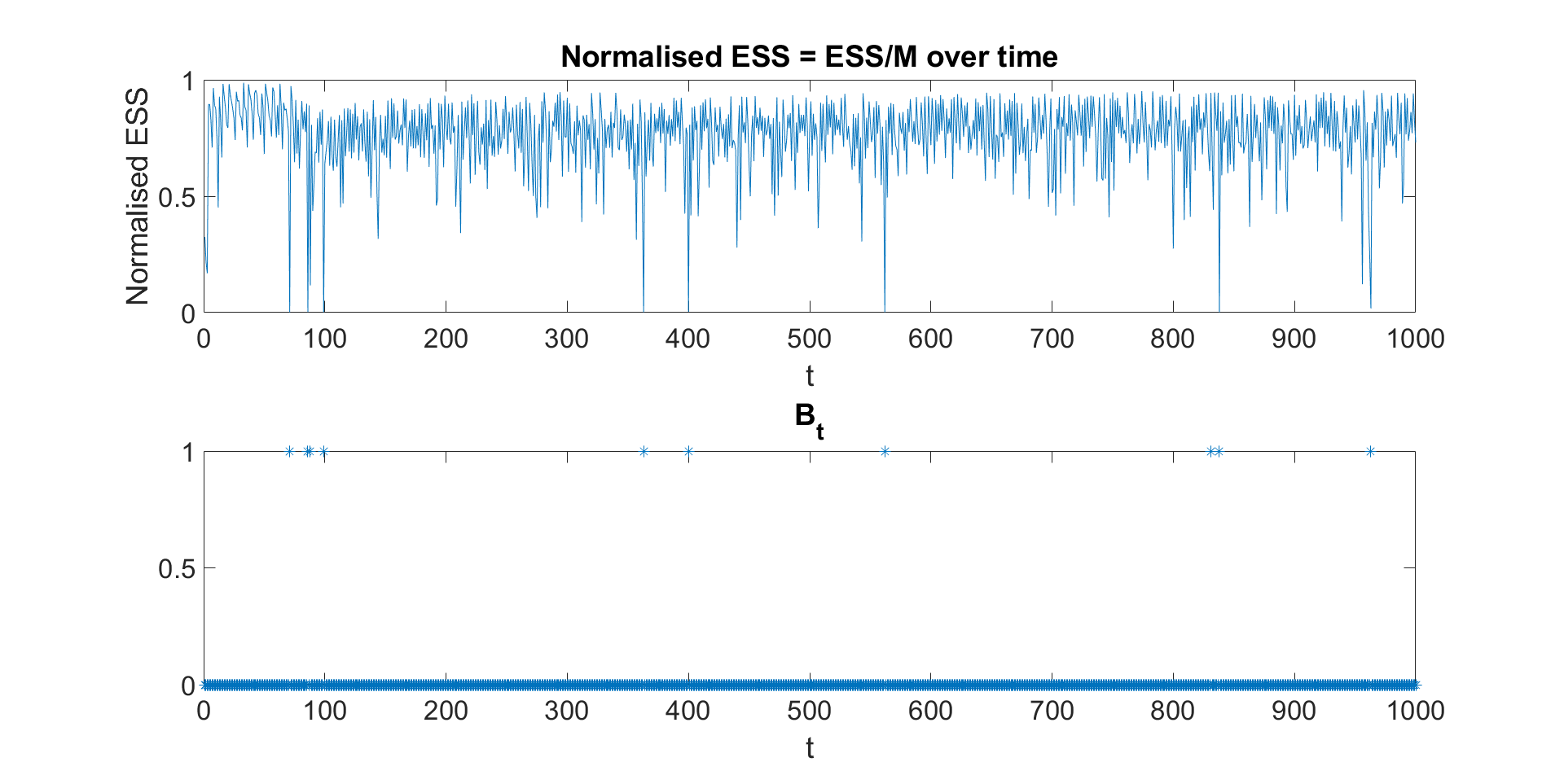}
\end{figure}

\begin{figure}[H]
\caption{Top panel: The number of annealing steps $P$ over time obtained from a run of SMC-PG-seqT
 for the SV model with $\kappa=25$; Middle panel: the absolute value
of the log of $\widehat{p}\left(y_{t}|\bs y_{1:t-1}\right)$ for $t=1,..,T$;
Bottom panel: The $B_{t}$, $t=1,...,T$ series; $B_{t}=1$ if the
observation is an outlier and $0$ otherwise. \label{fig:Top:-The-number of Markov steps}}

\centering{}\includegraphics[width=15cm,height=8cm]{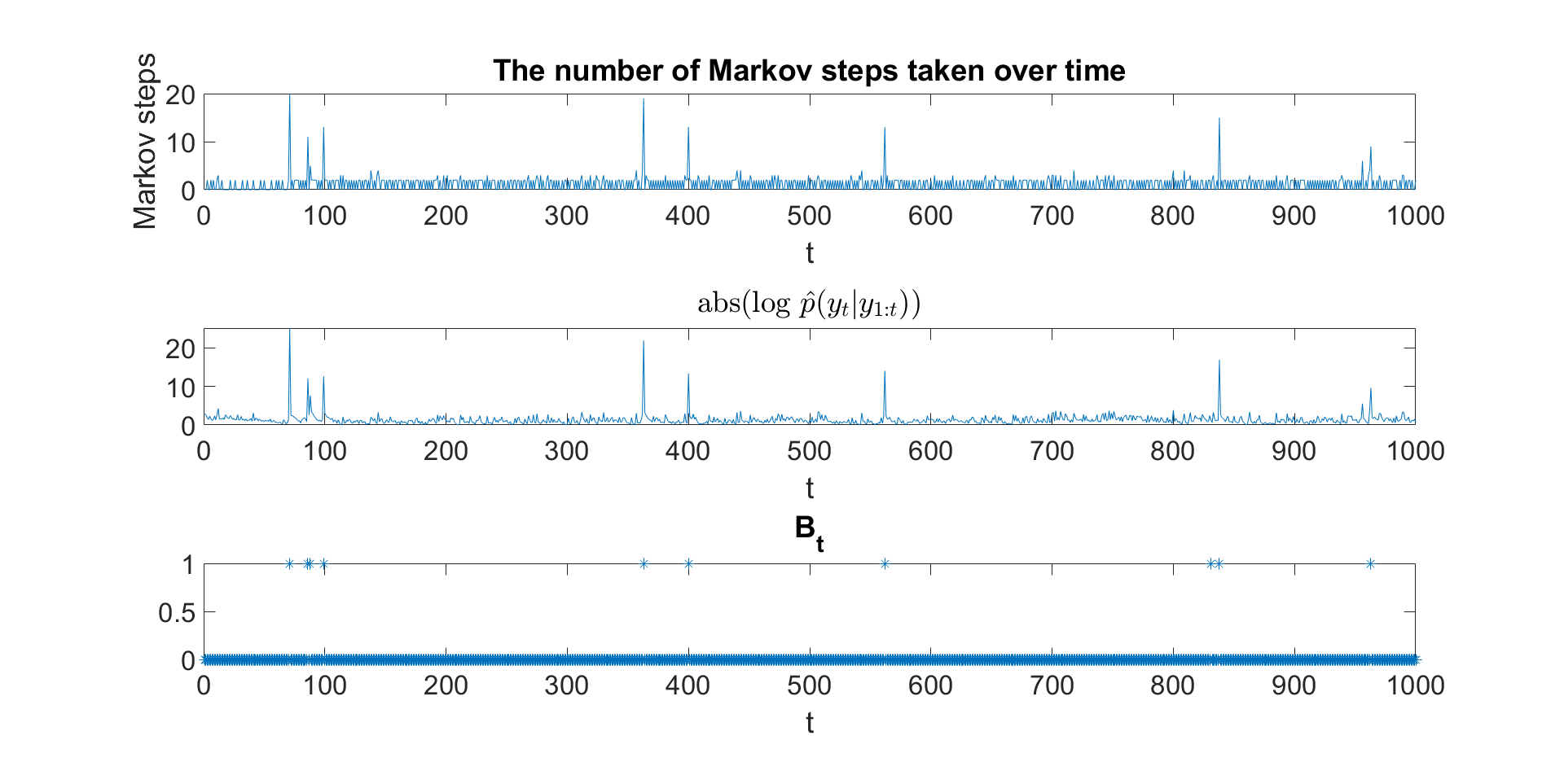}
\end{figure}

 \figref{fig:Top:-The-normalised ESS} shows
the normalised ESS over time obtained from a run of SMC-PG-seqNT
for the SV model with $\kappa=25$. The normalised ESS
is close to zero when the data point is the outlier. This suggests
that  SMC-PG-seqT is more stable than the SMC-PG-seqNT, 
as it ensures the ESS stays close to $\textrm{ESS}_\textrm{target}$. Fig.
\ref{fig:Top:-The-number of Markov steps} (top panel) shows the number
of Markov steps over time obtained from a run of SMC-PG-seqT algorithm.

The top panel of \figref{fig:Top:-The-number of Markov steps} shows that a larger number of Markov move steps are
taken when the data is an outlier. The middle panel of Fig. \ref{fig:Top:-The-number of Markov steps}
shows the absolute value of the log of $\widehat{p}\left(y_{t}|\bs y_{1:t-1}\right)$
for $t=1,..,T$. This figure can be used to identify outliers in the
data by investigating the unusually large absolute value
of the log of $\widehat{p}\left(y_{t}|\bs y_{1:t-1}\right)$ for $t=1,...,T$.

\subsection{Markov Switching Stochastic Volatility Models\label{SVbreaks}}
This section demonstrates the flexibility of SMC-PG-seqT by applying
it to the two state Markov switching SV model
\[
y_{t}=\mu_{s_{t}}^{p}+\exp\left(x_{t}/2\right)\epsilon_{t},\;x_{t}=\mu+\phi\left(x_{t-1}-\mu\right)+\tau\eta_{t},\;x_{1}\sim N\left(\mu,\tau^{2}/\left(1-\phi^{2}\right)\right);
\]
$\epsilon_{t}\sim N\left(0,1\right)$, $\eta_{t}\sim N\left(0,1\right)$,
$\bs s_{1:T}=\left(s_{1},...,s_{T}\right)^{\intercal}$, $s_{t}=1,2$ and $s_{t}=k$
indicates that $y_{t}$ is from regime $k$. This model is similar to that in \citet{Chib1998}.

The one-step ahead transition
probability matrix for $s_{t}$ is
\[
P=\left[\begin{array}{cc}
p_{11} & p_{12}\\
p_{21} & p_{22}
\end{array}\right],
\]
where $p_{lk}=\Pr\left(s_{t}=k|s_{t-1}=l\right)$ is the probability
of moving from regime $l$ at time $t-1$ to regime $k$ at time $t$.
The prior for $\mu^{p}_{1}$ and $\mu^{p}_{2}$ is $N\left(0,100\right)$; the prior $p_{11}$ and $p_{22}$ is $Beta\left(4,1\right)$. The same priors as in Section \ref{univariateexample} are used for the other SV parameters.

We compare the performance of SMC-PG-seqT   to SMC-PG-seqNT and SMC-PG for this model. A dataset is generated with $T=1000$ observations using parameter
values $\phi=0.98$, $\mu=-0.48$, $\tau^{2}=0.02$, $\mu_{1}=-1$,
$\mu_{2}=1$, and $p_{kk}=0.99$ for $k=1,2$. The
SMC-PG method is used as the ``gold standard'' to assess the accuracy
of the SMC-PG-seqT and SMC-PG-seqNT. All the SMC estimates
are obtained using 10 independent runs, each with $M=560$ samples.
For the SMC-PG and SMC-PG-seqT, the $\textrm{ESS}_\textrm{target}$ is
set to $0.8M$.

\begin{figure}[H]
\caption{Top panel: The normalised $ESS:=\left(ESS/M\right)$ over time obtained from
the SMC-PG-seqNT algorithm for the Markov switching SV model; Bottom panel:
the true series $s_{t}$ for $t=1,...,T$ \label{fig:Top:-The-normalised ESS-MS}}

\centering{}\includegraphics[width=15cm,height=8cm]{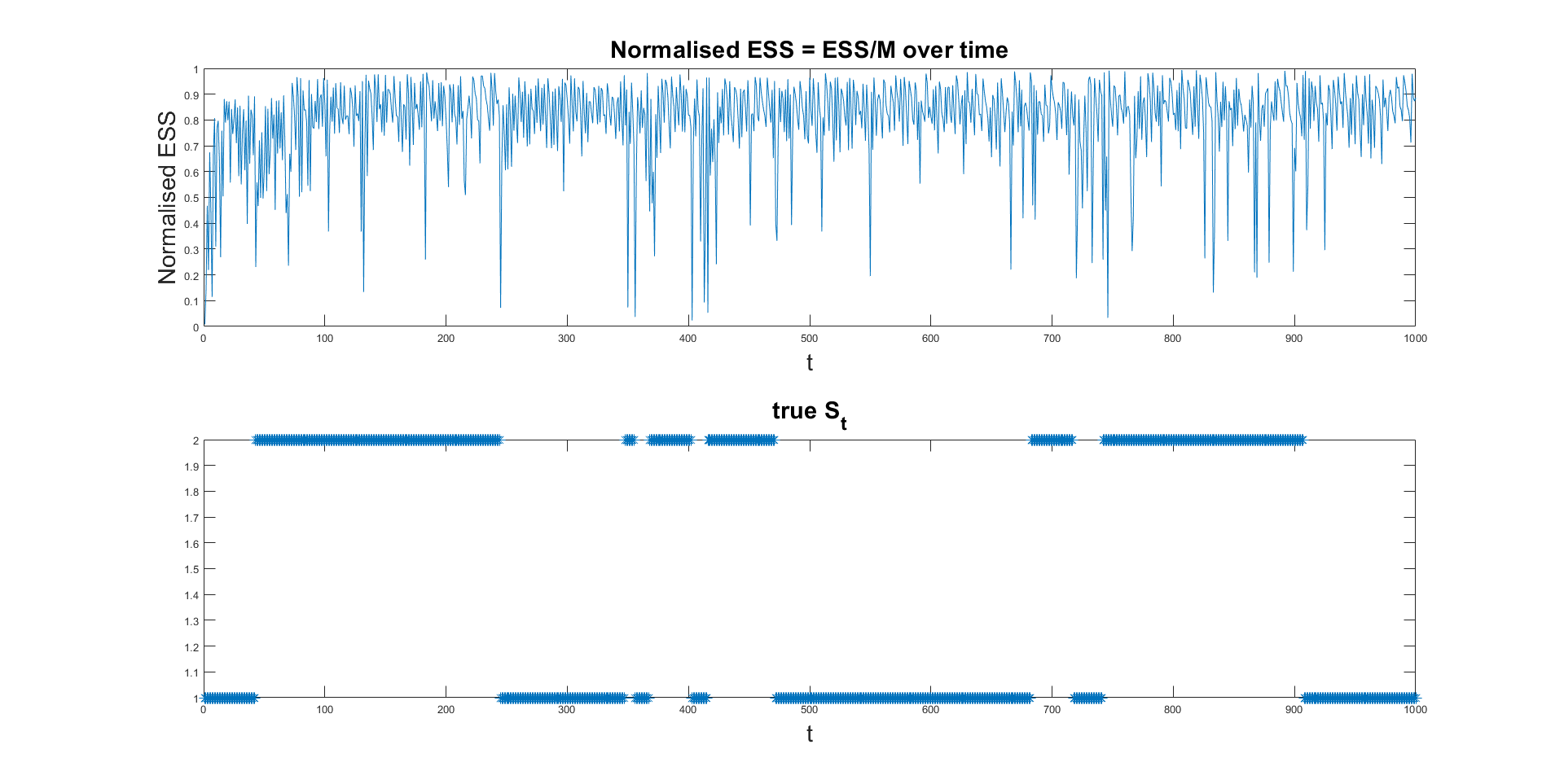}
\end{figure}

Table~\ref{tab:Simulation-results-for Markov switching SV models}
summarizes the simulation results for the Markov switching SV model. The
first seven rows report the standard errors of the posterior mean estimates
of the model parameters over the 10 runs; the eighth and ninth rows report the mean and standard error of the
log of the marginal likelihood estimates. In general, the standard
error of the posterior mean estimates obtained by SMC-PG-seqT and SMC-PG
is smaller than SMC-PG-seqNT. The standard error of the log of the marginal
likelihood estimates from PG-seqNT is about 3 times larger than from
SMC-PG-seqT and SMC-PG. Fig. \ref{fig:Top:-The-normalised ESS-MS} shows
the normalised ESS over time obtained from a run of PG-seqNT algorithm.
Clearly, the normalised ESS is quite small when the state  changes from 1 to 2 or from 2 to 1.
 



\begin{table}[H]
\caption{Simulation results for the Markov switching SV models. The results
are computed from 10 independent runs of each algorithm. The first
seven rows report the standard error of the posterior mean estimates
of the SV parameters $\left(\mu_{1}^{p},\mu_{2}^{p},\mu,\phi,\tau^{2}\right)$
over the 10 independent runs. The eighth and ninth rows report the
mean and standard error of the log of the marginal likelihood estimates
over 10 runs. \label{tab:Simulation-results-for Markov switching SV models}}

\centering{}%
\begin{tabular}{cccc}
\hline
 & SMC-PG-seqT & SMC-PG-seqNT & SMC-PG\tabularnewline
\hline
$\mu_{1}^{p}$ std err. & 0.0013 & 0.0013 & 0.0008\tabularnewline
$\mu_{2}^{p}$ std err. & 0.0015 & 0.0025 & 0.0013\tabularnewline
$p_{11}$ std err. & 0.0003 & 0.0003 & 0.0001\tabularnewline
$p_{22}$ std err. & 0.0004 & 0.0002 & 0.0003\tabularnewline
$\mu$ std err. & 0.0084 & 0.0066 & 0.0033\tabularnewline
$\phi$ std err. & 0.0007 & 0.0008 & 0.0004\tabularnewline
$\tau^{2}$ std err. & 0.0005 & 0.0005 & 0.0005\tabularnewline
$\log\widehat{p}\left(\bs y\right)$ mean & $-1434.83$ & $-1435.20$ & $-1437.83$\tabularnewline
$\log\widehat{p}\left(\bs y\right)$ std err. & 0.5100 & 1.5235 & 0.5071\tabularnewline
\hline
\end{tabular}
\end{table}


\section{Augmented Target Density for SMC-PG \label{targetdensityPG}}
Similarly to  \citet{Andrieu:2010}, the SMC-PG constructs a sequence of tempered densities
$\xi_{a_{p}}\left(\boldsymbol{\theta},\boldsymbol{x}_{1:T}\right)$,
$p=1,...,P$, based on the following augmented target density
\begin{equation} \label{eq:tempered augmented target density}
\begin{aligned}
& \widetilde{\xi}_{a_{p}}\left(\boldsymbol{x}_{1:T}^{\bs j_{1:T}},\bs j_{1:T},\boldsymbol{U}_{1:T}^{-\bs j_{1:T}},\boldsymbol{\theta}\right)  =\frac{\xi_{a_{p}}\left(\boldsymbol{\theta},\boldsymbol{x}_{1:T}^{\bs j_{1:T}}\right)}{N^{T}}\\
& \times \frac{\psi\left(\boldsymbol{U}_{1:T}|\boldsymbol{\theta}\right)}
{m_{1}\left({x}_{1}^{j_{1}}|\boldsymbol{\theta},{y}_{1}\right)\prod_{t=2}^{T}\widetilde{W}_{t-1}^{\widetilde{b}_{t-1}^{j_{t}}}m_{t}
\left({x}_{t}^{j_{t}}|{x}_{t-1}^{\widetilde{b}_{t-1}^{j_{t}}},\boldsymbol{\theta},{y}_{t}\right)}.
\end{aligned}
\end{equation}

The next lemma  shows that the marginal distribution of the augmented target distribution in \eqref{eq:tempered augmented target density} is proportional to the joint tempered posterior density of the parameters and states $\xi_{a_{p}}\left(\boldsymbol{\theta},\boldsymbol{x}_{1:T}\right)$

\begin{lem}\label{lemma: target density lemma for univariate SV}
The target distribution in \eqref{eq:tempered augmented target density}
has the marginal distribution
\[
\widetilde{\xi}_{a_{p}}\left(\boldsymbol{x}_{1:T}^{\bs j_{1:T}},\bs j_{1:T},\boldsymbol{\theta}\right)=N^{-T}\xi_{a_{p}}\left(\boldsymbol{\theta},\boldsymbol{x}_{1:T}^{\bs j_{1:T}}\right).
\]



\begin{proof}[Proof] 
We
 prove the lemma by carrying out the marginalisation similarly to the proof in  \citet{Andrieu:2010} for particle Gibbs MCMC. 
The marginal distribution $\xi_{a_{p}}\left(\boldsymbol{x}_{1:T}^{\bs j_{1:T}},\bs j_{1:T},\boldsymbol{\theta}\right)$
is obtained by integrating $\widetilde{\xi}_{a_{p}}\left(\boldsymbol{x}_{1:T}^{\bs j_{1:T}},\bs j_{1:T},\boldsymbol{U}_{1:T}^{-\bs j_{1:T}},\boldsymbol{\theta}\right)$
over $\left(\boldsymbol{x}_{1:T}^{\left(-\bs j_{1:T}\right)},\widetilde{\boldsymbol{b}}_{1:T-1}\right)$.
We begin by integrating over $\left(\bs{x}_{T}^{-j_{T}},\bs{\widetilde{b}}_{T-1}\right)$
to obtain
\begin{align}
& \widetilde{\xi}_{a_{p}}\left(d\boldsymbol{x}_{1:T}^{\bs j_{1:T}},\bs j_{1:T},
d\boldsymbol{x}_{1:T-1}^{\left(-\bs j_{1:T-1}\right)},d\widetilde{\boldsymbol{b}}_{1:T-2},d\boldsymbol{\theta}\right)  =\frac{\xi_{a_{p}}\left(d\boldsymbol{\theta},d\boldsymbol{x}_{1:T}^{\bs j_{1:T}}\right)}{N^{T}}\\
& \frac{\psi\left(d\boldsymbol{x}_{1:T-1}^{1:N},d\widetilde{\boldsymbol{b}}_{1:T-2}|\boldsymbol{\theta}\right)}
{m_{1}\left(d\boldsymbol{x}_{1}^{j_{1}}|\boldsymbol{\theta},\boldsymbol{y}_{1}\right)\prod_{t=2}^{T-1}
\widetilde{W}_{t-1}^{\widetilde{b}_{t-1}^{j_{t}}}m_{t}\left(d\boldsymbol{x}_{t}^{j_{t}}|
\boldsymbol{x}_{t-1}^{\widetilde{b}_{t-1}^{j_{t}}},\boldsymbol{\theta}\right)}\label{eq:tempered augmented target density-1}
 .\nonumber
\end{align}
We repeat this for $t=T-2,...,2$, to obtain,
\begin{align*}
\widetilde{\xi}_{a_{p}}\left(d\boldsymbol{x}_{1}^{\left(-j_{1}\right)},d\boldsymbol{x}_{1:T}^{\bs j_{1:T}},\bs j_{1:T},d\boldsymbol{\theta}\right)
=\frac{\xi_{a_{p}}\left(d\boldsymbol{\theta},d\boldsymbol{x}_{1:T}^{\bs j_{1:T}}\right)}{N^{T}}\times\frac{\psi\left(d\boldsymbol{x}_{1}^{1:N}\right)}{m_{1}\left(d\boldsymbol{x}_{1}^{j_{1}}|\boldsymbol{\theta},\boldsymbol{y}_{1}\right)}.
\end{align*}
Finally, we integrate over $\boldsymbol{x}_{1}^{j_{1}}$, to obtain
\begin{align*}
\widetilde{\xi}_{a_{p}}\left(d\boldsymbol{x}_{1:T}^{\bs j_{1:T}},\bs j_{1:T},d\boldsymbol{\theta}\right)
=\frac{\xi_{a_{p}}\left(d\boldsymbol{\theta},d\boldsymbol{x}_{1:T}^{\bs j_{1:T}}\right)}{N^{T}}.
\end{align*}
\end{proof}
\end{lem}

\section{Augmented Intermediate Target Density for the Factor SV model \label{augmentedtargetdensityfactorSVmodel}}
This section provides the augmented tempered target density
for the factor SV model; It includes
all the random variables produced by the $S+K$ univariate particle filter
methods  that generate the factor log-volatilities $\boldsymbol{\lambda}_{k,1:T}$,
for $k=1,...,K$, and the idiosyncratic log-volatilities $\boldsymbol{h}_{s,1:T}$
for $s=1,...,S$, as well as the latent factors $\boldsymbol{f}_{1:T}$
and the parameters $\boldsymbol{\theta}$. \eqref{eq:statetransitionidiosyncratic}
specifies the particle filters for the idiosyncratic SV
log-volatilities
$\boldsymbol{h}_{s,1:T}$ for $s=1,...,S$, and \eqref{eq:statetransitionfactor}
specifies the univariate particle filters that generate the factor
log-volatilities $\boldsymbol{\lambda}_{k,1:T}$, for $k=1,...,K$.
The $N$ weighted samples at time $t$ for the factor log-volatilities are denoted
by $\left(\boldsymbol{\lambda}_{kt}^{1:N},\widetilde{\bs W}_{fkt}^{1:N}\right)$
and the idiosyncratic error log-volatilities by  $\left(\boldsymbol{h}_{st}^{1:N},\widetilde{\bs W}_{\epsilon st}^{1:N}\right)$. The corresponding proposal
densities are $m_{fk1}\left(\lambda_{k1}|\boldsymbol{\theta}_{fk},\bs{y}_{1}\right)$,
$m_{fkt}\left(\lambda_{kt}|\lambda_{kt-1},\boldsymbol{\theta}_{fk},\bs{y}_{t}\right)$,
$m_{\epsilon s1}\left(h_{s1}|\boldsymbol{\theta}_{\epsilon s},\bs{y}_{1}\right)$,
and $m_{\epsilon st}\left(h_{st}|h_{st-1},\boldsymbol{\theta}_{\epsilon s},\bs{y}_{t}\right)$
for the $t=2,...,T$. The resampling schemes are denoted by $\M\left(\widetilde{\boldsymbol{b}}_{\epsilon st-1}^{1:N}|\widetilde{\bs W}_{\epsilon st-1}^{1:N}\right)$,
where each $\widetilde{b}_{\epsilon st-1}^{j}=k$ indexes a particle
in $\left(\boldsymbol{h}_{st-1}^{1:N},\widetilde{\bs W}_{\epsilon st-1}^{1:N}\right)$
and is chosen with probability $\widetilde{W}_{\epsilon st-1}^{k}$;
$\M\left(\widetilde{\boldsymbol{b}}_{fkt-1}^{1:N}|\widetilde{\bs W}_{fkt-1}^{1:N}\right)$
is defined similarly. 

The vectors of particles are denoted as
\[
\boldsymbol{U}_{\epsilon s,1:T}\coloneqq\left(\boldsymbol{h}_{s1}^{1:N},...,\boldsymbol{h}_{sT}^{1:N},\widetilde{\boldsymbol{b}}_{\epsilon s1}^{1:N},...,\widetilde{\boldsymbol{b}}_{\epsilon sT-1}^{1:N}\right)
\]
and
\[
\boldsymbol{U}_{fk,1:T}\coloneqq\left(\boldsymbol{\lambda}_{k1}^{1:N},...,\boldsymbol{\lambda}_{kT}^{1:N},\widetilde{\boldsymbol{b}}_{fk1}^{1:N},...,\widetilde{\boldsymbol{b}}_{fkT-1}^{1:N}\right).
\]

The joint distribution of the particles given the parameters is
\begin{multline*}
\psi_{\epsilon s}\left(\boldsymbol{U}_{\epsilon s,1:T}^{1:N}|\boldsymbol{\theta}_{\epsilon s}\right)\coloneqq\\
\prod_{j=1}^{N}m_{\epsilon s1}\left(h_{s1}^{j}|\bs{y}_{1},\boldsymbol{\theta}_{\epsilon s}\right)\prod_{t=2}^{T}\left\{ \M\left(\widetilde{\boldsymbol{b}}_{\epsilon st-1}^{1:N}|\widetilde{\bs W}_{\epsilon st-1}^{1:N}\right)\prod_{j=1}^{N}m_{\epsilon st}\left(h_{st}^{j}|h_{st-1}^{\widetilde{{b}}_{\epsilon st-1}^{j}},\boldsymbol{\theta}_{\epsilon s},\bs{y}_{t}\right)\right\} ,
\end{multline*}
for $s=1,...,S$ and
\begin{multline*}
\psi_{fk}\left(\boldsymbol{U}_{fk,1:T}^{1:N}|\boldsymbol{\theta}_{fk}\right)\coloneqq\\
\prod_{j=1}^{N}m_{fk1}\left(\lambda_{k1}^{j}|\bs{y}_{1},\boldsymbol{\theta}_{fk}\right)\prod_{t=2}^{T}\left\{ \M\left(\widetilde{\boldsymbol{b}}_{fkt-1}^{1:N}|\widetilde{\bs W}_{fkt-1}^{1:N}\right)\prod_{j=1}^{N}m_{fkt}\left(\lambda_{kt}^{j}|\lambda_{kt-1}^{\widetilde{{b}}_{fkt-1}^{j}},\boldsymbol{\theta}_{fk},\bs{y}_{t}\right)\right\} ,
\end{multline*}
for $k=1,...,K$. 

Next, we define indices $\bs j_{\epsilon s,1:T}$ for
$s=1,...,S$, the selected particle trajectory $\boldsymbol{h}_{s,1:T}^{\bs j_{\epsilon s,1:T}}=\left(h_{s1}^{j_{\epsilon s1}},...,h_{sT}^{j_{\epsilon sT}}\right)$,
indices $\bs j_{fk,1:T}$ for $k=1,...,K$ and the selected particle trajectory
$\boldsymbol{\lambda}_{k,1:T}^{\bs j_{fk,1:T}}=\left(\lambda_{k1}^{j_{fk1}},...,\lambda_{kT}^{j_{fkT}}\right)$.

The augmented intermediate target density in this case consists of
all the particle filter variables
\begin{equation}\label{eq:target distribution factor SV}
\begin{aligned}
&\tilde{\xi}_{a_{p}}\left(\boldsymbol{U}_{\epsilon s,1:T},\boldsymbol{U}_{fk,1:T},\boldsymbol{\theta},\boldsymbol{f}_{1:T}\right) \coloneqq
\frac{\xi_{a_{p}}\left(d\boldsymbol{\theta},d\boldsymbol{h}_{1:T}^{\bs j_{\epsilon s,1:T}},d\boldsymbol{\lambda}_{1:T}^{\bs j_{fk,1:T}},d\boldsymbol{f}_{1:T}\right)}{N^{T\left(S+K\right)}}\times\\
& \prod_{s=1}^{S}\frac{\psi_{\epsilon s}\left(\boldsymbol{U}_{\epsilon s,1:T}|\boldsymbol{\theta}_{\epsilon s}\right)}{m_{\epsilon s1}\left(h_{s1}^{j_{\epsilon s1}}|\boldsymbol{\theta}_{\epsilon s}\right)\prod_{t=2}^{T}\widetilde{W}_{\epsilon st-1}^{\widetilde{b}_{\epsilon st-1}^{j_{\epsilon st}}}m_{\epsilon st}\left(h_{st}^{j_{\epsilon st}}|h_{st-1}^{\widetilde{b}_{\epsilon st-1}^{j_{\epsilon st}}},\boldsymbol{\theta}_{\epsilon s}\right)}\\
& \prod_{k=1}^{K}\frac{\psi_{fk}\left(\boldsymbol{U}_{fk,1:T}|\boldsymbol{\theta}_{fk}\right)}{m_{fk1}
\left(\lambda_{k1}^{j_{fk1}}|\boldsymbol{\theta}_{fk}\right)\prod_{t=2}^{T}\widetilde{W}_{fkt-1}^{\widetilde{b}_{fkt-1}^{j_{fkt}}}m_{fkt}
\left(\lambda_{kt}^{j_{fkt}}|\lambda_{kt-1}^{\widetilde{b}_{fkt-1}^{j_{fkt}}},\boldsymbol{\theta}_{fk}\right)}.
\end{aligned}
\end{equation}

The next lemma shows that the target distribution in \eqref{eq:target distribution factor SV}
has the required marginal distribution. \citet{Mendes2020} give 
a similar lemma and proof for describing the augmented target density of the factor SV models for PMCMC methods.

\begin{lem}\label{lemma: taget distn for factor sv}
The target distribution in \eqref{eq:target distribution factor SV}
has the marginal distribution
\begin{equation*}
\tilde{\xi}_{a_{p}}\left(\boldsymbol{h}_{1:T}^{\bs j_{\epsilon s,1:T}},\bs j_{\epsilon s,1:T},\boldsymbol{\lambda}_{1:T}^{\bs j_{fk,1:T}},\bs j_{fk,1:T},\boldsymbol{\theta},\boldsymbol{f}_{1:T}\right)
=\frac{\xi_{a_{p}}\left(d\boldsymbol{\theta},d\boldsymbol{h}_{1:T}^{\bs j_{\epsilon s,1:T}},d\boldsymbol{\lambda}_{1:T}^{\bs j_{fk,1:T}},d\boldsymbol{f}_{1:T}\right)}{N^{T\left(S+K\right)}}.
\end{equation*}

\begin{proof}[Proof] Similarly
 to the proof of Lemma \ref{lemma: target density lemma for univariate SV}, we can show that the marginal distribution
\[
\tilde{\xi}_{a_{p}}\left(d\boldsymbol{h}_{1:T}^{\bs j_{\epsilon s,1:T}},\bs j_{\epsilon s,1:T},d\boldsymbol{\lambda}_{1:T}^{\bs j_{fk,1:T}},\bs j_{fk,1:T},d\boldsymbol{\theta},d\boldsymbol{f}_{1:T}\right)=\frac{\xi_{a_{p}}\left(d\boldsymbol{\theta},d\boldsymbol{h}_{1:T}^{\bs j_{\epsilon s,1:T}},d\boldsymbol{\lambda}_{1:T}^{\bs j_{fk,1:T}},d\boldsymbol{f}_{1:T}\right)}{N^{T\left(S+K\right)}},
\]
is obtained by integrating $\tilde{\xi}_{a_{p}}\left(\boldsymbol{U}_{\epsilon s,1:T},\boldsymbol{U}_{fk,1:T},\boldsymbol{\theta},\boldsymbol{f}_{1:T}\right)$
over $\left(\boldsymbol{h}_{s1:T}^{\left(-\bs j_{\epsilon s1:T}\right)},\widetilde{\boldsymbol{b}}_{\epsilon s1:T-1}\right)$
for $s=1,...,S$, and $\left(\boldsymbol{\lambda}_{k1:T}^{\left(-\bs j_{fk1:T}\right)},\widetilde{\boldsymbol{b}}_{fk1:T-1}\right)$
for $k=1,...,K$.
\end{proof}
\end{lem}

\subsection{Application of the SMC-HMC method to the Multivariate Factor Stochastic
Volatility Model \label{subsec:Application-of-AISIL-HMC factor SV}}

This section discusses the application of the SMC-HMC method to the multivariate
factor SV model described in \secref{subsec:Model} in a batch context. We first
define the appropriate sequence of intermediate target densities,
\begin{equation}
\begin{aligned}\label{eq:intermediate target density factor SV}
\xi_{a_{p}}\left(\boldsymbol{\theta},\boldsymbol{h}_{1:T},\boldsymbol{\lambda}_{1:T},\boldsymbol{f}_{1:T}\right) & \propto p\left(\boldsymbol{y}_{1:T}|\boldsymbol{f}_{1:T},\boldsymbol{h}_{1:T},\boldsymbol{\lambda}_{1:T},\boldsymbol{\theta}\right)^{a_{p}}
p\left(\boldsymbol{f}_{1:T}|\boldsymbol{h}_{1:T},\boldsymbol{\lambda}_{1:T},\boldsymbol{\theta}\right)\\
 & \times  p\left(\boldsymbol{\lambda}_{1:T},\boldsymbol{h}_{1:T}|\boldsymbol{\theta}\right)p\left(\boldsymbol{\theta}\right),\\
\text{where} \quad \boldsymbol{\theta}& =\left\{ \boldsymbol{\theta}_{\epsilon s}  =\left\{ \mu_{\epsilon s},\phi_{\epsilon s},\tau_{\epsilon s}^{2}\right\} _{s=1}^{S},{\bs \theta}_{fk}=\left\{ \phi_{fk},\tau_{fk}^{2}\right\} _{k=1}^{K},{\beta}\right\}, \\
p\left(\boldsymbol{y}_{1:T}|\boldsymbol{f}_{1:T},\boldsymbol{h}_{1:T},\boldsymbol{\lambda}_{1:T},\boldsymbol{\theta}\right)^{a_{p}}
& =\prod_{t=1}^{T}\prod_{s=1}^{S}p\left(y_{st}|\boldsymbol{\beta}_{s},\boldsymbol{f}_{t},h_{st}\right)^{a_{p}},\\
p\left(\boldsymbol{f}_{1:T}|\boldsymbol{\lambda}_{1:T}\right)& =\prod_{t=1}^{T}\prod_{k=1}^{K}p\left(f_{kt}|\lambda_{kt}\right),\\
p\left(\boldsymbol{\lambda}_{1:T}|\boldsymbol{\theta}_{f}\right) &=\prod_{k=1}^{K}\prod_{t=2}^{T}p\left(\lambda_{kt}|\lambda_{kt-1},\boldsymbol{\theta}_{fk}\right)p\left(\lambda_{k1}|\boldsymbol{\theta}_{fk}\right),\\
 p\left(\boldsymbol{h}_{1:T}|\boldsymbol{\theta}_{\epsilon}\right) &=\prod_{s=1}^{S}\prod_{t=2}^{T}p\left(h_{st}|h_{st-1},\boldsymbol{\theta}_{\epsilon s}\right)p\left(h_{s1}|\boldsymbol{\theta}_{\epsilon s}\right).
 \end{aligned}
 \end{equation}

Algorithm \ref{alg:Markov-moves-based HMC for multivariate factor SV model} gives the Markov move step based on Hamiltonian Monte Carlo for the multivariate factor SV model with the target densities $ \xi_{a_{p}}\left(\boldsymbol{\theta},\boldsymbol{h}_{1:T},\boldsymbol{\lambda}_{1:T},\boldsymbol{f}_{1:T}\right)$. Steps 1 to 4 update the idiosyncratic log-volatilities parameters $(\mu_{\epsilon s},\phi_{\epsilon s},\tau^{2}_{\epsilon s})$ for $s=1,...,S$, factor log-volatilities parameters $(\phi_{fk},\tau^{2}_{fk})$ for $k=1,...,K$, the factor loading matrix $\beta$ and the latent factors $\boldsymbol{f}_{t}$, respectively. \secref{SS: samplingparamSV} discusses updating the idiosyncratic and factor log-volatilities parameters.
\secref{sec:Sampling-factor-loading} discusses
 sampling of the factor loading matrix ${\beta}$ and the latent factors $\boldsymbol{f}_{t}$ . Steps 5 and 6 update $S$ idiosyncratic log-volatilities, $\bs{h}_{s,1:T}$ and $K$ factor log-volatilities $\bs{\lambda}_{k,1:T}$ using Hamiltonian Monte Carlo, respectively.
\secref{SS: Markov moves for HMC} presents the details
on  updating the idiosyncratic and factor log-volatilities.

\begin{algorithm}[H]
\caption{Markov moves based on Hamiltonian Monte Carlo for multivariate factor
SV model\label{alg:Markov-moves-based HMC for multivariate factor SV model}}

for each $i=1,...,M$,
\begin{enumerate}
\item for $s=1:S$

\begin{enumerate}
\item Sample $\mu_{i,\epsilon s}|\boldsymbol{h}_{is,1:T},\boldsymbol{\theta}_{i,-\mu_{i,\epsilon s}}$
\item Sample $\phi_{i,\epsilon s}|\boldsymbol{h}_{is,1:T},\boldsymbol{\theta}_{i,-\phi_{i,\epsilon s}}$
\item Sample $\tau_{i,\epsilon s}^{2}|\boldsymbol{h}_{is,1:T},\boldsymbol{\theta}_{i,-\tau_{i,\epsilon s}^{2}}$
\end{enumerate}
\item For $k=1,...,K$

\begin{enumerate}
\item Sample $\phi_{i,fk}|\boldsymbol{\lambda}_{ik,1:T},\boldsymbol{\theta}_{i,-\phi_{i,fk}}$
\item Sample $\tau_{i,fk}^{2}|\boldsymbol{\lambda}_{ik,1:T},\boldsymbol{\theta}_{i,-\tau_{i,fk}^{2}}$
\end{enumerate}
\item Sample the factor loading matrix $\boldsymbol{\beta}_{i}|\boldsymbol{\theta}_{i,-\boldsymbol{\beta}},\boldsymbol{f}_{i,1:T},\boldsymbol{h}_{i,1:T},\boldsymbol{\lambda}_{i,1:T},\boldsymbol{y}_{1:T}$
\item Sample the latent factors $\boldsymbol{f}_{i,t}|\boldsymbol{\theta}_{i},\boldsymbol{h}_{i,1:T},\boldsymbol{\lambda}_{i,1:T},\boldsymbol{y}_{1:T}$
\item For $s=1,...,S$, sample $\boldsymbol{h}_{is,1:T}|\boldsymbol{\theta}_{i},\boldsymbol{f}_{i,1:T},\boldsymbol{y}_{1:T}$
using Hamiltonian Monte Carlo.
\item For $k=1,...,K$, sample $\boldsymbol{\lambda}_{ik,1:T}|\boldsymbol{\theta}_{i},\boldsymbol{f}_{i,1:T},\boldsymbol{y}_{1:T}$
using Hamiltonian Monte Carlo.
\end{enumerate}
\end{algorithm}

\subsection{Application of the SMC-PG and SMC-PHS methods to the Multivariate Factor Stochastic
Volatility Model \label{subsec:Application-of-AISIL-PG factor SV}}

This section discusses the application of the SMC-PG method to a multivariate
factor SV model in a batch context. Similarly to Section \ref{subsec:Application-of-AISIL-HMC factor SV},
\eqref{eq:intermediate target density factor SV} gives the appropriate sequence of intermediate target densities. Section \ref{augmentedtargetdensityfactorSVmodel} discusses the augmented intermediate target density for the multivariate factor SV model.



Algorithm \ref{alg:Markov-moves-based PG for multivariate factor SV model} describes the SMC-PG Markov moves for the multivariate factor stochastic volatility model. Steps 1 to 4 update the idiosyncratic log-volatilty parameters $(\mu_{\epsilon s},\phi_{\epsilon s},\tau^{2}_{\epsilon s})$ for $s=1,...,S$, factor log-volatility parameters $(\phi_{fk},\tau^{2}_{fk})$ for $k=1,...,K$, the factor loading matrix ${\beta}$ and the latent factors $\boldsymbol{f}_{t}$, respectively. \secref{SS: samplingparamSV} discusses updating for the idiosyncratic and factor log-volatility parameters.
\secref{sec:Sampling-factor-loading} shows how  to sample the factor loading matrix ${\beta}$ and the latent factors $\boldsymbol{f}_{t}$. Steps 5 and 7 update the $\bs{U}^{-\bs j_{\epsilon s,1:T}}_{\epsilon s,1:T}$ for $s=1,...,S$ and $\bs{U}^{-\bs j_{\epsilon fk,1:T}}_{\epsilon fk,1:T}$ for $k=1,...,K$, respectively, by running $(S+K)$ conditional sequential Monte Carlo algorithms. Steps 6 and 8 update the indices $\bs j_{\epsilon s,1:T}$ and  $\bs j_{fk,1:T}$, by running $(S+K)$ backward simulation algorithms. SMC-PHS is similar to Algorithm \ref{alg:Markov-moves-based PG for multivariate factor SV model}, but generates  $\tau_{\epsilon,s}^{2}$ and $\tau_{fk}^{2}$  using a PMMH step with a random walk proposal for $s=1,...,S$ and $k=1,...,K$.

\begin{algorithm}[H]
\caption{Markov moves based on particle Gibbs for the multivariate factor SV model
\label{alg:Markov-moves-based PG for multivariate factor SV model}}

for $i=1,...,M$,
\begin{enumerate}
\item For $s=1,...,S$,

\begin{enumerate}
\item Sample $\mu_{i\epsilon s}|\boldsymbol{h}_{is,1:T}^{\bs j_{i\epsilon s,1:T}},\bs j_{i\epsilon s,1:T},\boldsymbol{\theta}_{i,-\mu_{i\epsilon s}},\boldsymbol{f}_{i,1:T},\boldsymbol{y}_{s,1:T}$
\item Sample $\phi_{i\epsilon s}|\boldsymbol{h}_{is,1:T}^{\bs j_{i\epsilon s,1:T}},\bs j_{i\epsilon s,1:T},\boldsymbol{\theta}_{i,-\phi_{i\epsilon s}},\boldsymbol{f}_{i,1:T},\boldsymbol{y}_{s,1:T}$
\item Sample $\tau_{i\epsilon s}^{2}|\boldsymbol{h}_{is,1:T}^{\bs j_{i\epsilon s,1:T}},\bs j_{i\epsilon s,1:T},\boldsymbol{\theta}_{i,-\tau_{i\epsilon s}^{2}},\boldsymbol{f}_{i,1:T},\boldsymbol{y}_{s,1:T}$
\end{enumerate}
\item For $k=1,...,K$,

\begin{enumerate}
\item Sample $\phi_{ifk}|\boldsymbol{\lambda}_{ik,1:T}^{\bs j_{ifk,1:T}},\bs j_{ifk,1:T},\boldsymbol{\theta}_{i,-\phi_{ifk}},\boldsymbol{f}_{ik,1:T}$
\item Sample $\tau_{ifk}^{2}|\boldsymbol{\lambda}_{ik,1:T}^{\bs j_{ifk,1:T}},\bs j_{ifk,1:T},\boldsymbol{\theta}_{i,-\tau_{ifk}^{2}},\boldsymbol{f}_{ik,1:T}$
\end{enumerate}
\item Sample the factor loading matrix $\boldsymbol{\beta}_{i}|\boldsymbol{h}_{is,1:T}^{\bs j_{i\epsilon s,1:T}},\bs j_{i\epsilon s,1:T},\boldsymbol{\lambda}_{ik,1:T}^{\bs j_{ifk,1:T}},\bs j_{ifk,1:T},\boldsymbol{\theta}_{-\boldsymbol{\beta}_{i}},\boldsymbol{f}_{i,1:T},\boldsymbol{y}_{1:T}$.
\item Sample the latent factors $\boldsymbol{f}_{i,t}|\boldsymbol{h}_{is,1:T}^{\bs j_{i\epsilon s,1:T}},\bs j_{i\epsilon s,1:T},\boldsymbol{\lambda}_{ik,1:T}^{\bs j_{ifk,1:T}},\bs j_{ifk,1:T},\boldsymbol{\theta}_{i},\boldsymbol{y}_{1:T}$
for $t=1,..,T$.
\item For $s=1,...,S$, sample
\begin{align*}
\boldsymbol{U}_{i\epsilon s,1:T}^{\left(-\bs j_{i\epsilon s,1:T}\right)}\sim\tilde{\xi}_{a_{p}}^{N}\left(\boldsymbol{U}_{i\epsilon s,1:T}^{\left(-\bs j_{i\epsilon s,1:T}\right)}|\boldsymbol{h}_{is,1:T}^{\bs j_{i\epsilon s,1:T}},\bs j_{i\epsilon s,1:T},\boldsymbol{\theta}_{i},\boldsymbol{f}_{i1:T},\boldsymbol{y}_{1:T}\right)
\end{align*}
\item For $s=1,...,S$, sample\\
$J_{i\epsilon s,t}=j_{i\epsilon s,t}\sim\xi_{a_{p}}\left(j_{i\epsilon s,t}|\boldsymbol{\theta}_{i},\boldsymbol{h}_{is,1:t}^{1:N},\widetilde{\boldsymbol{a}}_{i\epsilon s,1:t-1}^{1:N},\boldsymbol{h}_{is,t+1:T}^{\bs j_{i\epsilon s,t+1:T}},\bs j_{i\epsilon s,t+1:T}\right)$
\\for $t=T-1,....,1$ and $J_{i\epsilon s,T}=j_{i\epsilon s,T}\sim\xi_{a_{p}}\left(j_{i\epsilon s,T}|\boldsymbol{\theta}_{i},\boldsymbol{h}_{is,1:T}^{1:N},\widetilde{\boldsymbol{a}}_{i\epsilon s,1:T-1}^{1:N}\right)$
\item For $k=1,...,K$, sample
\begin{align*}
   \boldsymbol{U}_{ifk,1:T}^{\left(-\bs j_{ifk,1:T}\right)}\sim\tilde{\xi}_{a_{p}}^{N}\left(\boldsymbol{U}_{ifk,1:T}^{\left(-\bs j_{ifk,1:T}\right)}|
   \boldsymbol{\lambda}_{ik,1:T}^{\bs j_{i,fk,1:T}},\bs j_{ifk,1:T},\boldsymbol{\theta}_{i},\boldsymbol{f}_{ik1:T}\right)
   \end{align*}
\item For $k=1,...,K$, sample \\$J_{ifk,t}=j_{ifk,t}\sim\xi_{a_{p}}\left(j_{ifk,t}|\boldsymbol{\theta}_{i},\boldsymbol{\lambda}_{ik,1:t}^{1:N},\widetilde{\boldsymbol{a}}_{ifk,1:t-1}^{1:N},\boldsymbol{\lambda}_{ik,t+1:T}^{\bs j_{i\epsilon s,t+1:T}},\bs j_{ifk,t+1:T}\right)$
\\for $t=T-1,....,1$ and $J_{ifk,T}=j_{ifk,T}\sim\xi_{a_{p}}\left(j_{ifk,T}|\boldsymbol{\theta}_{i},\boldsymbol{\lambda}_{ik,1:T}^{1:N},\widetilde{\boldsymbol{a}}_{ifk,1:T-1}^{1:N}\right)$.
\end{enumerate}
Steps 1 and 2 are discussed in \secref{SS: samplingparamSV}.
Steps 3 and 4 are discussed in \secref{sec:Sampling-factor-loading}.
\end{algorithm}

\section{Markov moves for the Factor SV model\label{S: markov moves for factor SV}}
\subsection{Sampling the latent volatilities using Hamiltonian Monte Carlo \label{SS: Markov moves for HMC}}

We obtain the gradient  $\nabla_{\boldsymbol{h}_{s}}\mathcal{L}\left(\boldsymbol{h}_{s}\right), s=1, \dots, S$
with respect to each of the latent volatilities $\boldsymbol{h}_{s,1:T}$.
The  gradient for $t=1$ is 
\begin{align*}
\nabla_{\boldsymbol{h}_{s1}}\mathcal{L}\left(\boldsymbol{h}_{s}\right) & =a_{p}\left(-0.5+0.5\left(y_{s1}-{\beta}_{s}\boldsymbol{f}_{1}\right)^{2}\exp\left(-h_{s1}\right)\right)-\frac{\left(1-\phi_{\epsilon s}^{2}\right)}{\tau_{\epsilon s}^{2}}\left(h_{s1}-\mu_{\epsilon s}\right)\\
 & +\frac{\phi_{\epsilon s}}{\tau_{\epsilon s}^{2}}\left(h_{s2}-\mu_{\epsilon s}-\phi_{\epsilon s}\left(h_{s1}-\mu_{\epsilon s}\right)\right); 
\end{align*}
the gradient for $1<t<T$ is
\begin{align*}
\nabla_{h_{st}}\mathcal{L}\left(\boldsymbol{h}_{s}\right) & =a_{p}\left(-0.5+0.5\left(y_{st}-{\beta}_{s}\boldsymbol{f}_{t}\right)^{2}\exp\left(-h_{st}\right)\right)+\frac{\phi_{\epsilon s}}{\tau_{\epsilon s}^{2}}\left(h_{st+1}-\mu_{\epsilon s}-\phi_{\epsilon s}\left(h_{st}-\mu_{\epsilon s}\right)\right)-\\
 & \frac{1}{\tau_{\epsilon s}^{2}}\left(h_{st}-\mu_{\epsilon s}-\phi_{\epsilon s}\left(h_{st-1}-\mu_{\epsilon s}\right)\right);
\end{align*}
and the gradient for $t=T$ is
\begin{align*}
\nabla_{h_{sT}}\mathcal{L}\left(\boldsymbol{h}_{s}\right) & =a_{p}\left(-0.5+0.5\left(y_{sT}-{\beta}_{s}\boldsymbol{f}_{T}\right)^{2}\exp\left(-h_{sT}\right)\right)-\frac{1}{\tau_{\epsilon s}^{2}}\left(h_{sT}-\mu_{\epsilon s}-\phi_{\epsilon s}\left(h_{sT-1}-\mu_{\epsilon s}\right)\right).
\end{align*}

We now obtain the gradient $\nabla_{\boldsymbol{\lambda}_{k,1:T}}\mathcal{L}\left(\boldsymbol{\lambda}_{k}\right)$ 
with respect to each of the latent volatilities $\boldsymbol{\lambda}_{k,1:T}, k=1, \dots, K$.
 The gradient  for $t=1$ is 
\[
\nabla_{\boldsymbol{\lambda}_{k1}}\mathcal{L}\left(\boldsymbol{\lambda}_{k}\right)=a_{p}\left(-0.5+0.5f_{k1}^{2}\exp\left(-\lambda_{k1}\right)\right)-\frac{\left(1-\phi_{fk}^{2}\right)}{\tau_{fk}^{2}}\lambda_{k1}+\frac{\phi_{fk}}{\tau_{fk}^{2}}\left(\lambda_{k2}-\phi_{fk}\lambda_{k1}\right);
\]
the gradient for $1<t<T$ is
\begin{align*}
\nabla_{\lambda_{kt}}\mathcal{L}\left(\boldsymbol{\lambda}_{k}\right) & =a_{p}\left(-0.5+0.5f_{kt}^{2}\exp\left(-\lambda_{kt}\right)\right)+\frac{\phi_{fk}}{\tau_{fk}^{2}}\left(\lambda_{kt+1}-\phi_{fk}\lambda_{kt}\right)-\\
 & \frac{1}{\tau_{fk}^{2}}\left(\lambda_{kt}-\phi_{fk}\lambda_{kt-1}\right);
\end{align*}
the gradient for $t=T$ is
\begin{align*}
\nabla_{\lambda_{kT}}\mathcal{L}\left(\boldsymbol{\lambda}_{k}\right) & =a_{p}\left(-0.5+0.5f_{kT}^{2}\exp\left(-\lambda_{kT}\right)\right)-\frac{1}{\tau_{fk}^{2}}\left(\lambda_{kT}-\phi_{fk}\lambda_{kT-1}\right).
\end{align*}

\subsection{Sampling the idiosyncratic and factor log-volatilities parameters \label{SS: samplingparamSV}}
For $s=1,...,S$, sample $\mu_{\epsilon s}$ from $N\left(\mu_{\mu,\epsilon s},\sigma_{\mu,\epsilon s}^{2}\right)$
truncated within $\left(-10,10\right)$, where
\[
\sigma_{\mu,\epsilon s}^{2}=\frac{\tau_{\epsilon s}^{2}}{1-\phi_{\epsilon s}^{2}+\left(T-1\right)\left(1-\phi_{\epsilon s}\right)^{2}}
\]
and
\[
\mu_{\mu,\epsilon s}=\sigma_{\mu,\epsilon s}^{2}\frac{h_{s1}\left(1-\phi_{\epsilon s}^{2}\right)+\sum_{t=2}^{T}h_{st}-\phi_{\epsilon s}h_{st}+\phi_{\epsilon s}^{2}h_{st-1}-\phi_{\epsilon s}h_{st-1}}{\tau_{\epsilon s}^{2}}.
\]
For $s=1,...,S$, we sample $\phi_{\epsilon s}$ by drawing a proposed
value $\phi_{\epsilon s}^{*}$ from $N\left(\mu_{\phi,\epsilon s},\sigma_{\phi,\epsilon s}^{2}\right)$
truncated within $\left(-1,1\right)$, where
\[
\sigma_{\phi,\epsilon s}^{2}=\frac{\tau_{\epsilon s}^{2}}{\sum_{t=2}^{T}\left(h_{st-1}-\mu_{\epsilon s}\right)^{2}-\left(h_{s1}-\mu_{\epsilon s}\right)^{2}}
\]
and
\[
\mu_{\phi,\epsilon s}=\sigma_{\phi,\epsilon s}^{2}\frac{\sum_{t=2}^{T}\left(h_{st}-\mu_{\epsilon s}\right)\left(h_{st-1}-\mu_{\epsilon s}\right)}{\tau_{\epsilon s}^{2}}.
\]
The candidate is accepted with probability
\begin{align*}
\min\left(1,\frac{p\left(\phi_{\epsilon s}^{*}\right)\sqrt{1-\phi_{\epsilon s}^{2*}}}{p\left(\phi_{\epsilon s}\right)\sqrt{1-\phi_{\epsilon s}^{2}}}\right).
\end{align*}

For $s=1,...,S$, we sample $\tau_{\epsilon s}^{2}$ from $\textrm{IG}\left(v_{1,\epsilon s}/2,s_{1,\epsilon s}/2\right)$,
where $v_{1,\epsilon s}=v_{0,\epsilon s}+T$ and $s_{1,\epsilon s}=s_{0,\epsilon s}+\left(1-\phi_{\epsilon s}^{2}\right)\left(h_{s1}-\mu_{\epsilon s}\right)^{2}+\sum_{t=2}^{T}\left(h_{st}-\mu_{\epsilon s}-\phi_{\epsilon s}\left(h_{st-1}-\mu_{\epsilon s}\right)\right)^{2}$.
For $k=1,...,K$, we sample $\phi_{fk}$ by drawing a proposed value
$\phi_{fk}^{*}$ from $N\left(\mu_{\phi,fk},\sigma_{\phi,fk}^{2}\right)$
truncated within $\left(-1,1\right)$, where
\[
\sigma_{\phi,fk}^{2}=\frac{\tau_{fk}^{2}}{\sum_{t=2}^{T-1}\lambda_{kt}^{2}},\quad \text{and} \quad
\mu_{\phi,fk}=\frac{\sum_{t=2}^{T}\lambda_{kt}\lambda_{kt-1}}{\sum_{t=2}^{T-1}\lambda_{kt}^{2}}.
\]
For $k=1,...,K$, we sample $\tau_{fk}^{2}$ from $\textrm{IG}\left(v_{1,fk}/2,s_{1,fk}/2\right)$,
where $v_{1,fk}=v_{0,fk}+T$ and $s_{1,fk}=s_{0,fk}+\left(1-\phi_{fk}^{2}\right)\lambda_{k1}^{2}+\sum_{t=2}^{T}\left(\lambda_{kt}-\phi_{\epsilon s}\lambda_{kt-1}\right)^{2}$.

\section{Sampling the factor loading matrix ${\beta}$ and the latent factors
$\bs f_{1:T}$\label{sec:Sampling-factor-loading}}

This section discusses the parameterisation of the factor loading
matrix and the latent factors, and how to sample from their
full conditional distribution.

To identify the parameters of the  factor loading matrix
${\beta}$, it is necessary to impose some further constraints.
Usually, the factor loading matrix ${\beta}$ is assumed
lower triangular in the sense that $\beta_{sk}=0$ for $k>s$ and
furthermore, one of two constraints is used. i) The first is that the ${f}_{kt}$ have unit variance \citep{Geweke1996}; or, alternatively, 
ii)  assume that $\beta_{ss}=1$, for $s=1,...,S$,
and the variance of $\boldsymbol{f}_{t}$ is diagonal but unconstrained.
The main drawback of the lower triangular assumption on ${\beta}$
is that the resulting inference can depend on the order in which the
components of $\bs y_{t}$ are chosen \citep{Chan:2017}.
However, a unique identification of the loading matrix is unnecessary \citep{Kastner:2019} if the aim is to estimate and predict the covariance structure. This allows the factor loading matrix to be  unrestricted. In our empirical application, we leave the factor loading matrix  unrestricted so that the results are invariant with respect to the ordering of the series.





We now describe how to sample the factor loadings and latent factors from their full conditional distribution based on \citet{Kastner:2017}.
Let $z_{s}$ denote the
number of unrestricted elements in row $s$ and define
\[
F_{s}:=\left[\begin{array}{ccc}
f_{11} & \cdots & f_{z_{s}1}\\
\vdots &  & \vdots\\
f_{1T} & \cdots & f_{z_{s}T}
\end{array}\right],
\]
and
\[
V_{s}:=\left[\begin{array}{ccc}
\exp\left(h_{s1}\right) & \cdots & 0\\
0 & \ddots & \vdots\\
0 & \cdots & \exp\left(h_{sT}\right)
\end{array}\right]. 
\]
Then sample the factor loadings $\boldsymbol{\beta}_{s,.}^{\top}=\left(B_{s1},...,B_{sz_{s}}\right)$
for $s=1,...,S$, conditionally on $\boldsymbol{f}_{1:T}$  independently for each $s$, by performing a Gibbs-update from
\begin{equation}
\boldsymbol{\beta}_{s,.}^{\top}|\boldsymbol{f}_{1:T},\boldsymbol{y}_{s,1:T}\sim N_{z_{s}}\left(\bs{a}_{sT},\bs{b}_{sT}\right),\label{eq:Bfactor-1}
\end{equation}
where
\[
\bs{b}_{sT}=\left(a_{p}\left({F}_{s}^{\top}{V}_{s}^{-1}{F}_{s}\right)+{B}_{0}^{-1}{I}_{z_{s}}\right)^{-1},
\]
and
\[
\bs{a}_{sT}=b_{sT}{F}_{s}^{\top}\left(a_{p}{V}_{s}^{-1}\boldsymbol{y}_{s,1:T}\right).
\]
We now describe how to sample  $\left\{ \bs{f}_{t}\right\} |\bs{y}_{t},\left\{ \bs{h}_{t}\right\} ,\left\{ \bs{\lambda}_{t}\right\} ,\boldsymbol{\beta}$.
After some algebra, we can show that $\left\{ \bs{f}_{t}\right\} $
is sampled from
\begin{equation}
\left\{\bs{f}_{t}\right\} |\bs{y}_{t},\left\{ \bs{h}_{t}\right\}, \left\{ \bs{\lambda}_{t}\right\}, {\beta}\sim N\left(a_{t},b_{t}\right),\label{eq:factordraws}
\end{equation}
where
\[
\bs{b}_{t}=\left(a_{p}\left({\beta}^{\top}{V}_{t}^{-1}{\beta}\right)+{D}_{t}^{-1}\right)^{-1}
\quad 
\text{and} \quad
\bs{a}_{t}=b_{t}{\beta}^{\top}\left(a_{p}{V}_{t}^{-1}\boldsymbol{y}_{t}\right).
\]



\section{Algorithms\label{sec:Particle-Filter-and CPF}}
 Algorithm \ref{alg:Sequential Monte-Carlo-Algorithm-1-1}
is the standard particle filter.  Algorithm \ref{alg:Conditional-Sequential-Monte carlo} is the conditional sequential Monte Carlo used in the SMC-PG Markov move. Algorithm \ref{alg:The-backward simulation algorithm} is the backward simulation used in SMC-PG and SMC-PHS. 

\begin{algorithm}[H]
\caption{Standard Particle Filter Algorithm \label{alg:Sequential Monte-Carlo-Algorithm-1-1}}

Inputs: $\boldsymbol{y}_{1:T}$, $N$, $\boldsymbol{\theta}$

Outputs: $\boldsymbol{x}_{1:T}^{1:N}$, $\boldsymbol{\widetilde{b}}_{1:T-1}^{1:N}$,$\widetilde{\bs w}_{1:T}^{1:N}$
\begin{enumerate}
\item For $t=1$

\begin{enumerate}
\item Sample $\boldsymbol{x}_{1}^{j}$ from $m_{1}\left(\boldsymbol{x}_{1}|\boldsymbol{y}_{1},\boldsymbol{\theta}\right)$,
for $j=1,...,N$
\item Calculate the importance weights
\[
\widetilde{w}_{1}^{j}=\frac{p\left(\boldsymbol{y}_{1}|\boldsymbol{x}_{1}^{j},\boldsymbol{\theta}\right)^{a_{p}}p\left(\boldsymbol{x}_{1}^{j}|\boldsymbol{\theta}\right)}{m_{1}\left(\boldsymbol{x}_{1}^{j}|\boldsymbol{y}_{1},\boldsymbol{\theta}\right)},j=1,...,N.
\]
and normalise to obtain $\widetilde{\bs W}_{1}^{1:N}$.
\end{enumerate}
\item For $t>1$

\begin{enumerate}
\item Sample the ancestral indices $\widetilde{\boldsymbol{b}}_{t-1}^{1:N}\sim\M\left(\widetilde{\boldsymbol{b}}_{t-1}^{1:N}|\widetilde{\bs W}_{t-1}^{1:N}\right)$.
\item Sample $\boldsymbol{x}_{t}^{j}$ from $m_{t}\left(\boldsymbol{x}_{t}|\boldsymbol{x}_{t-1}^{\widetilde{b}_{t-1}^{j}},\boldsymbol{\theta}\right)$,
$j=1,...,N$.
\item Calculate the importance weights
\[
\widetilde{w}_{t}^{j}=\frac{p\left(\boldsymbol{y}_{t}|\boldsymbol{x}_{t}^{j},\boldsymbol{\theta}\right)^{a_{p}}p\left(\boldsymbol{x}_{t}^{j}|\boldsymbol{x}_{t-1}^{\widetilde{b}_{t-1}^{j}},\boldsymbol{\theta}\right)}{m_{t}\left(\boldsymbol{x}_{t}^{j}|\boldsymbol{x}_{t-1}^{\widetilde{b}_{t-1}^{j}},\boldsymbol{\theta}\right)},j=1,...,N.
\]
and normalise to obtain $\widetilde{\bs W}_{t}^{1:N}$.
\end{enumerate}
\end{enumerate}
\end{algorithm}

\begin{algorithm}[H]
\caption{Conditional Sequential Monte Carlo algorithm \label{alg:Conditional-Sequential-Monte carlo}}

Inputs: $N$, $\boldsymbol{\theta}$, $\boldsymbol{y}_{1:T}$, $\boldsymbol{x}_{1:T}^{\bs j_{1:T}}$,
and $\bs j_{1:T}$

Outputs: $\boldsymbol{x}_{1:T}^{1:N}$, $\boldsymbol{\widetilde{b}}_{1:T-1}^{1:N}$,$\widetilde{\bs w}_{1:T}^{1:N}$
\begin{enumerate}
\item For $t=1$

\begin{enumerate}
\item Sample $\boldsymbol{x}_{1}^{j}$ from $m_{1}\left(\boldsymbol{x}_{1}|\boldsymbol{y}_{1},\boldsymbol{\theta}\right)$,
for $j\in\left\{ 1,...,N\right\} \setminus\left\{ j_{1}\right\} $.
\item Calculate the weights
\[
\widetilde{w}_{1}^{j}=\frac{p\left(\boldsymbol{y}_{1}|\boldsymbol{x}_{1}^{j},\boldsymbol{\theta}\right)^{a_{p}}p\left(\boldsymbol{x}_{1}^{j}|\boldsymbol{\theta}\right)}{m_{1}\left(\boldsymbol{x}_{1}^{j}|\boldsymbol{y}_{1},\boldsymbol{\theta}\right)},j=1,...,N.
\]
and normalise to obtain $\widetilde{\bs W}_{1}^{1:N}$.
\end{enumerate}
\item For $t>1$

\begin{enumerate}
\item Sample the ancestral indices $\widetilde{\boldsymbol{b}}_{t-1}^{\left(-j_{t}\right)}\sim\M\left(\widetilde{\boldsymbol{b}}_{t-1}^{\left(-j_{t}\right)}|\widetilde{\bs W}_{t-1}^{1:N}\right)$.
\item Sample $\boldsymbol{x}_{t}^{j}$ from $m_{t}\left(\boldsymbol{x}_{t}|\boldsymbol{x}_{t-1}^{\widetilde{b}_{t-1}^{j}},\boldsymbol{\theta}\right)$,
$j=1,...,N\setminus\left\{ j_{t}\right\} $.
\item Calculate the weights
\[
\widetilde{w}_{t}^{j}=\frac{p\left(\boldsymbol{y}_{t}|\boldsymbol{x}_{t}^{j},\boldsymbol{\theta}\right)^{a_{p}}p\left(\boldsymbol{x}_{t}^{j}|\boldsymbol{x}_{t-1}^{\widetilde{b}_{t-1}^{j}},\boldsymbol{\theta}\right)}{m_{t}\left(\boldsymbol{x}_{t}^{j}|\boldsymbol{x}_{t-1}^{\widetilde{b}_{t-1}^{j}},\boldsymbol{\theta}\right)},j=1,...,N.
\]
and normalise to obtain $\widetilde{\bs W}_{t}^{1:N}$.
\end{enumerate}
\end{enumerate}
\end{algorithm}


\begin{algorithm}[H]
\caption{Backward simulation \label{alg:The-backward simulation algorithm} }
\begin{enumerate}
\item Sample $J_{T}=j_{T},$ conditional on $\left(\bs U_{1:T},\bs\theta\right)$,
with probability proportional to $w_{T}^{j_{T}}$, and choose $x_{T}^{j_{T}}$;
\item For $t=T-1,...,1$, sample $J_{t}=j_{t},$ conditional on $\left(\bs u_{1:t},\bs j_{t+1:T},x_{t+1}^{j_{t+1}},...,x_{T}^{j_{T}}\right)$,
and with probability proportional to $w_{t}^{j_{t}}p\left(x_{t+1}^{j_{t+1}}|\bs\theta,x_{t}^{j_{t}}\right)$,
and choose $x_{t}^{j_{t}}$.
\end{enumerate}
\end{algorithm}

\subsection{Constrained Conditional Sequential Monte Carlo \label{sub:Conditional-Sequential-Monte Carlo constrained}}
This section discusses constrained conditional sequential Monte Carlo (CCSMC)  (Algorithm~\ref{alg:The-conditional Sequential-Monte carlo algorithm}),
which is used in Part 4 of SMC-PHS (Algorithm~\ref{alg:Sampling-Scheme:-The correlated PMMH+PG}).

Let $\bs{v}$ be the pseudo-random vector used in the particle filter; $\bs{v}$ has two
components $\bs{v}^{1:N}_{x,1:T}$ and $\bs{v}^{1:N}_{B,1:T-1}$. 
Let $v_{x,t}^{j}$ be the vector
random variable used to generate the particles $x_{t}^{j}$ given
$\bs\theta$ and $x_{t-1}^{\wt{b}_{t-1}^{j}}$, where $\wt{b}_{t-1}^{i}$ is the ancestor index of $x_{t}^{j}$. The distribution of $v_{x,t}^{i}$ is $\N(0,1)$. For $t\geq2$, let $v_{B,t-1}$ be the vector of random variables used to generate the ancestor
indices $\wt{\bs{b}}_{t-1}^{1:N}$  using
the resampling scheme $\mathcal{M}\left(\wt{\bs{b}}_{t-1}^{1:N}|\bs{\wt{W}}_{t-1}^{1:N},\bs{x}_{t-1}^{1:N}\right)$ and  the distribution of $v_{B,t-1}$ is $U(0,1)$.

CCSMC is a sequential Monte Carlo algorithm in which a particle
trajectory $\bs{x}_{1:T}^{j_{1:T}}=\left(x_{1}^{j_{1}},...,x_{T}^{j_{T}}\right)$
and the associated sequence of indices $j_{1:T}$ are kept unchanged, which means that some elements of $\bs{v}_{x,1:T}^{1:N}$ and $\bs{v}_{B,1:T-1}$
are constrained.  It is a constrained version of the conditional SMC sampler in Algorithm \ref{alg:Conditional-Sequential-Monte carlo}. \citet{Gunawan2020PHS} give further details.

The CCSMC algorithm takes the number of particles $N$, the parameters $\bs\theta$, and the reference trajectory $\bs{x}_{1:T}^{j_{1:T}}$ as the input.
It produces the set of particles $\bs{x}_{1:T}^{1:N}$, ancestor indices $\bs{\widetilde{b}}_{1:T-1}^{1:N}$, and weights $\bs{\widetilde{W}}_{1:T}^{1:N}$.
It also produces the random variables used to propagate state particles $\bs{v}_{x,1:T}^{1:N}$ and the random numbers used in the resampling steps $\bs{v}_{B,1:T-1}$.

At $t=1$, Step (1a) samples the basic random numbers $v_{x1}^{i}$ from the standard normal distribution $N(0,1)$ for $i=1,...,N$ and obtains the set of particles $x_{1}^{1:N}$, except for the reference particles $x_{1}^{j_{1}}$. We obtain the basic random number  $v_{x1}^{j_{1}}$ associated with the reference particle  $x_{1}^{j_{1}}$ in Step (1b) using 
\begin{align*}
\quad v_{x1}^{j_{1}} & =\bigg ( \left ( 1-\phi^{2}\right)/\tau^{2}\bigg)^{\frac12}  \left(x_{1}^{j_{1}}-\mu\right) 
\end{align*}
for the univariate SV model  using the bootstrap filter and compute the weights of all particles in step (1c).  

Step (2a) sorts the particles from smallest to largest using the Euclidean sorting procedure of \citet{Choppala2016} and obtains the sorted particles and weights.
The particles are then resampled using constrained multinomial resampling  (Algorithm \ref{alg:Multinomial-Resampling-Algorithm for CCSMC})
and the ancestor index $\bs{\widetilde{b}}_{1:T-1}^{1:N}$ obtained in the original order of the particles in step (2b).
Step (2c) samples the basic random numbers $v_{xt}^{i}$ from the standard normal distribution $N(0,1)$ for $i=1,...,N$ and obtains the set of particles $\bs{x}_{t}^{1:N}$, except for the reference particles $x_{t}^{j_{t}}$.
We obtain the basic random number  $v_{xt}^{j_{t}}$ associated with the reference particle  $x_{t}^{j_{t}}$ in Step (2d) using
\begin{align*}
\quad  v_{xt}^{j_{t}} & = \frac{ \bigg ( x_{t}^{j_{t}}-\mu-\phi\left(x_{t-1}^{\widetilde{b}_{t-1}^{j_{t}}}-
\mu\right)\bigg)}   { \bigg (\tau^{2}\bigg )^{\frac12}}
\end{align*}
for the univariate SV model  using the bootstrap filter and compute the weights of all the particles in step (2e). 

\begin{algorithm}[]
\caption{The constrained conditional sequential Monte Carlo algorithm \label{alg:The-conditional Sequential-Monte carlo algorithm} }

Inputs: $N$, $\bs\theta$, $\bs{x}_{1:T}^{j_{1:T}}$, and $\bs{j}_{1:T}$

Outputs: $\bs{x}_{1:T}^{1:N}$, $\bs{\widetilde{b}}_{1:T-1}^{1:N}$, $\bs{\widetilde{W}}_{1:T}^{1:N}$, $\bs{v}_{x,1:T}^{1:N}$, and $\bs{v}_{B,1:T-1}$

Fix $\bs{x}_{1:T}^{\bs{j}_{1:T}}$, $\bs{\widetilde{b}}_{1:T-1}^{J}=\bs{j}_{1:T-1}$,
and $J_{T}=j_{T}$.
\begin{enumerate}
\item For $t=1$

\begin{enumerate}
\item Sample $v_{x1}^{i}\sim N(0,1)$ and set $x_1^{i} = m_{1}\left(\cdot;v_{x1}^{i}, \bs\theta\right)$
for $i=1,...,N\setminus\left\{ j_{1}\right\}$.
\item Obtain $v_{x1}^{j_{1}}$ such that $x_{1}^{j_{1}}=m_{1}\left(\cdot;v_{x1}^{j_{1}},\bs\theta\right)$.
\item Compute the importance  weights $\widetilde{w}_{1}^{i}=\frac{p\left(x_{1}^{i}|\bs\theta\right)p\left(y_{1}|x_{1}^{i},\bs\theta\right)}{m_{1}\left(x_{1}^{i}|v_{x1}^{i},\bs\theta\right)}$,
for $i=1,...,N$,
and normalize to obtain $\bs{\widetilde{W}}_{1}^{1:N}$.
\end{enumerate}

\item For $t\geq2$

\begin{enumerate}
\item Sort the particles $x_{t-1}^{i}$ using the Euclidean sort of \citet{Choppala2016} and obtain the sorted index $\zeta_{i}$ for $i=1,...,N,$ and the sorted particles and weights $\wt x_{\textrm{sorted},t-1}^i = x_{t-1}^{\zeta_i} $ and $\wt {W}^i_{\textrm{sorted},t-1} =  \wt{W}_{t-1}^{\zeta_i}$, for $i=1, \dots, N$.

\item Use a  constrained sampling algorithm, for example the constrained multinomial sampler
(Algorithm~\ref{alg:Multinomial-Resampling-Algorithm for CCSMC}),
\begin{enumerate} \item Generate the random variables used in the resampling step $\bs{v}_{B,t-1}^{1:N}$ and obtain the ancestor indices based on the sorted particles $\bs{\widetilde{b}}_{\textrm{sorted},t-1}^{1:N\setminus\left(j_{t}\right)}$
\item Obtain the ancestor indices in the original order of the particles $\bs{\wt{b}}_{t-1}^{1:N}$.
\end{enumerate}
\item Sample $v_{xt}^{i}\sim N(0,1)$ for $i=1,...,N\setminus\left\{ j_{t}\right\} $
and obtain $v_{xt}^{j_{t}}$ such that $x_{t}^{j_{t}}=m_{t}(\cdot;v_{xt}^{j_{t}},\bs\theta,x_{t-1}^{\wt{b}_{t-1}^{j_{t}}})$
\item Set $x_{t}^{i}=m_{t}\left(\cdot|v_{xt}^{i},\bs\theta,x_{t-1}^{\wt{b}_{t-1}^{i}}\right)$
for $i=1,...,N\setminus\left\{ j_{t}\right\} $
\item Compute the importance weights,
\begin{align*}
    \wt{w}_{t}^{i} & =\frac{p\left(x_{t}^{i}|x_{t-1}^{\wt{b}_{t-1}^{i}},\bs{\theta}\right)p
   \left(y_{t}|x_{t}^{i},\bs{\theta}\right)}{m_{t}\left(x_{t}^{i}|x_{t-1}^{\wt{b}_{t-1}^{i}},\bs\theta,v_{xt}^{i}\right)},
\quad \text{for} \quad i=1,...,N,
\end{align*}
and  normalize to obtain $\bs{\widetilde{W}}_{t}^{1:N}$.
\end{enumerate}
\end{enumerate}
\end{algorithm}

\newpage
Algorithm \ref{alg:Multinomial-Resampling-Algorithm for CCSMC} describes constrained multinomial resampling  used in  CCSMC.
It takes the sorted particles $\bs{\widetilde{x}}_{\textrm{sorted},t-1}^{1:N}$ and weights $\bs{\widetilde{W}}_{\textrm{sorted},t-1}^{1:N}$ as the inputs and produces the basic random numbers $\bs{v}_{B,1:T-1}^{1:N}$
and the ancestor indices based on the sorted particles and weights $\bs{\widetilde{b}}_{\textrm{sorted},t-1}^{1:N}$.
The first step computes the cumulative weights based on the sorted particles; the second step generates the random numbers $\bs{v}_{B,1:T-1}^{1:N}$ and ancestor indices $\bs{\widetilde{b}}_{\textrm{sorted},t-1}^{1:N}$. 

\begin{algorithm}[H]
\caption{Constrained Multinomial Resampling Algorithm for CCSMC\label{alg:Multinomial-Resampling-Algorithm for CCSMC}}

Input: $\bs{\widetilde{x}}_{\textrm{sorted},t-1}^{1:N}$, and $\bs{\widetilde{W}}_{\textrm{sorted},t-1}^{1:N}$

Output: $\bs{v}_{B,1:T-1}^{1:N}$ and $\bs{\widetilde{b}}_{t-1}^{1:N}$.
\begin{enumerate}
\item Compute the cumulative weights based on the sorted particles $\left\{\bs{\widetilde{x}}_{\textrm{sorted},t-1}^{1:N},\bs{\widetilde{W}}_{\textrm{sorted},t-1}^{1:N}\right\} $,
\[
\widehat{F}_{t-1}^{N}\left(j\right)=\sum_{i=1}^{j}\bs{\widetilde{W}}_{\textrm{sorted},t-1}^{i}.
\]
\item
Generate $N-1$ uniform $(0,1)$ random numbers $v_{Bt-1}^i\sim U(0,1)$ 
for $i=1,...,N$, such that $i\neq j_{t}$, and set $\widetilde{b}_{\textrm{sorted},t-1}^{i}=\underset{j}{\min}\,\, \widehat{F}_{t-1}^{N}\left(j\right)\geq v_{Bt-1}^{i}$.
For $i=j_{t}$,
\begin{align*}
v_{Bt-1}^{j_{t}}\sim U\left(\widehat{F}_{t-1}^{N}\left(j_{t}-1\right),\widehat{F}_{t-1}^{N}\left(j_{t}\right)\right),
\quad
\text{where} \\
\widehat{F}_{t-1}^{N}\left(j_{t}-1\right)=\sum_{i=1}^{j_{t}-1}\bs{\widetilde{W}}_{\textrm{sorted}t-1}^{i}\quad
\text{and} \quad
\widehat{F}_{t-1}^{N}\left(j_{t}\right)=\sum_{i=1}^{j_{t}}\bs{\widetilde{W}}_{\textrm{sorted},t-1}^{i}.
\end{align*}
\end{enumerate}
\end{algorithm}

\subsection{The particle filter using a Euclidean sort\label{SS: SMC algorithms22}}
We now discuss the  particle filter  (Algorithm~\ref{alg:The-correlated particle filter algorithm}) using a Euclidean sort,
which is used in Part 1 of the SMC-PHS (Algorithm~\ref{alg:Sampling-Scheme:-The correlated PMMH+PG}).

This particle filter  takes the number of particles $N$, the parameters $\bs\theta$, the random variables used to propagate state particles $\bs{v}_{x,1:T}^{1:N}$,
and the random numbers used in the resampling steps $\bs{v}_{B,1:T-1}$ as the inputs; it outputs the set of particles $\bs{x}_{1:T}^{1:N}$, ancestor indices $\bs{\wt{b}}_{1:T-1}^{1:N}$, and weights $\bs{\wt{W}}_{1:T}^{1:N}$. At $t=1$,
we obtain the particles $\bs{x}_{1}^{1:N}$ as a function of the basic random numbers $\bs{v}_{x1}^{1:N}$;  the weights of all particles are computed in step (2). 


Step (3a) sorts the particles from smallest to largest using the  Euclidean sorting procedure of \citet{Choppala2016} to obtain the sorted particles and weights.
Algorithm \ref{alg:Multinomial-Resampling-Algorithm} resamples the particles using multinomial sampling 
to obtain the ancestor index $\bs{\wt{b}}_{1:T-1}^{1:N}$ in the original order of the particles in Steps (3b) and (3c).
Steps (3d) generates the particles $\bs{x}_{t}^{1:N}$ as a function of the basic random numbers $\bs{v}_{xt}^{1:N}$ 
and then computes the weights of all particles in step (3e).

\begin{algorithm}[]
\caption{The particle filter with Euclidean sort \label{alg:The-correlated particle filter algorithm}}
Inputs: $N, \bs\theta$, $\bs{v}_{x,1:T}^{1:N}$ and $\bs{v}_{B,1:T-1}$.\\  
Outputs: $\bs{x}_{1:T}^{1:N}, \bs{\wt{b}}_{1:T-1}^{1:N},\bs{\wt{W}}_{1:T}^{1:N}$.
\begin{enumerate}
\item
For $t=1$,  set $x_1^{i} = m_{1}\left(\cdot;v_{x1}^{i}, \bs\theta\right)$
for $i=1,...,N$.
\item Compute the importance  weights $\widetilde{w}_{1}^{i}=\frac{p\left(x_{1}^{i}|\bs\theta\right)p\left(y_{1}|x_{1}^{i},\bs\theta\right)}{m_{1}\left(x_{1}^{i}|v_{x1}^{i},\bs\theta\right)}$,
for $i=1,...,N$,
and normalize to obtain $\bs{\widetilde{W}}_{1}^{1:N}$.

\item

For $t = 2, \dots, T$,
\begin{enumerate}
\item Sort the particles $x_{t-1}^{i}$ using the Euclidean sorting procedure of \citet{Choppala2016} and obtain the sorted index $\zeta_{i}$ for $i=1,...,N$ and the sorted particles and weights $\wt x_{\textrm{sorted},t-1}^i = x_{t-1}^{\zeta_i} $ and $\wt {W}^i_{\textrm{sorted},t-1} =  \wt{W}_{t-1}^{\zeta_i}$, for $i=1, \dots, N$.

\item Obtain the ancestor index based on the sorted particles $ \bs{\wt{b}}_{\textrm{sorted},t-1}^{1:N}$ using   multinomial resampling  (Algorithm~\ref{alg:Multinomial-Resampling-Algorithm}).

\item Obtain the ancestor indices based on the original order of the particles $\bs{\wt{b}}_{t-1}^{i}$ for $i=1,...,N$.

\item  Set $x_t^{i} = m_{t}\left(\cdot;v_{xt}^{i}, \bs\theta, \bs{x}_{t-1}^{\wt{b}_{t-1}^i}\right)$
for $i=1,...,N$.

\item    Compute the importance weights,
\begin{align*}
    \wt{w}_{t}^{i} & =\frac{p\left(x_{t}^{i}|x_{t-1}^{\wt{b}_{t-1}^{i}},\bs{\theta}\right)p
   \left(y_{t}|x_{t}^{i},\bs{\theta}\right)}{m_{t}\left(x_{t}^{i}|x_{t-1}^{\wt{b}_{t-1}^{i}},\bs\theta,v_{xt}^{i}\right)},
\quad \text{for} \quad i=1,...,N,
\end{align*}
and  normalize to obtain $\bs{\widetilde{W}}_{t}^{1:N}$.
\end{enumerate}
\end{enumerate}
\end{algorithm}



\begin{algorithm}[]
\caption{Multinomial Resampling Algorithm \label{alg:Multinomial-Resampling-Algorithm}}

Input: $\bs{v}_{Bt-1}$, $\bs{\widetilde{x}}_{\textrm{sorted},t-1}^{1:N}$, and $\bs{\widetilde{{W}}}_{\textrm{sorted},t-1}^{1:N}$

Output: $\bs{\widetilde{b}}_{\textrm{sorted},t-1}^{1:N}$
\begin{enumerate}
\item Compute the cumulative weights based on the sorted particles $\left\{ \bs{\widetilde{x}}_{\textrm{sorted},t-1}^{1:N},\bs{\widetilde{{W}}}_{\textrm{sorted},t-1}^{1:N}\right\} $
\[
\widehat{F}_{t-1}^{N}\left(j\right)=\sum_{i=1}^{j}\bs{\widetilde{{W}}}_{t-1}^{i}.
\]
\item Set $\widetilde{b}_{\textrm{sorted},t-1}^{i}=\underset{j}{\min}\,\, \widehat{F}_{t-1}^{N}\left(j\right)\geq v_{Bt-1}^{i}$
for $i=1,...N$, and note that $\widetilde{b}_{\textrm{sorted},t-1}^{i}$ for $i=1,...,N$
is the ancestor index based on the sorted particles.
\end{enumerate}
\end{algorithm}

\section{Assumptions\label{sec:Assumptions}}

 \secref{sec:Particle-Filter-and CPF}
uses the particle filter to approximate the joint filtering densities
$\left\{ p\left(\bs{x}_{t}|\boldsymbol{y}_{1:t}\right):t=1,...,T\right\} $
sequentially using $N$ particles, $\left\{ \boldsymbol{x}_{t}^{1:N},\widetilde{W}_{t}^{1:N}\right\} $,
drawn from some proposal densities $m_{1}\left(\boldsymbol{x}_{1}|\boldsymbol{y}_{1},\boldsymbol{\theta}\right)$
and $m_{t}\left(\boldsymbol{x}_{t}|\boldsymbol{x}_{t-1},\boldsymbol{y}_{1:t},\boldsymbol{\theta}\right)$
for $t\geq2$. For $t\geq1$, we follow \citep{Andrieu:2010} and define
\begin{align*}
S_{t}^{\boldsymbol{\theta}} &  \coloneqq\left(\boldsymbol{x}_{1:t}\in\boldsymbol{\chi}^{t}:\pi\left(\boldsymbol{x}_{1:t}|\boldsymbol{\theta}\right)>0\right)  \textrm{and} \\ Q_{t}^{\boldsymbol{\theta}}& \coloneqq\left\{ \boldsymbol{x}_{1:t}\in\boldsymbol{\chi}^{t}:\pi
\left(\boldsymbol{x}_{1:t-1}|\boldsymbol{\theta}\right)m_{t}\left(\boldsymbol{x}_{t}|
\boldsymbol{\theta},\boldsymbol{x}_{1:t-1},\boldsymbol{y}_{1:t}\right)>0\right\} .
\end{align*}
\begin{assumption}
\citep{Andrieu:2010}  $S_{t}^{\boldsymbol{\theta}}\subseteq Q_{t}^{\boldsymbol{\theta}}$
for any $\boldsymbol{\theta}\in\boldsymbol{\Theta}$ and $t=1,...,T$
\label{assu:propstatespace}
\end{assumption}
Assumption \ref{assu:propstatespace} is always satisfied in our implementation
because we use the bootstrap filter with $p\left(\boldsymbol{x}_{t}|\boldsymbol{x}_{t-1},\boldsymbol{\theta}\right)$
as a proposal density, and $p\left(\boldsymbol{y}_{t}|\boldsymbol{x}_{t},\boldsymbol{\theta}\right)>0$
for all $\boldsymbol{\theta}$.
\begin{assumption}\label{assu:resampling}
For any $j=1,...,N$ and $t=1,..,T$, the resampling scheme $\M\left(\widetilde{\boldsymbol{b}}_{t-1}^{1:N}|\widetilde{\bs W}_{t-1}^{1:N}\right)$
satisfies $\Pr\left(\widetilde{B}_{t-1}^{k}=j|\widetilde{\bs W}_{t-1}^{1:N}\right)=\widetilde{W}_{t-1}^{j}$.
\citep{Chopin2015,Andrieu:2010}.
\end{assumption}
Assumption 2 is satisfied by the popular resampling schemes, such
as the multinomial, systematic, and residual resampling.

\section{Bayesian Forecasting \label{subsec:BayesianForecasting}}

Let $Y_{T+1}$ be the  unobserved value of the dependent variable at time
$T+1$. Given the joint posterior distribution of all the parameters
$\bs\theta$ and latent state variables $\bs{x}_{1:T}$
of the multivariate factor SV model up to time $T$,
the forecast density of $Y_{T+1}=y_{T+1}$ that takes into account the uncertainty about $\bs\theta$ and $\bs{x}_{1:T}$ is
\begin{equation}
p\left(y_{T+1}|y_{1:T}\right)=\int p\left(y_{T+1}|\bs\theta,x_{T+1}\right)p\left(x_{T+1}|x_{T},\bs\theta\right)p\left(\bs\theta,\bs{x}_{1:T}|\bs{y}_{1:T}\right)d\bs\theta d\bs{x}_{1:T}; \label{eq:predictivedenseqn}
\end{equation}
$p\left(\bs{\theta},\bs{x}_{1:T}|\bs{y}_{1:T}\right)$ is the exact posterior
density that can be estimated  from MCMC or particle MCMC.
The draws from MCMC or particle MCMC can be used to produce the simulation-consistent
estimate of this predictive density
\begin{equation}
\widehat{p\left(y_{T+1}|\bs{y}_{1:T}\right)}=\frac{1}{M}\sum_{m=1}^{M}
p\left(y_{T+1}|\bs{y}_{1:T},\bs{\theta}^{\left(m\right)},x_{T+1}^{\left(m\right)}\right);\label{eq:preddensRB}
\end{equation}
it is necessary to know the conditional density
$p\left(y_{T+1}|\bs{y}_{1:T,}\bs\theta^{\left(m\right)},x_{T+1}^{\left(m\right)}\right)$
in closed form to obtain the estimates in \eqref{eq:preddensRB}.  
We can alternatively obtain $M$ draws
of $y_{T+1}$ from $p\left(y_{T+1}|\bs{y}_{1:T,}\bs{\theta}^{\left(m\right)},x_{T+1}^{\left(m\right)}\right)$
that can be used to obtain the kernel density estimate $\widehat{p\left(y_{T+1}|y_{1:T}\right)}$
of $p\left(y_{T+1}|\bs{y}_{1:T}\right)$.
Assuming that the MCMC converged, these are two simulation-consistent estimates
of the exact  Bayesian  predictive density. Similarly, it is also straightforward to obtain multiple-step ahead predictive densities
$p\left(y_{T+h}|y_{1:T}\right)$, for $h>1$.


\section{List of Industry Portfolios\label{sec:List-of-Industry}}

\begin{table}[H]
\caption{The list of industry portfolios\label{tab:The-list-of industry portfolios}}
\centering{}%
\begin{tabular}{cc}
\hline
 & Stocks\tabularnewline
\hline
1 &  Coal \tabularnewline
2 &  Health Care and Equipment\tabularnewline
3 &  Retail\tabularnewline
4 & Tobacco\tabularnewline
5 & Steel Works\tabularnewline
6 & Food Products\tabularnewline
7 & Recreation\tabularnewline
8 & Printing and Publishing\tabularnewline
9 & Consumer Goods\tabularnewline
10 & Apparel\tabularnewline
11 & Chemicals\tabularnewline
12 & Textiles\tabularnewline
13 & Fabricated Products\tabularnewline
14 & Electrical Equipment\tabularnewline
15 & Automobiles and Trucks\tabularnewline
16 & Aircraft, ships, and Railroad Equipment\tabularnewline
17 & Industrial Mining\tabularnewline
18 & Petroleum and Natural Gas\tabularnewline
19 & Utilities\tabularnewline
20 & Telecommunication\tabularnewline
21 & Personal and Business Services\tabularnewline
22 & Business Equipment\tabularnewline
23 & Transportation\tabularnewline
24 & Wholesale\tabularnewline
25 & Restaurants, Hotels, and Motels\tabularnewline
26 & Banking, Insurance, Real Estate\tabularnewline
\hline
\end{tabular}
\end{table}


\section{Additional Empirical Results for the Factor SV model \label{additionalresultfactorsvmodel}}

This section provides some additional empirical results for the factor SV model. 
As in Section \ref{univariateexample}, for SMC-PG,
we denote SMC-PG-batch-seq for a combination
of batch and sequential estimation. 
For the SMC-PG-batch-seq method, batch estimation is used for the first 975 observations and sequential estimation is used for the last
25 observations. The sequential part of SMC-PG-batch-seq is done using tempering. 

\begin{figure}[H]
\begin{centering}
\caption{Factor SV model: Sequential one-step ahead predictive density estimates for the equally weighted
portfolio of US industry stock return from 25/10/2005 to 29/11/2005 estimated using SMC-PG-batch-seq
\label{fig:One-step-ahead-predictive factorSV}}
 \includegraphics[width=15cm,height=8cm]{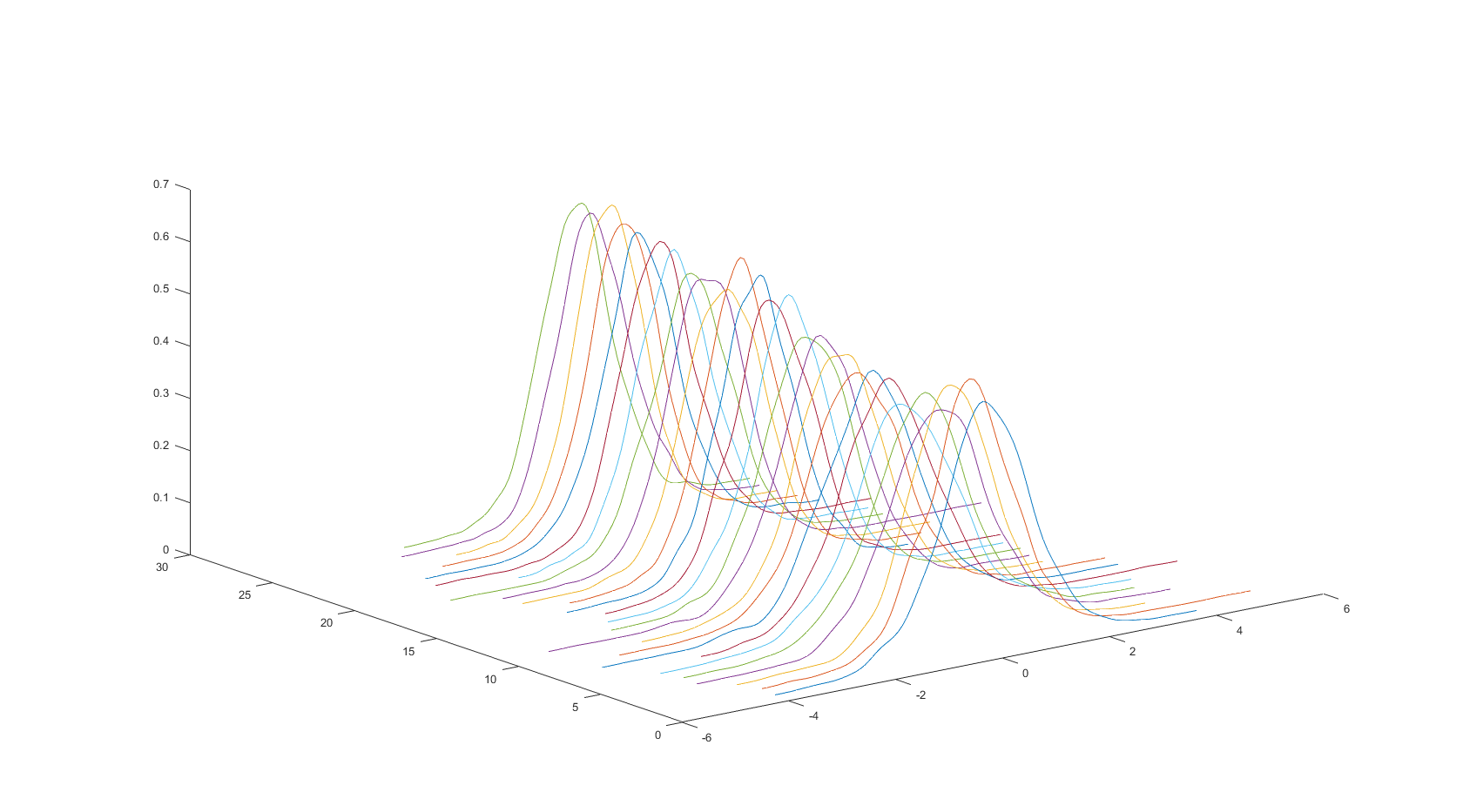}
\par\end{centering}
\end{figure}

As in the univariate case, it is also possible to obtain sequential
one-step ahead predictive densities of the portfolio return at time
$t+1$ given the information up to time $t$. Suppose, for example,
that we have an equally weighted portfolio.
Figure \ref{fig:One-step-ahead-predictive factorSV} shows the sequential
one-step ahead predictive densities of the equally weighted portfolio of US industry
stock returns from 25/10/2005 to 29/11/2005 estimated using SMC-PG-batch-seq. This demonstrates the usefulness of the SMC-PG-batch-seq to obtain sequential one-step ahead predictive densities of the portfolio return which is expensive to obtain using MCMC methods. 



Finally,  we use the log of the marginal likelihood estimates  to select the number of factors. We estimate factor models with 1-6 factors using the SMC-PG method with an unrestricted factor loading matrix $\beta$. 
Table~\ref{tab:Selecting-number-of factors}  gives  the logs of the marginal likelihoods for the six models, with standard errors in brackets, obtained with $10$ independent runs for each factor model. The standard errors are reasonably small  relative to the absolute values of the logs of the marginal likelihood estimates, indicating that the logs of the marginal likelihoods are estimated  accurately enough to discriminate between models. The log of the marginal likelihood increases appreciably  from $k=1$ to $k=3$ factors, then the improvement with  additional factors  quickly flattens out. It is therefore reasonable to say that the three-factor or four-factor models are the best.




\begin{table}[H]

\caption{Factor SV model: Selecting the number of factors using estimates of the  log of the marginal likelihood
(with standard errors in brackets) obtained by SMC-PG. \label{tab:Selecting-number-of factors}}

\begin{centering}
\begin{tabular}{cc}
\hline
$K$ & $\log\widehat{p}\left(\bs y\right)$\tabularnewline
\hline
1 & $-26149.66$ (30.29)\tabularnewline
2 & $-25569.64$ (41.53)\tabularnewline
3 & $-25289.10$ (34.20)\tabularnewline
4 & $-25231.49$ (38.74)\tabularnewline
5 & $-25247.06$ (47.93)\tabularnewline
6 & $-25228.52$ (48.79)\tabularnewline
\hline
\end{tabular}
\par\end{centering}
\end{table}

\begin{figure}[H]
\caption{The marginal posterior density plots of the parameters $\tau^2_{\epsilon,s}$ for $s=1,...,S$ estimated using PMCMC (PHS) and MCMC of \citet{Kastner:2017} methods. \label{tau2FactorSVall}}

\centering{}\includegraphics[width=15cm,height=10cm]{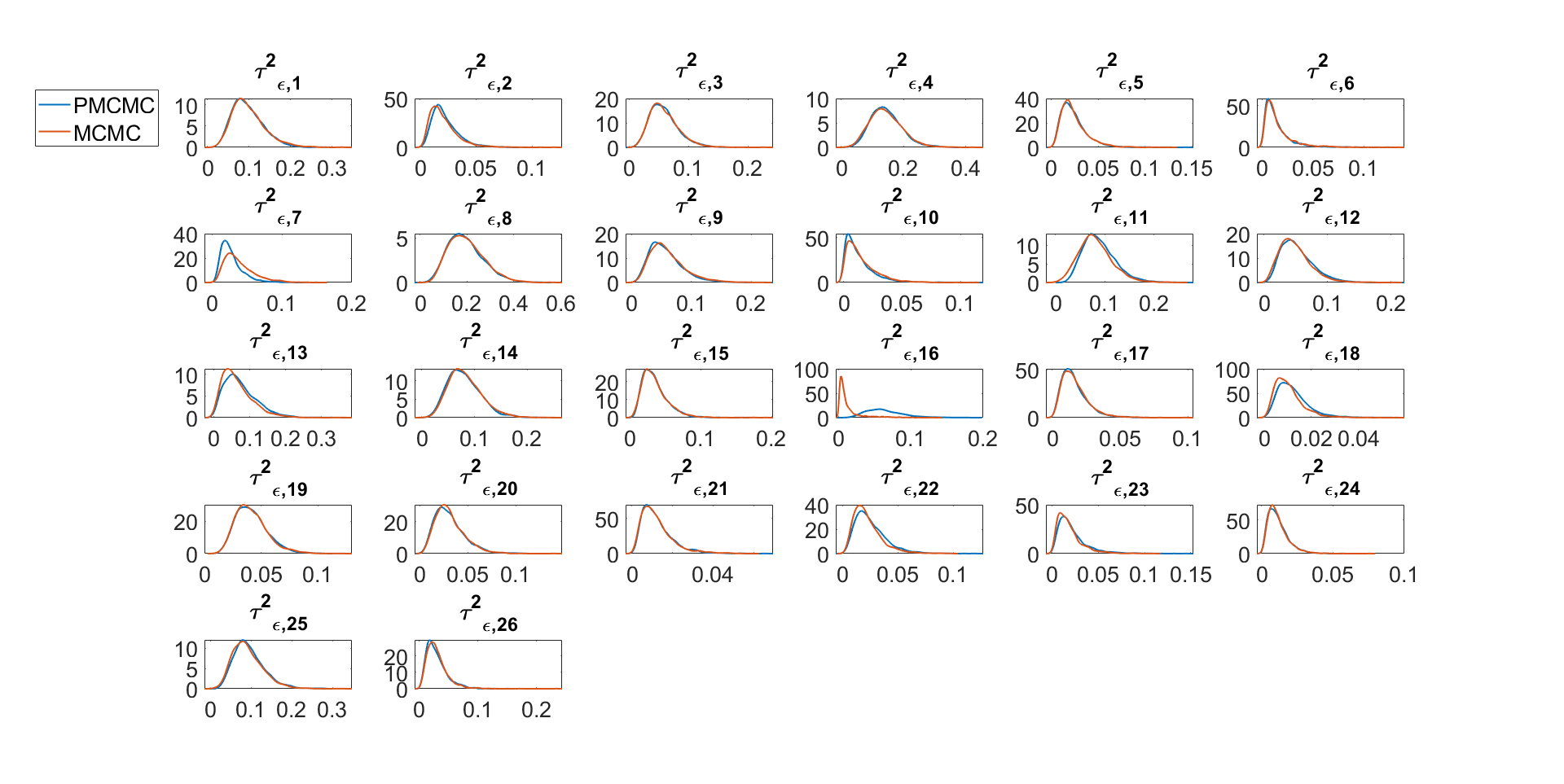}
\end{figure}

\begin{figure}[H]
\caption{The marginal posterior density plots of the parameters $\mu_{\epsilon,s}$ for $s=1,...,S$ estimated using PMCMC (PHS) and MCMC of \citet{Kastner:2017} methods. \label{muFactorSVall}}

\centering{}\includegraphics[width=15cm,height=10cm]{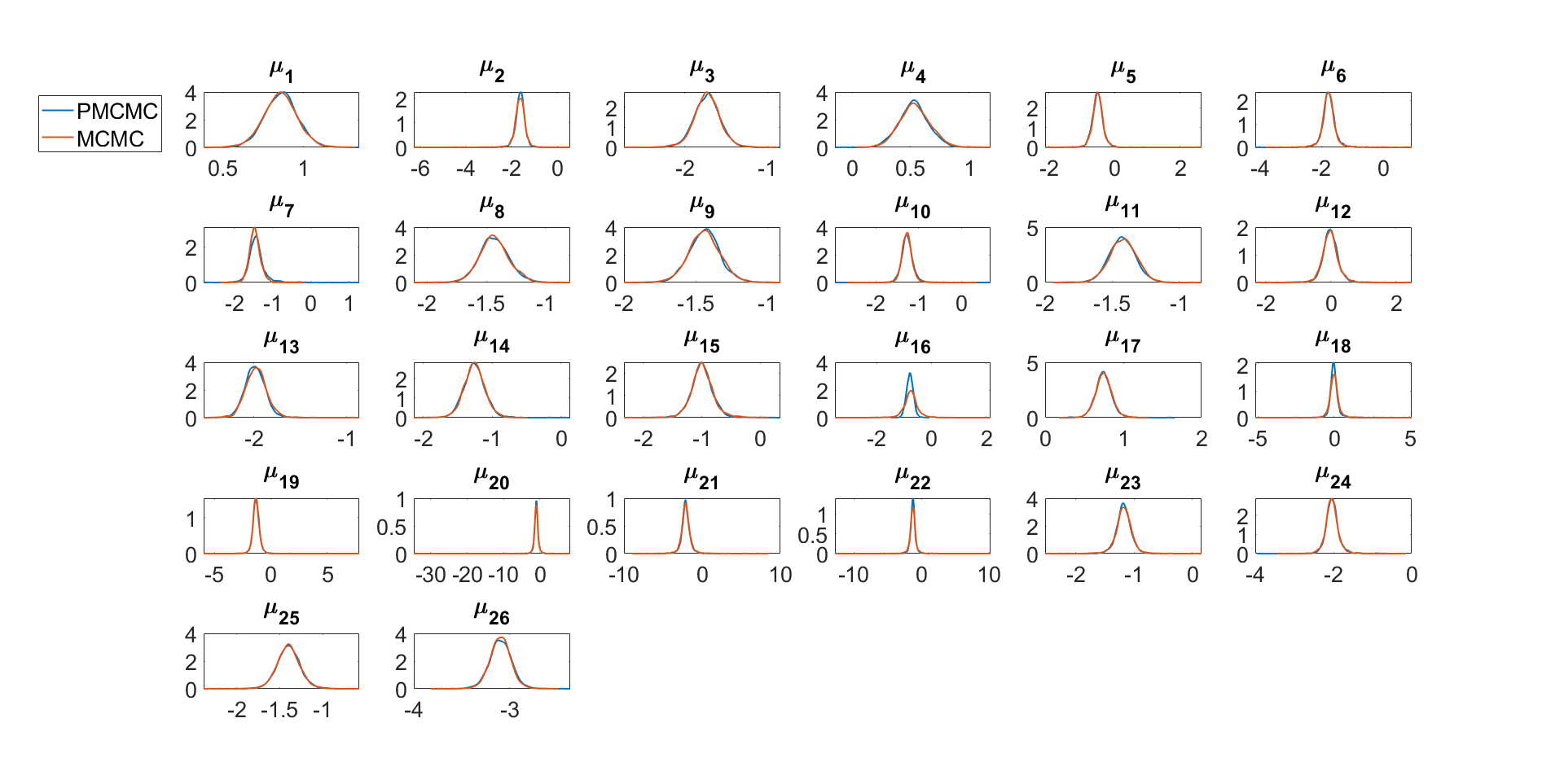}
\end{figure}

\begin{figure}[H]
\caption{The marginal posterior density plots of the parameters $\phi_{\epsilon,s}$ for $s=1,...,S$ estimated using PMCMC (PHS) and MCMC of \citet{Kastner:2017} methods. \label{phiFactorSVall}}

\centering{}\includegraphics[width=15cm,height=10cm]{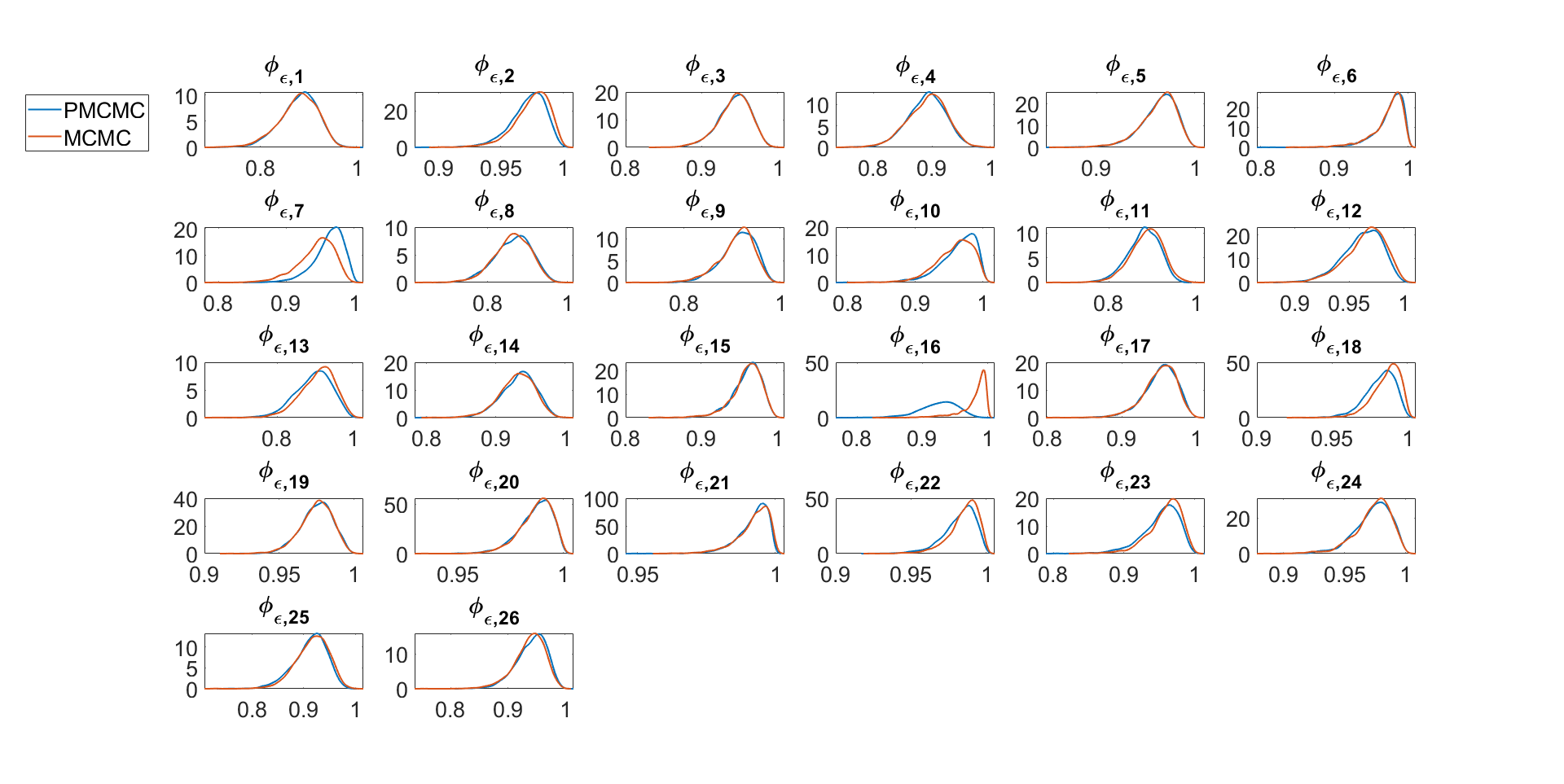}
\end{figure}

\begin{figure}[H]
\caption{The marginal posterior density plots of the parameters $\beta$ estimated using PMCMC (PHS) and MCMC of \citet{Kastner:2017} methods. \label{B1FactorSVall}}

\centering{}\includegraphics[width=15cm,height=10cm]{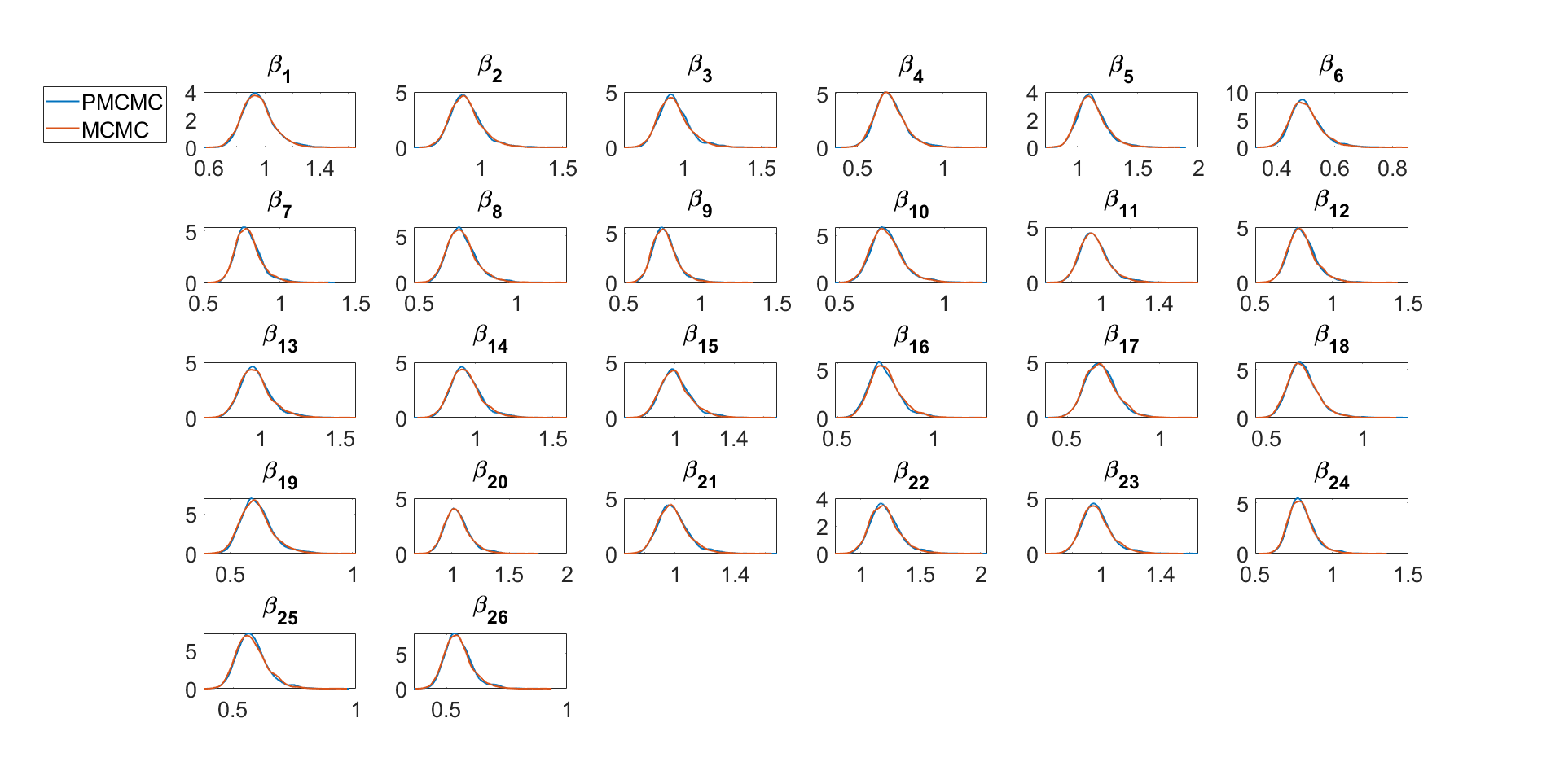}
\end{figure}

Section \ref{subsec:Multivariate-Factor-Stochastic examples-1}  compares  PHS and  the approximate MCMC sampler of \citet{Kastner:2017} for estimating the factor SV model.    Figures~\ref{tau2FactorSVall} to \ref{B1FactorSVall} give results for the parameters $\tau^{2}_{\epsilon,s}$, $\mu_{\epsilon,s}$, and $\phi_{\epsilon,s}$, for $s=1,...,S$, and the factor loading matrix $\beta$ of the factor SV model. The figures show that although some  parameters estimated using the approximate MCMC sampler are close to PHS, other parameters are quite different. The approximate MCMC sampler can be unreliable when estimating the parameters of the factor SV model.     

\begin{figure}[H]
\caption{Factor SV model: The marginal posterior density plots of the parameters $\tau^2_{\epsilon,s}$ for $s=1,...,S$ estimated using PMCMC (PHS), SMC-PG, and SMC-PHS methods. \label{tau2FactorSVallSMCPG}}

\centering{}\includegraphics[width=15cm,height=10cm]{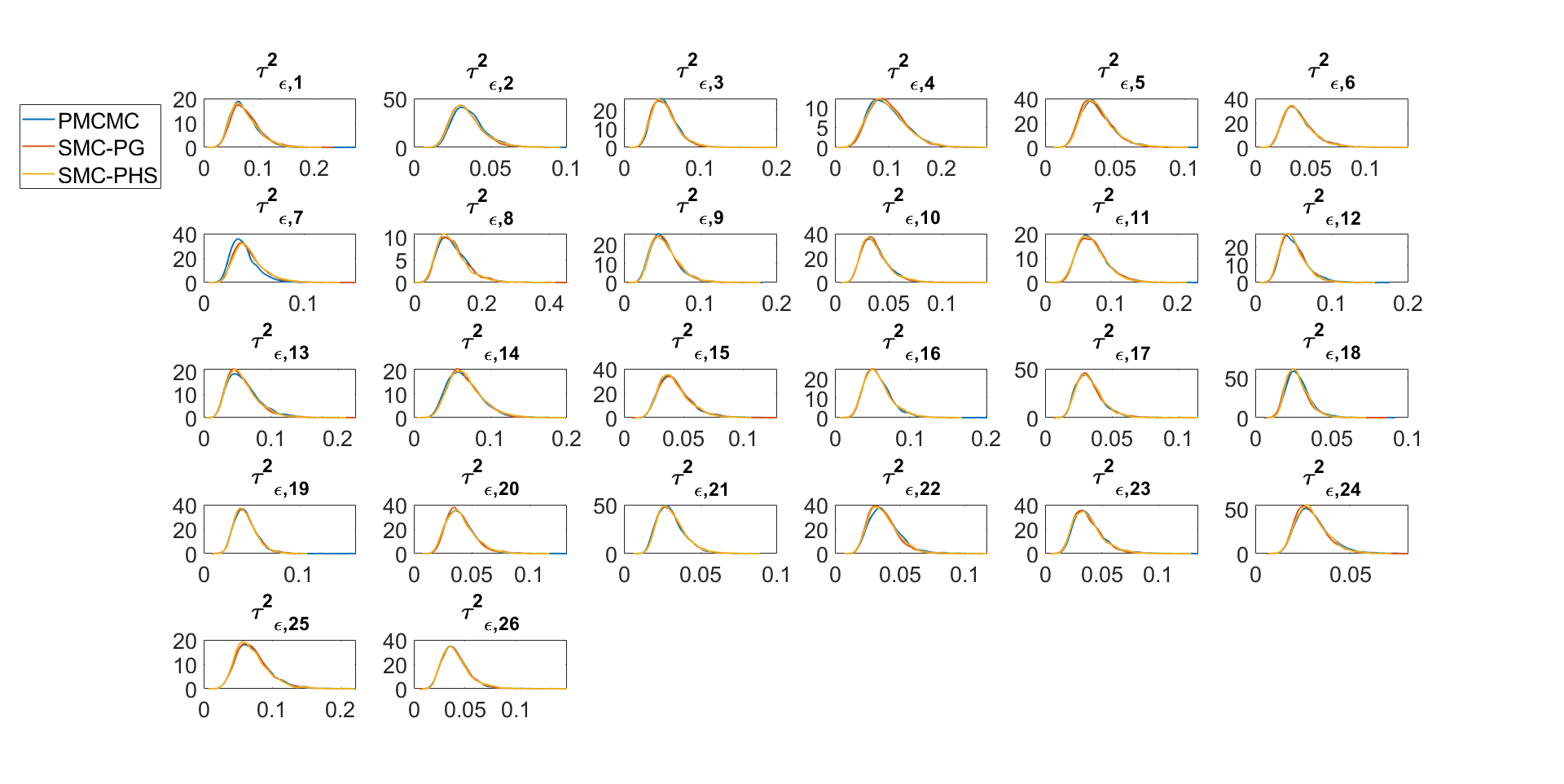}
\end{figure}

\begin{figure}[H]
\caption{The marginal posterior density plots of the parameters $\mu_{\epsilon,s}$ for $s=1,...,S$ estimated using PMCMC (PHS), SMC-PG, and SMC-PHS methods. \label{muFactorSVallSMCPG}}

\centering{}\includegraphics[width=15cm,height=10cm]{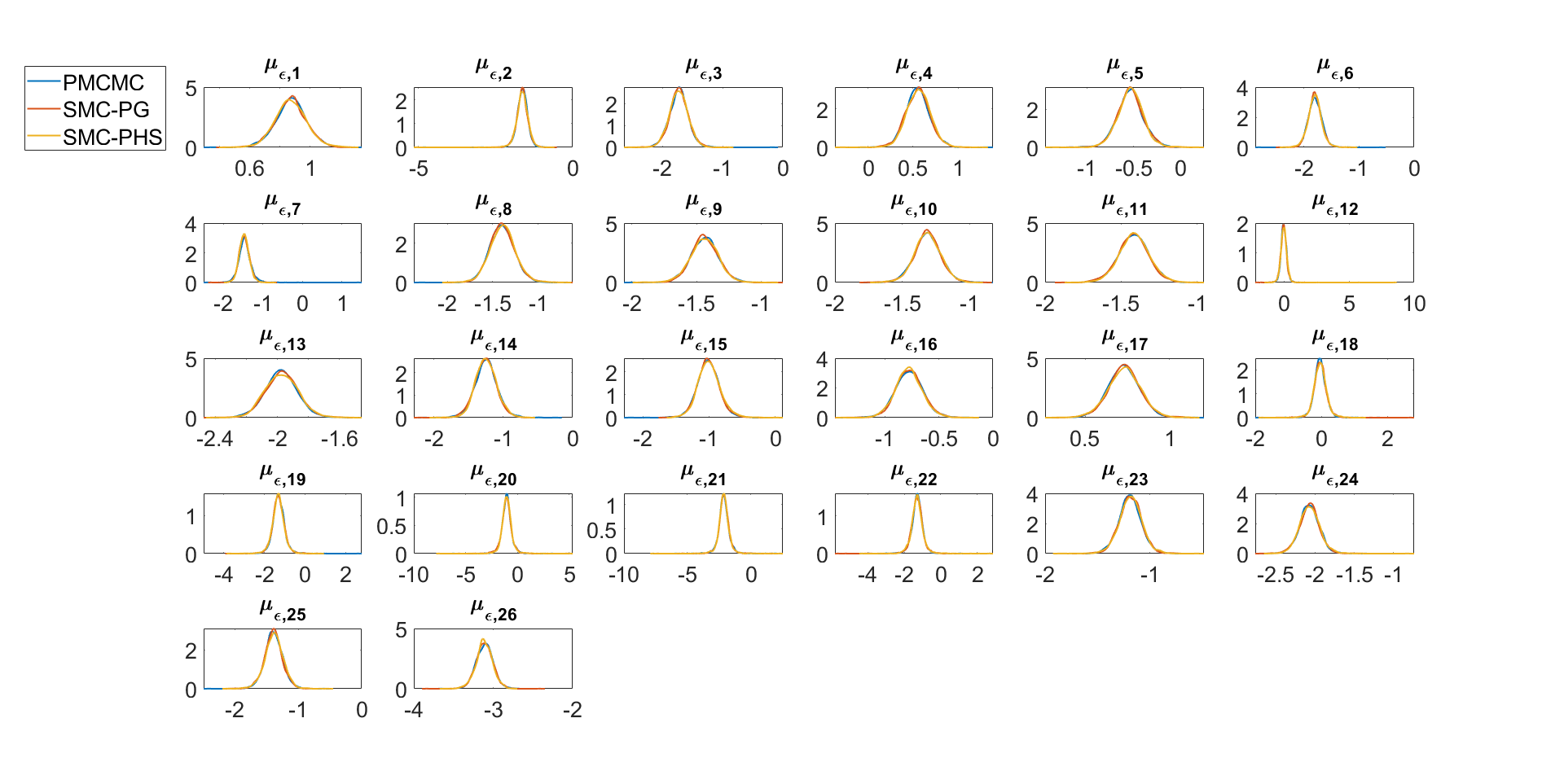}
\end{figure}

\begin{figure}[H]
\caption{The marginal posterior density plots of the parameters $\phi_{\epsilon,s}$ for $s=1,...,S$ estimated using PMCMC (PHS), SMC-PG, and SMC-PHS methods. \label{phiFactorSVallSMCPG}}

\centering{}\includegraphics[width=15cm,height=10cm]{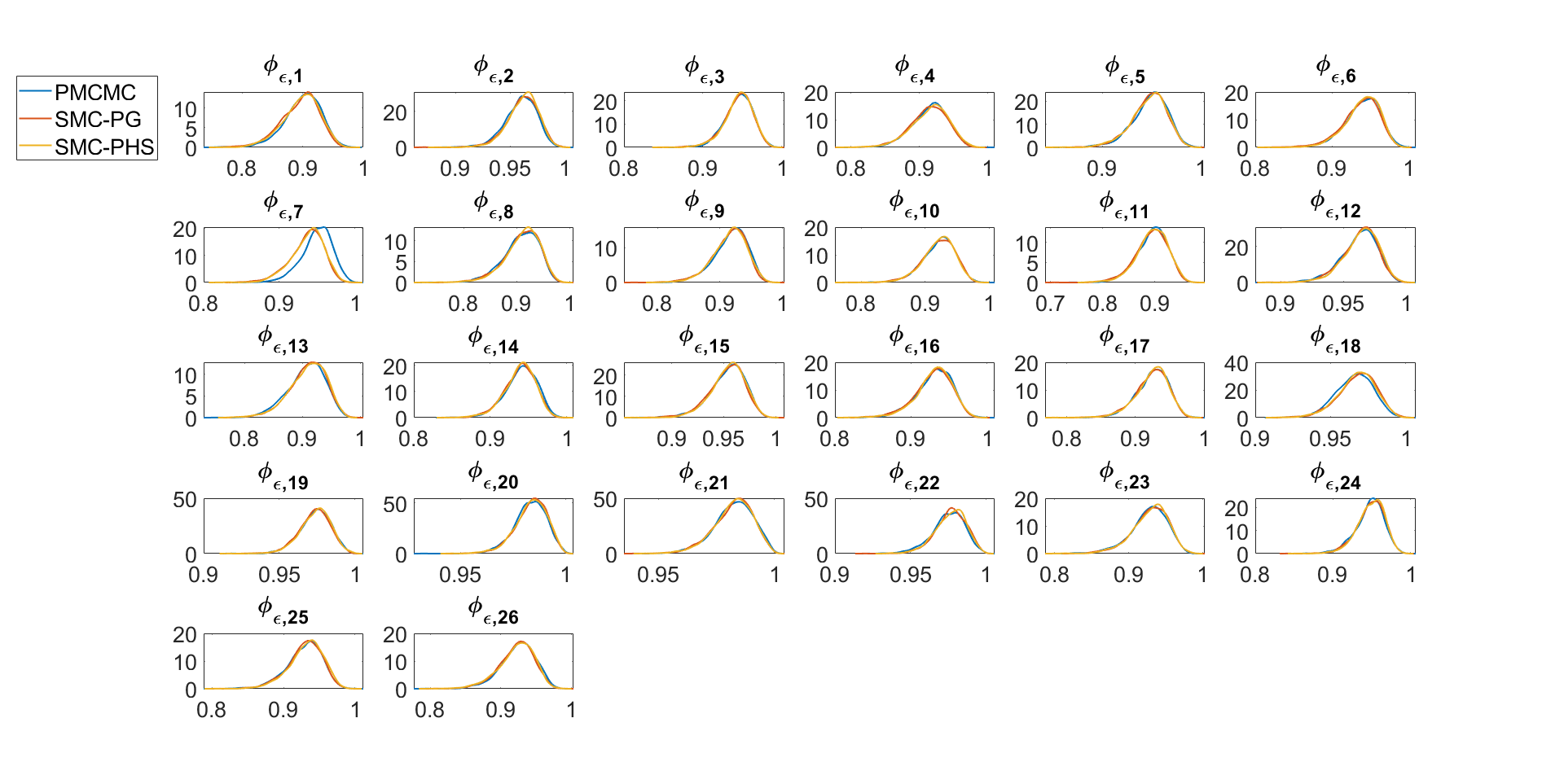}
\end{figure}

\begin{figure}[H]
\caption{The marginal posterior density plots of the parameters $\beta$ estimated using PMCMC (PHS), SMC-PG and SMC-PHS methods. \label{B1FactorSVallSMCPG}}

\centering{}\includegraphics[width=15cm,height=10cm]{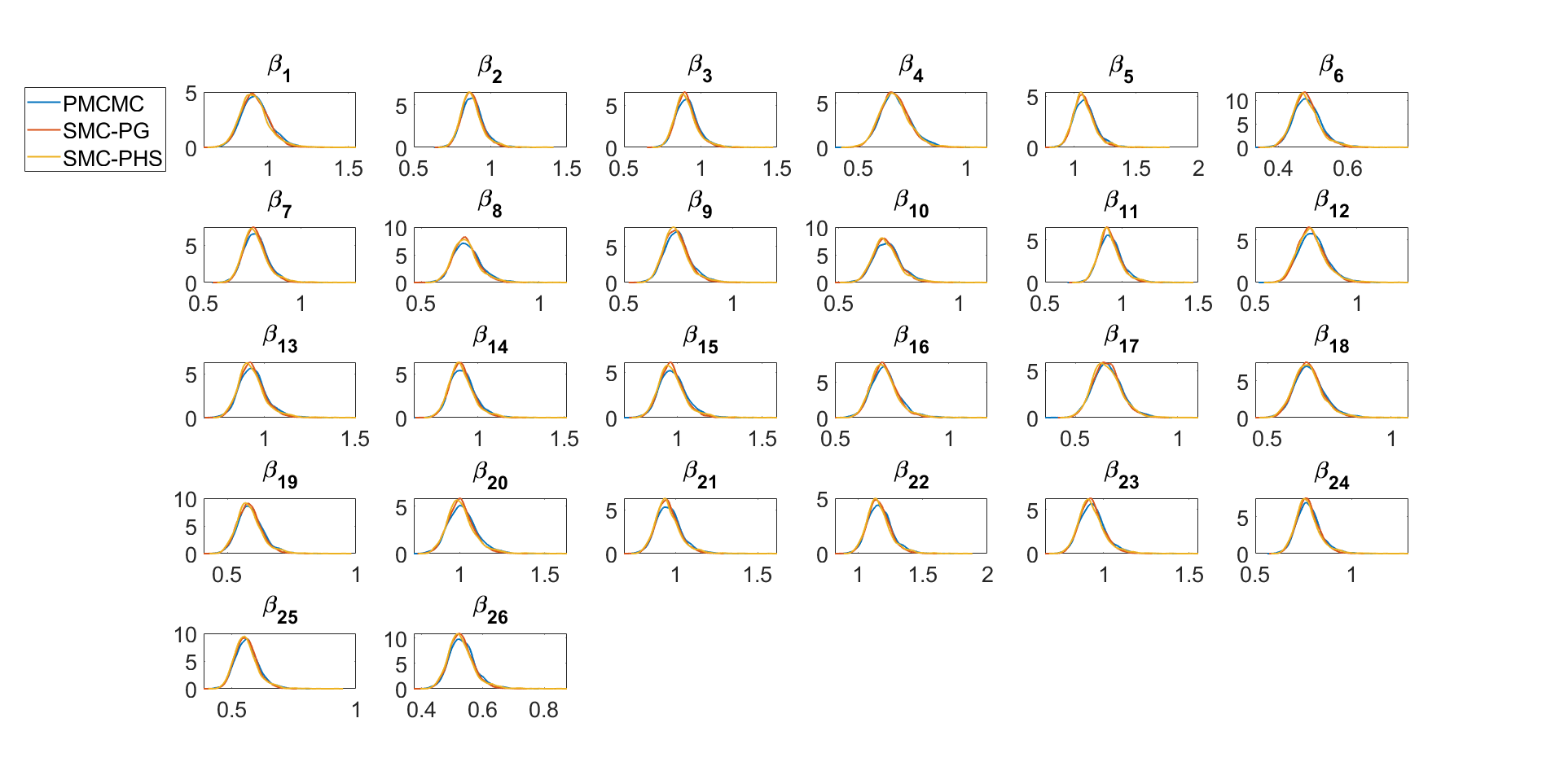}
\end{figure}

Section \ref{subsec:Multivariate-Factor-Stochastic examples-1} also investigates the performance of SMC-PG and SMC-PHS and compares them to PHS for estimating the factor SV model. 
Figures~\ref{tau2FactorSVallSMCPG} to \ref{B1FactorSVallSMCPG} show the estimates of $\tau^{2}_{\epsilon,s}$, $\mu_{\epsilon,s}$, and $\phi_{\epsilon,s}$, for $s=1,...,S$, and the factor loading matrix $\beta$ of the factor SV model. In general, the SMC-PG and SMC-PHS estimates are very close to the PHS estimates for all parameters, suggesting that SMC-PG is as accurate as SMC-PHS and faster for estimating the standard factor SV model.

\begin{figure}[H]
\caption{The marginal posterior density plots of the  $\bs{\alpha}_{\epsilon}$ of the factor SV model, where the idiosyncratic errors follow OU processed estimated using PHS, PG, SMC-PG, and SMC-PHS. \label{aFactordiffusion}}

\centering{}\includegraphics[width=15cm,height=8cm]{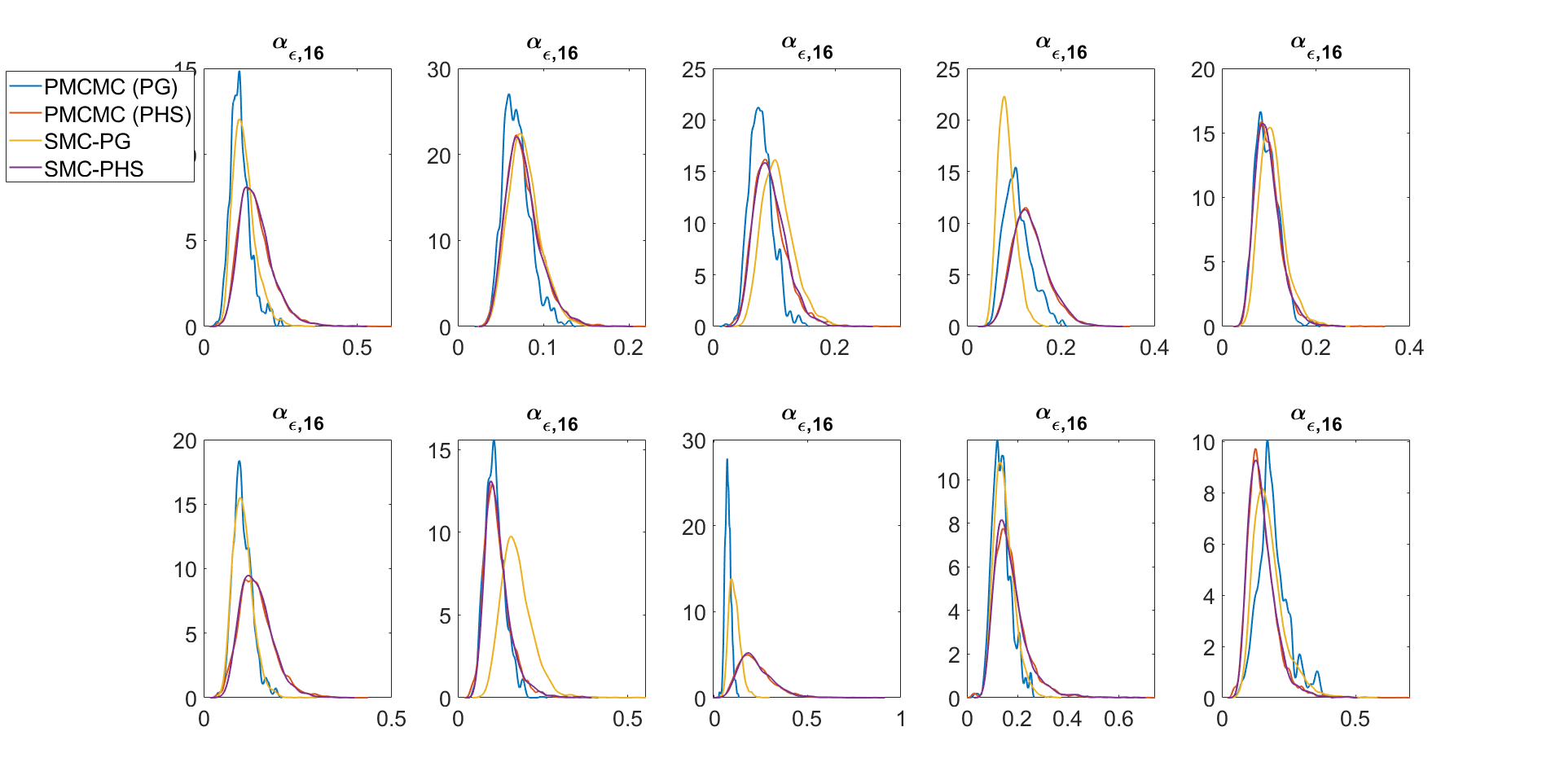}
\end{figure}


\bibliographystyle{apalike}
\bibliography{references_v1}
\end{document}